%% file: bootes_hba_c24_v6_final.tex
\documentclass[fleqn,usenatbib]{mnras}
\usepackage{times}
\usepackage{mathptmx}

\usepackage{graphicx}	
\usepackage{amsmath}	
\usepackage{amssymb}	

\usepackage{color}

\newcommand{\degree}{\hbox{$^{\circ}$}}
\renewcommand{\arcsec}{\hbox{arcsec}}
\renewcommand{\arcmin}{\hbox{arcmin}}

\newcommand\dd{\hbox{$^{\romn d}$}}
\newcommand\hh{\hbox{$^{\romn h}$}}
\newcommand\mm{\hbox{$^{\romn m}$}}
\newcommand\s{\hbox{$^{\romn s}$}}

\newcommand\muJybeam{\hbox{$\umu$Jy\,beam$^{-1}$}}
\newcommand\mJy{\hbox{mJy}}
\newcommand\mJybeam{\hbox{mJy\,beam$^{-1}$}}

\title[LOFAR HBA Bo\"{o}tes]{LOFAR 150-MHz observations of the Bo\"otes field: Catalogue and Source Counts}

\author[Williams et al.]{ 
{\parbox{\textwidth}{
W.~L.~Williams\thanks{E-mail: w.williams5@herts.ac.uk (WLW)}$^{1,2,3}$, 
R.~J.~van~Weeren\thanks{Clay Fellow}$^{4}$, 
H.~J.~A.~R\"ottgering$^{1}$,
P.~Best$^{5}$,   
T.~J.~Dijkema$^{2}$, 
F.~de~Gasperin$^{6}$, 
M.~J.~Hardcastle$^{3}$, 
G.~Heald$^{2,7}$, 
I.~Prandoni$^{8}$, 
J.~Sabater$^{5}$, 
T.~W.~Shimwell$^{1}$, 
C.~Tasse$^{9}$, 
I.~M.~van~Bemmel$^{10}$,
M.~Br\"uggen$^{6}$, 
G.~Brunetti$^{11}$,
J.~E.~Conway$^{12}$, 
T.~En{\ss}lin$^{13}$,
D.~Engels$^{6}$,
H.~Falcke$^{14,2}$,
C.~Ferrari$^{15}$,
M.~Haverkorn$^{14,1}$,
N.~Jackson$^{16}$,
M.~J.~Jarvis$^{17,18}$,
A.~D.~Kapi\'{n}ska$^{19,20,21}$,
E.~K.~Mahony$^{2}$,
G.~K.~Miley$^{1}$, 
L.~K.~Morabito$^{1}$, 
R.~Morganti$^{2,7}$,
E.~Orr\'u$^{2}$, 
E.~Retana-Montenegro$^{1}$, 
S.~S.~Sridhar$^{2,7}$, 
M.~C.~Toribio$^{1}$,
G.~J.~White$^{22,23}$, 
M.~W.~Wise$^{2,24}$, 
J.~T.~L.~Zwart$^{18,25}$
} 
}
\vspace{0.4cm}
\\
{ 
\parbox{\textwidth}{
$^{1}$Leiden Observatory, Leiden University, P.O. Box 9513, 2300 RA Leiden, The Netherlands\\
$^{2}$ASTRON, the Netherlands Institute for Radio Astronomy, Postbus 2, 7990 AA, Dwingeloo, The Netherlands\\
$^{3}$School of Physics, Astronomy and Mathematics, University of Hertfordshire, College Lane, Hatfield AL10 9AB, UK\\
$^{4}$Harvard-Smithsonian Center for Astrophysics, 60 Garden Street, Cambridge, MA 02138, USA\\
$^{5}$SUPA, Institute for Astronomy, Royal Observatory, Blackford Hill, Edinburgh, EH9 3HJ, UK\\
$^{6}$Sternwarte Hamburg, University of Hamburg, Gojenbergsweg 112, 21029 Hamburg, Germany\\
$^{7}$Kapteyn Astronomical Institute, University of Groningen, PO Box 800, 9700AV Groningen, The Netherlands\\
$^{8}$INAF - Observatory of Radioastronomy, Via P. Gobetti 101, 40129 Bologna, Italy\\
$^{9}$GEPI, Observatoire de Paris, CNRS, Universite Paris Diderot, 5 place Jules Janssen, F-92190 Meudon, France\\
$^{10}$Joint Institute for VLBI ERIC, PO Box 2, 7990 AA Dwingeloo, The Netherlands\\
$^{11}$INAF-Institute of Radioastronomy, Via P. Gobetti 101, 40129 Bologna, Italy\\
$^{12}$Department of Earth and Space Sciences, Chalmers University of Technology, Onsala Space Observatory, 439 92 Onsala, Sweden\\
$^{13}$Max Planck Institute for Astrophysics, Karl-Schwarzschild-Str. 1, 85741 Garching, Germany\\
$^{14}$Department of Astrophysics/IMAPP, Radboud University, Nijmegen, PO Box 9010, 6500 GL Nijmegen, the Netherlands\\
$^{15}$Laboratoire Lagrange, Universit\'{e} Côte d'Azur, Observatoire de la C\^{o}te d'Azur, CNRS, Blvd de l'Observatoire, CS 34229, 06304 Nice cedex 4, France\\
$^{16}$Jodrell Bank Centre for Astrophysics, School of Physics and Astronomy, University of Manchester, Oxford Road, Manchester M13 9PL, UK\\
$^{17}$Oxford Astrophysics, Department of Physics, Keble Road, Oxford, OX1 3RH, UK\\
$^{18}$Physics Department, University of the Western Cape, Bellville 7535, South Africa\\
$^{19}$International Centre for Radio Astronomy Research (ICRAR), The University of Western Australia, 35 Stirling Hwy, Crawley 6009, Western Australia\\
$^{20}$Institute of Cosmology and Gravitation, University of Portsmouth, Portsmouth, PO1 3FX, United Kingdom\\
$^{21}$ARC Centre of Excellence for All-sky Astrophysics (CAASTRO)\\
$^{22}$Department of Physical Sciences, The Open University, Milton Keynes MK7 6AA, England\\
$^{23}$RAL Space, The Rutherford Appleton Laboratory, Chilton, Didcot, Oxfordshire OX11 0NL, England\\
$^{24}$Astronomical Institute `Anton Pannekoek', University of Amsterdam, Postbus 94249, 1090 GE Amsterdam, The Netherlands\\
$^{25}$Department of Astronomy, University of Cape Town, Rondebosch 7701, South Africa\\
}} 
}  

\date{Accepted XXX. Received YYY; in original form ZZZ}

\pubyear{2015}

\defcitealias{2016ApJS..223....2V}{vW16}
\defcitealias{2012MNRAS.423L..30S}{SH12}

\begin{document}
\label{firstpage}
\pagerange{\pageref{firstpage}--\pageref{lastpage}}

\maketitle
\clearpage

\begin{abstract}
We present the first wide area ($19$\,deg$^2$), deep ($\approx120$--$150$\,{\muJybeam}), high resolution ($5.6 \times 7.4$\,{\arcsec}) LOFAR High Band Antenna image of the Bo\"otes field made at $130$-$169$\,MHz.  This image is at least an order of magnitude deeper and $3-5$ times higher in angular resolution than previously achieved for this field at low frequencies. The observations and data reduction, which includes full direction-dependent calibration, are described here. We present a radio source catalogue containing $6\,276$\ sources detected over an area of $19$\,deg$^2$,  with a peak flux density threshold of $5\sigma$. As the first thorough test of the facet calibration strategy, introduced by \citeauthor{2016ApJS..223....2V}, we investigate the flux and positional accuracy of the catalogue. We present differential source counts that reach an order of magnitude deeper in flux density than previously achieved at these low frequencies, and show flattening at $150$\,MHz flux densities below  $10$\,mJy associated with the rise of the low flux density star-forming galaxies and radio-quiet AGN.
\end{abstract}


\section{Introduction}
\label{sect:intro}

The LOw Frequency ARray (LOFAR) is a new generation radio telescope operating at $10$--$240$\,MHz \citep{2013A&A...556A...2V}. Its large instantaneous field of view, combined with multi-beaming capabilities, high-spatial resolution, and large fractional bandwidth make LOFAR an ideal instrument for carrying out large surveys of the sky which will have long-lasting legacy value. As such, `Surveys' is one of the six LOFAR Key Science Projects (KSP). The science goals of the Surveys KSP are broad, covering aspects from  the formation and evolution of large-scale structure of the Universe; the physics of the origin, evolution and end-stages of radio sources;  the magnetic field and interstellar medium in nearby galaxies and galaxy clusters; to Galactic sources. The deep LOFAR surveys will be crucial in the study of AGN evolution and the history of black-hole accretion. In particular, the Surveys KSP  aims to answer questions related to the nature of the different accretion processes, the properties of the host galaxies, the role of AGN feedback in galaxy growth and evolution, the radio-source duty cycle and the relation of the AGN with their environment \citep[e.g.][and references therein]{2014ARA&A..52..589H}. The radio-source population has not been well-studied at low flux densities and low frequencies. To achieve the diverse goals of the LOFAR surveys, which will be carried out over the next five years, a tiered approach is being used: Tier-1 covers the largest area at the lowest sensitivity, and will include low-band (LBA; $15$--$65$\,MHz) and high-band (HBA; $110$--$180$\,MHz) observations across the whole 2$\pi$ steradians of the northern sky with a targeted rms noise of $\approx0.1$\,{\mJybeam} and a resolution of $\approx5\,{\arcsec}$. Deeper Tier-2 and Tier-3 observations will cover smaller areas, focussing on fields with the highest quality multi-wavelength datasets available \cite[for details see][]{2010iska.meetE..50R}. 

Several low-frequency surveys have been performed in the past, such as the Cambridge surveys 3C, 4C, 6C and 7C at $159$, $178$, $151$ and $151$\,MHz, respectively \citep{1959MmRAS..68...37E,1962MmRAS..68..163B,1965MmRAS..69..183P,1967MmRAS..71...49G,1988MNRAS.234..919H,2007MNRAS.382.1639H}, the UTR-2 sky survey between $10-25$\,MHz \citep{2002Ap&SS.280..235B}, and the VLSS at $74$\,MHz \citep{2007AJ....134.1245C,2014MNRAS.440..327L}. The GMRT  significantly improved low frequency imaging, particularly in terms of sensitivity and angular resolution,  and several GMRT surveys have now been performed at $150$\,MHz \citep[e.g.][]{2007ASPC..380..237I,2009MNRAS.392.1403S,2010MNRAS.405..436I,2011A&A...535A..38I}, including in particular a full-sky survey \citep{2016arXiv160304368I}, and further surveys at $325$\,MHz \citep[e.g][]{2013MNRAS.435..650M} and $610$\,MHz \citep[e.g][]{2007MNRAS.376.1251G,2008MNRAS.383...75G,2008MNRAS.387.1037G}.
Recently, the Murchison Widefield Array \citep[MWA;][]{2009IEEEP..97.1497L,2013PASA...30....7T}, operating at $72$--$231$\,MHz, has yielded the GaLactic and Extragalactic All-sky MWA survey  \citep[GLEAM;][]{2015PASA...32...25W}. GLEAM covers the entire Southern sky ($\delta < 25{\degree}$) with a noise level of a few {\mJybeam} and angular resolution of a few arcminutes. {The  Multifrequency Snapshot Sky Survey \citep[MSSS;][]{2015A&A...582A.123H} is an initial LOFAR  survey at low-resolution and a few {\mJybeam} that is complementary to GLEAM covering the Northern sky.} However, for extragalactic science in particular, the high resolution LOFAR surveys will provide a significant advantage in both image resolution and sensitivity.

Advanced calibration and processing techniques are needed to obtain deep high-fidelity images at low radio frequencies. In particular, direction-dependent effects (DDEs) caused by the ionosphere and imperfect knowledge of the station beam shapes need to be corrected for. \citet[][herafter vW16]{2016ApJS..223....2V} have recently presented a new scheme for calibrating the direction-dependent effects and imaging  LOFAR data that combines elements from existing direction-dependent calibration methods such as {\scshape SPAM} \citep{2009A&A...501.1185I} and {\scshape SAGECal}  \citep{2013A&A...550A.136Y,2011MNRAS.414.1656K}. The Bo\"otes field observations presented here serve as a testbed for this  calibration strategy, which allows us to produce science quality images at the required Tier-1 survey depth.

Here we report on the first LOFAR Cycle 2 High Band Antenna (HBA) observations of the Bo\"otes field. The  Bo\"otes field is one of the Tier-3 Survey fields and the aim is to eventually survey this field to the extreme rms depth of $12$\,{\muJybeam} ($1\sigma$) at $150$\,{MHz}.  The Bo\"otes field  has been extensively studied at higher radio frequencies and in other parts of the electromagnetic spectrum.  Radio observations have been carried out at $153$\,MHz with the GMRT, both as a single deep $10$\,deg$^2$ pointing \citep{2011A&A...535A..38I} and as a seven-pointing $30$\,deg$^2$ mosaic \citep{2013A&A...549A..55W}. Further observations include those at $325$\,MHz with the VLA \citep{2008AJ....135.1793C,2015MNRAS.450.1477C}, and deep, $28$\,{\muJybeam} rms,  $1.4$\,GHz observations with WSRT \\ \citep{2002AJ....123.1784D}. The field has also been observed with the LOFAR Low Band Antennae (LBA) at $62$\,MHz \citep{2014ApJ...793...82V}.

The Bo\"{o}tes field is one of the largest of the well-characterised extragalactic deep fields and was originally targeted as part of the NOAO Deep Wide Field Survey \citep[NDWFS;][]{1999AAS...195.1207J} covering $\approx9$~deg$^2$ in the optical (\textit{B$_W$}, \textit{R}, \textit{I}) and near infra-red  (\textit{K}) bands. There is a wealth of ancillary data available for this field, including X-ray \citep{2005ApJS..161....1M,2005ApJS..161....9K}, UV \citep[GALEX;][]{2003SPIE.4854..336M}, and mid infrared \citep{2004ApJS..154...48E}. The AGN and Galaxy Evolution Survey (AGES) has provided redshifts for $23,745$ galaxies and AGN across $7.7$~deg$^2$ of the Bo\"otes field \citep{2012ApJS..200....8K}.
This  rich multiwavelength dataset, combined with the new low frequency radio data presented here, will be important for determining the evolution of black-hole accretion over cosmic time.

The outline of this paper is as follows. In Section~\ref{sect:obs} we describe the LOFAR observations covering the NOAO Bo\"otes field. In Section~\ref{sect:red} we describe the data reduction techniques employed to achieve the deepest possible images. Our data reduction relies on the `Facet' calibration scheme \citepalias{2016ApJS..223....2V} which corrects for direction-dependent ionospheric phase corruption as well as LOFAR beam amplitude corruption. In Section~\ref{sect:imcat} we present the final image and describe the source-detection method and the compilation of the source catalogue. This section also includes an analysis of the quality of the catalogue. The spectral index distribution and differential source counts  are presented in Section \ref{sect:results}. Finally, Section \ref{sect:concl} summarises and concludes this work. Throughout this paper, the spectral index, $\alpha$, is defined as $S_{\nu} \propto \nu^\alpha$, where $S$ is the source flux density and $\nu$ is the observing frequency.  We assume a spectral index of $-0.8$ unless otherwise stated.

\section{Observations}
\label{sect:obs}

\begin{figure}
 \centering
 \includegraphics[width=0.495\textwidth]{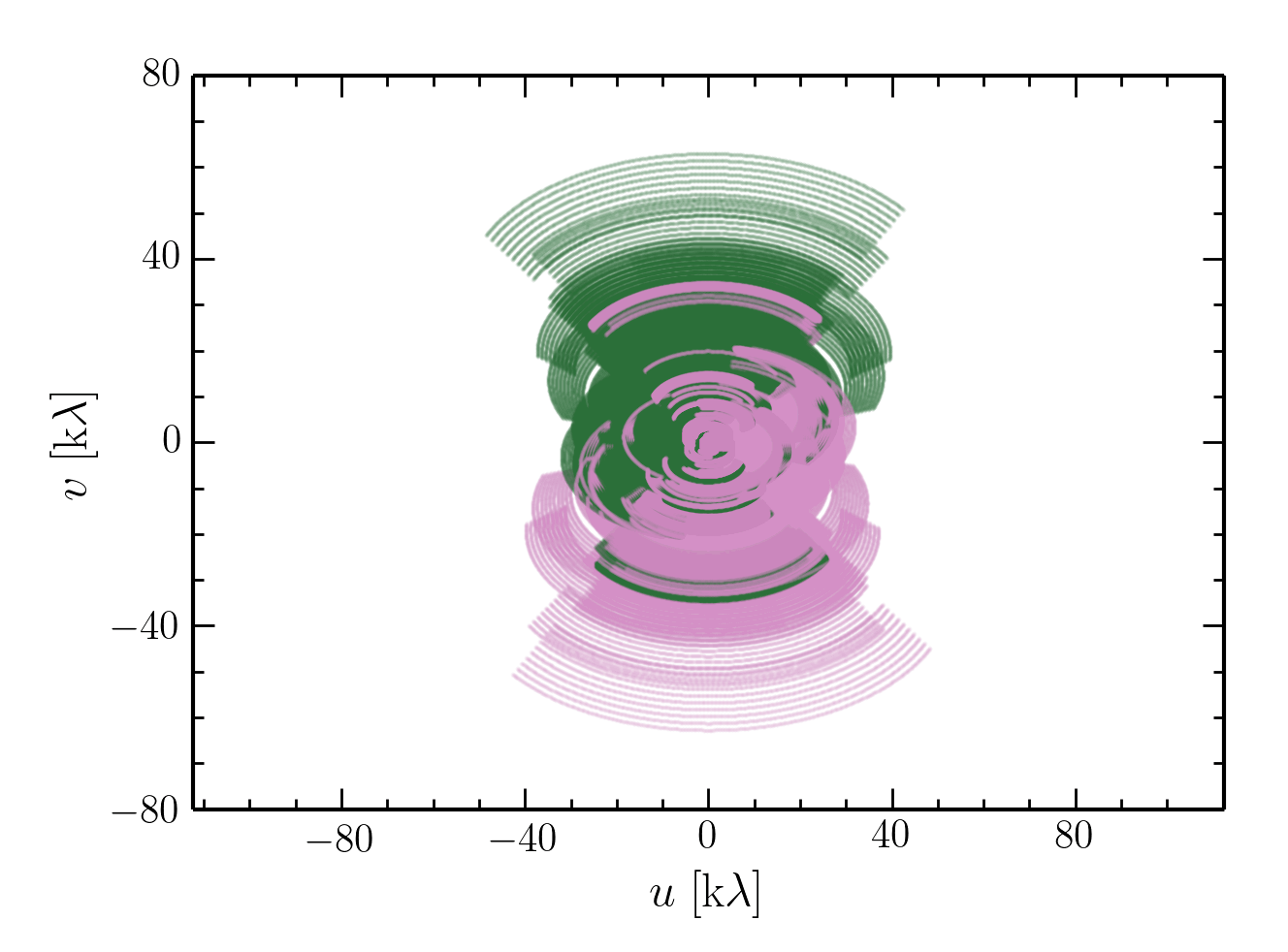}
\caption{$uv$-coverage for the Bo\"otes field at $130$--$169$\,MHz. The maximum baseline is $120$\,km (or $60$\,k$\lambda$). Only one out of every ten $uv$-points in time and one out of every $40$ points in frequency are plotted: the plot nevertheless shows how the large fractional bandwidth fills the $uv$-plane radially. The two colours show the symmetric $uv$ points obtained from the conjugate visibilities.}
\label{fig:uvcover}
\end{figure}

The Bo\"{o}tes field was observed on 2014 August 10 with the LOFAR High Band Antenna (HBA) stations.  An overview of the observations is given in Table~\ref{tab:observations}. By default, all four correlation products were recorded with the frequency band divided into $195.3125$\,kHz--wide subbands (SBs). Each SB was further divided into $64$ channels. The integration time used was $1$\,s in order to facilitate the removal of radio frequency interference (RFI) at high time resolution. The maximum number of SBs for the system in $8$ bit mode is $488$ and the chosen strategy was to use $366$ for the Bo\"otes field giving a total bandwidth of $72$\,MHz between $112$--$181$\,MHz. The remaining $122$ SBs were used to observe the nearby bright calibrator source,  3C\,294, located $5.2${\degree} away,  with a simultaneous station beam, with SBs {semi-regularly} spread between $112$--$181$\,MHz, {avoiding SBs with known strong RFI -- the exact frequency coverage is available through the LOFAR Long Term Archive\footnote{\url{http://lofar.target.rug.nl/}} (LTA)}. The main observations were preceded and succeeded by $10$\,min observations of the primary flux calibrators 3C\,196 and 3C\,295, respectively, with identical SB setup to the Bo\"otes observation, i.e.\ $366$ SBs ($72$\,MHz bandwidth) between $112$--$181$\,MHz. For the observations $14$ Dutch remote and $24$ Dutch core stations were used. This setup results in baselines that range between $40$\,m and $120$\,km. The $uv$-coverage for the Bo\"{o}tes field observation is displayed in Fig.~\ref{fig:uvcover}.  The `HBA\_DUAL\_INNER' configuration was employed. In this configuration, the core stations are each split into two substations ($48$ total), and only the inner $\approx30.8$\,m of the remote stations (which have a total diameter of $41$\,m) are used to obtain similar station beam sizes to the core stations (which have a diameter of $30.8$\,m). The resulting half-power beam width (HPBW) is $\approx4.2${\degree}\footnote{Based on the calculated average primary beam for the Bo\"otes observation (see Section~\ref{sect:pbcor})} at $150$\,MHz.

\begin{table}
 \begin{center}
 \caption{LOFAR HBA observation parameters.}
 \label{tab:observations}
\begin{tabular}{ll}
\hline 
Observation IDs &  L240762  (3C\,196)  \\
               &  L240764  (Bo\"otes, 3C\,294) \\
               &  L240766  (3C\,295)  \\
Pointing centres & $08{\hh}13{\mm}36{\s}$ $+48{\dd}13{\mm}03{\s}$ (3C\,196)\\
                & $14{\hh}32{\mm}00{\s}$ $+34{\dd}30{\mm}00{\s}$ (Bo\"otes)\\
                & $14{\hh}06{\mm}44{\s}$ $+34{\dd}11{\mm}25{\s}$ (3C\,294)\\
                & $14{\hh}11{\mm}20{\s}$ $+52{\dd}12{\mm}10{\s}$ (3C\,295)\\
Integration time & 1\,s \\
Observation date & 2014 August 10 \\
Total on-source time & $10$\,min (3C\,196, 3C\,295) \\
                     & $8$\,hr (Bo\"otes, 3C\,294) \\
Correlations &  XX, XY, YX, YY\\
Sampling mode & $8$-bit \\
Sampling clock frequency  & $200$\,MHz \\
Frequency range & $112$--$181$\,MHz\\
Bandwidth & $71.48$\,MHz (Bo\"otes, 3C\,196, 3C\,295) \\
          & $23.83$\,MHz (3C\,294)\\
Subbands (SBs) & $366$ contiguous (Bo\"otes, 3C\,196, 3C\,295)\\
         & $122$ {semi-regularly spaced}$^{\mathrm{a}}$ (3C\,294)\\
Bandwidth per SB &  $195.3125$\,kHz \\
Channels per SB & $64$ \\
Stations & $62$ total \\
         & $14$ remote \\
         & $24$ core ($48$ split) \\
\hline
\multicolumn{2}{p{0.48\textwidth}}{$^{\mathrm{a}}${avoiding SBs with strong RFI, but spanning the range  $112$--$181$\,MHz.}}
 \end{tabular}
 \end{center}
\end{table}

\section{Data Reduction}
\label{sect:red}
In this section we describe the calibration method that was used to obtain the required deep  high-fidelity high-resolution images. The data reduction and calibration consists of two stages: a non-directional and a directional part  to correct DDEs caused by the ionosphere and imperfect station beam models. The non-directional part includes the following steps: 
\begin{enumerate}
 \item initial flagging and removal of RFI; 
 \item solving for the calibrator complex gains, including `clock-TEC (Total Electron Content) separation', and transfer of the amplitudes, median clock offsets and the XX--YY phase offset from calibrator to the target field;
 \item removal of bright off-axis sources;
 \item averaging; and
 \item amplitude and phase (self-)calibration of the target field at medium ($20$--$30$\,{\arcsec}) resolution
\end{enumerate}
This is then followed by a scheme to correct for DDEs in order to reach near thermal-noise-limited images using the full resolution offered by the longest `Dutch-LOFAR' baselines of about $120$\,km. All calibration steps are performed with the {\scshape BlackBoard Selfcal} ({\scshape BBS}) software system \citep{2009ASPC..407..384P} and other data handling steps were undertaken with the LOFAR Default Pre-Processing Pipeline ({\scshape DPPP}). These steps are explained in more detail below. The direction-dependent calibration scheme is described in full by \citetalias[][]{2016ApJS..223....2V}.

We used the  full frequency coverage of the $10$\,min 3C\,196 observation as the primary calibrator observation to derive the time-independent instrumental calibration including amplitudes, median clock offsets and the XX--YY phase offset. We did not use the simultaneous, but sparse, frequency coverage on the calibrator 3C\,294, {nor did we use the second calibrator, 3C\,295}.  For the  Bo\"otes field we selected  $200$ out of the total $366$  observed SBs  ($55$~per~cent)  covering the frequency range $130$--$169$\,MHz for further processing. The main limitation to the number of subbands processed was computational time; the main data reduction was carried out on a node with $20$ virtual cores on ASTRON's CEP3 cluster\footnote{Each node has $2$ ten-core Intel Xeon e5 2660v2 ($25$M Cache, $2.20$\,GHz) processors with $128$\,GB RAM} and most of the parallelisation in the reduction is performed over $10$-SB blocks, so that the full `facet' reduction could be achieved in a reasonable time.

\subsection{Direction-independent calibration}

\subsubsection{Flagging and RFI removal}
The initial preprocessing of the data was carried out using the Radio Observatory pipeline and consisted of RFI excision \citep[using AOFlagger;][]{2010MNRAS.405..155O, 2012A&A...539A..95O}, flagging the noisy first channel and last three channels of each SB, and averaging in time and frequency to $2$\,s and $8$\,channels per SB. The data were stored at this resolution in the LOFAR LTA at this point. One core station (CS007) and one half of another core station (CS501HBA1) were flagged entirely due to malfunction (failure to record data or low gains). CS013 was also flagged entirely due its different design (the dipoles are rotated by $45{\degree}$).

\subsubsection{Calibration transfer from Primary Calibrator 3C\,196}
Using {\scshape BBS} we obtained parallel hand (XX and YY) gain solutions for 3C\,196, on timescales of $2$\,s for each frequency channel independently. In this step we also solved simultaneously for `Rotation Angle' per station per channel to remove the effects of differential Faraday Rotation from the parallel hand amplitudes. The solutions were computed with the LOFAR station beam applied to separate the beam effects from  the gain solutions. We used a second-order spectral model for 3C\,196 consisting of $4$ point sources separated by $3-6$\,{\arcsec} each with a spectral index and curvature term \citetext{V.~N.~Pandey (ASTRON), priv.~comm.}. {The total flux density of this model differs from the \citetalias{2012MNRAS.423L..30S} value by a factor of $1.074 \pm 0.024$}. We used these calibration solutions to determine the direction-independent and time-invariant instrumental calibrations, including amplitude calibration, correction of clock delays between the remote and core stations and an offset between the XX and YY phases.

The remote LOFAR stations each have their own GPS-corrected rubidium clocks, which are not perfectly synchronized with the single clock that is used for all the core stations. The offsets between the remote station clocks and the core can be  of the order of $100$\,ns, which is large enough to cause strong phase delays within a single SB for the remote-remote and core-remote baselines. However, this offset appears to be fairly constant and similar for observations separated by days to weeks, and can thus safely be assumed to be time-invariant. In addition to these clock offsets, the individual clocks can drift within $\pm15$\,ns over the course of a few tens of minutes, before being reset to the offset values (i.e. the drifts do not accumulate over the course of several hours). The primary calibrator phase solutions were used to calculate the clock offsets to be applied to the target observation (we do not correct for the $15$\,ns clock drifts {because the effects they induce are not time-invariant}). We used the `Clock-TEC separation' method, described in detail by \citetalias[][]{2016ApJS..223....2V}, which uses the frequency-dependent phase information across the full frequency range to separate the direction-independent clock errors from the direction-dependent ionospheric effects. The clock phase errors, or delays, vary linearly with frequency (phase $\propto \delta t \times \nu$, where $\delta t$ is the clock difference), while the ionospheric phases vary  inversely with frequency  (phase $\propto {\rm dTEC} \times  \nu^{-1}$, where dTEC is the differential Total Electron Content). Fitting was performed on
a solution interval timescale of $5$\,s and smoothed with a running median filter with a local window size of $15$\,s.
{The clock offsets for the remote stations in our observation were between $-90$\,ns and $80$\,ns with respect to the core.} The ionospheric conditions during the  $10$\,min calibrator observation were good, showing relatively smooth variations in the differential TEC of $\approx0.2 \times 10^{16}$\,m$^{-2}$.

For a few stations we found small but constant offsets between the XX and YY phases. We determined these offsets by taking the median phase difference between the XX and YY phases during the 3C\,196 observation for each station. 

The amplitudes were inspected for outliers and smoothed in the frequency axis with a running median filter with window size of $3$\,SBs ($\approx0.6$\,MHz), and a single median value in the time axis of the $10$\,min observation.

These calculated median clock offsets, XX--YY phase offsets and amplitude values were transferred from the $10$\,min observation of 3C\,196 to the target-field data. The resulting target-field visibilities have amplitudes in Jy and are free of {time-invariant} clock offsets and XX--YY phase offsets.

\subsubsection{Removal of bright off-axis sources}
A few radio sources are sufficiently bright to contribute flux through the sidelobes of the station beams, the amplitudes of which are strongly modulated in frequency, time and baseline as they move in and out of the station-beam sidelobes.  To remove these effects we simply predicted the visibilities of the brightest of these `A-team'  sources (Cyg\,A, Cas\,A, Vir\,A, and Tau\,A) with the station beam applied in {\scshape BBS}, and flagged all times, frequencies and baselines where the contributed apparent flux density from these sources exceeded $5$\,Jy. {The $5$\,Jy limit was set based on visual inspection of the predicted visibility amplitudes and experience with this and other LOFAR HBA datasets. It was found that the contributed flux was $\lesssim 5$\,Jy in the majority of the  frequency-time-baseline stamps, and exceeding several tens of Jy in only a few per~cent of the frequency-time-baseline stamps}. The amount of data flagged in this step was typically $2$--$5$\,per~cent per SB.

\subsubsection{Averaging}
The data were then averaged in time to a more manageable $8$\,s and two channels per SB ($98$\,kHz channelwidth). The main limit on the time resolution is set by the requirement to avoid decorrelation due to rapid ionospheric phase variations. Even at this  time and frequency resolution we expect some smearing at large radial distances from the field centre due to time and bandwidth smearing, of the order of $7$ and $10$~per~cent respectively  at the half-power point of the primary beam ($2.1${\degree} from the pointing centre) and $13$ and $15$~per~cent respectively at $2.5${\degree} at $150$\,MHz. The individual corrected SBs were combined in groups of ten, providing datasets of $\approx2$\,MHz bandwidth and $20$ channels, each of which was $\approx30$\,GB in size. 

\subsubsection{Self-calibration of target}
\label{sect:selfcaltarget}
We used a single $10$-SB block at $148$--$150$\,MHz on which to perform direction-independent self-calibration. {This block was largely free of RFI and lies at the peak of the response of the HBA tiles.} {The $2$\,MHz bandwidth means that the signal-to-noise ratio within the $10$-SB block was high enough to obtain coherent solutions in each timestep, while the bandwidth was not so high as to experience decorrelation due to ionospheric effects.} We started with a model derived from the $30$\,deg$^2$ GMRT image of the Bo\"otes field at $153$\,MHz \citep{2013A&A...549A..55W}. The resolution of this initial sky model is $25 \times 25$\,{\arcsec} -- the native resolution of the GMRT image -- and it includes all sources in the GMRT image greater than $20$\,{\mJy} ($\approx6\sigma$). The brightest $10$ sources in the GMRT model were replaced by Gaussian components taken from the  $\approx5$\,{\arcsec} resolution FIRST catalogue \citep{1995ApJ...450..559B}, with their total GMRT $150$\,MHz flux density and flux density ratios taken from FIRST. {While this was not strictly necessary, given the resolution of the imaging in this self-calibration step, it was found that it did improve the calibration. This may be due to the fact that the brightest source in the GMRT model is only barely resolved, while in FIRST it consists of four components.  }

\begin{figure*}
 \centering
 \includegraphics[width=\textwidth, trim=0.2cm 0.6cm 0.2cm 0.2cm, clip]{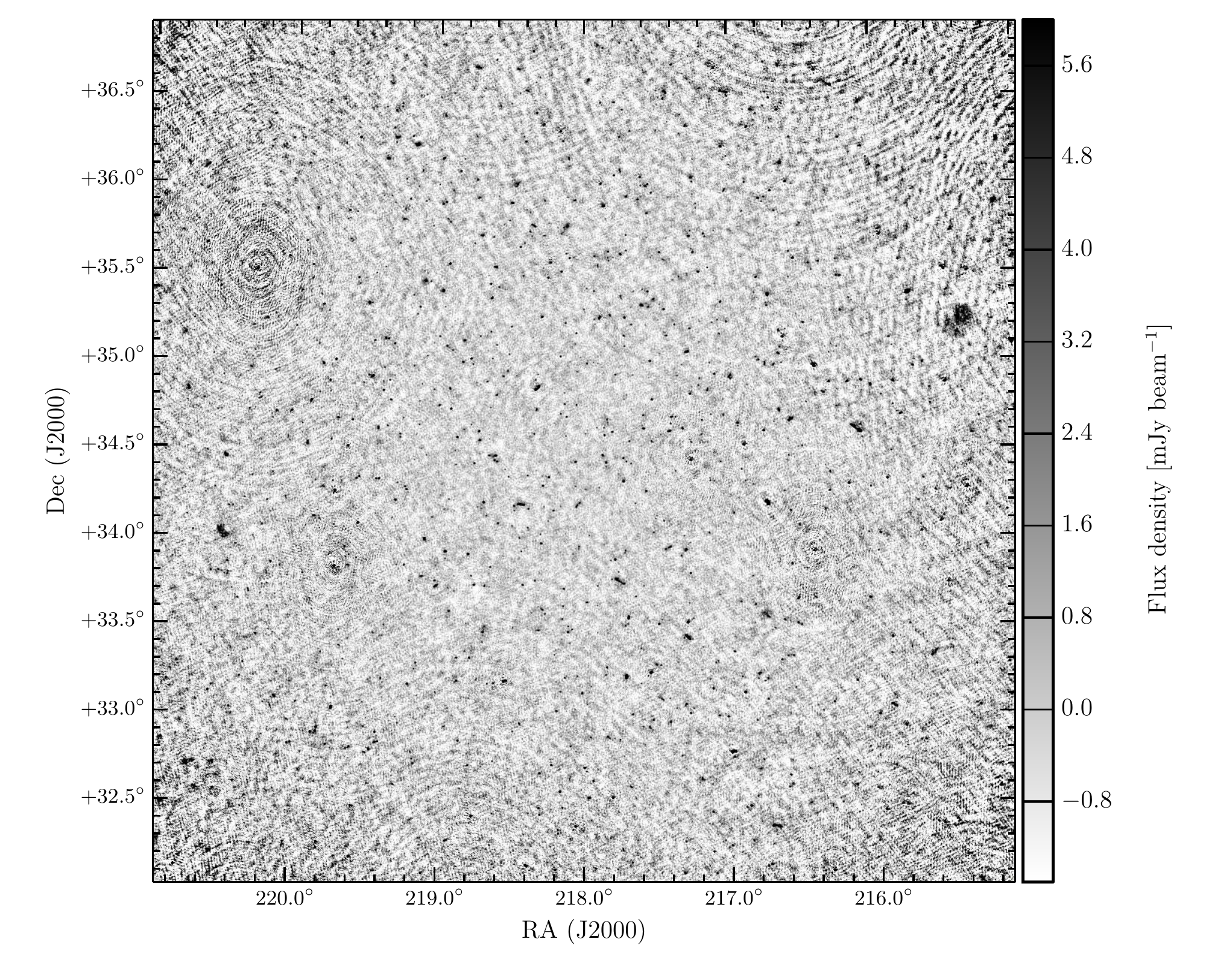}
\caption{Final self-calibrated image for the $10$-SB block at $148$--$150$\,MHz. The resolution is $23 \times 20$\,{\arcsec}.  The greyscale shows the flux density {from $-1.5\sigma$ to $6\sigma$} where $\sigma = 1$\,{\mJybeam} is the approximate rms noise in the central part of the image. Calibration artefacts are clearly visible around the brightest sources as only direction-independent self-calibration has been performed. }
\label{fig:selfcalimage}
\end{figure*}

The self-calibration was performed with two iterations with phase-only solutions followed by two iterations with amplitude and phase solutions. The solution interval in all iterations was $8$\,s and we obtained a single solution in frequency, neglecting phase changes within the $2$\,MHz bandwidth.  In each calibration step, the station beam model was applied in the `solve' step in {\scshape BBS}, i.e. the model visibilities were corrupted by the station beams before the gain solutions were derived. In this way the solutions do not contain the beam terms. {The solutions were smoothed using a running median filter  $30$\,s in width to remove outliers.} The data were corrected for the gain solutions, as well as the station beam in the phase centre. Imaging was carried out using  {\scshape AWImager} \citep{2013A&A...553A.105T}, which accounts for the non-coplanar nature of the array and performs a proper beam correction across the field of view.  Each imaging step in the self-calibration cycle was carried out with the same parameters: a field of view of $6.4${\degree}, a resolution of $\approx22$\,{\arcsec} imposed by the combination of a maximum $w$ term of $10$\,k$\lambda$, an outer $uv$-limit of $10$\,k$\lambda$, and \cite{1995AAS...18711202B} robust weighting ({\tt robust=0.25}). This weighting results in a slightly lower resolution, with less emphasis on the calibration artefacts. The imaging  was always performed in two stages: first without a mask and then with a mask. The {\scshape clean} masks for each image were generated automatically  using {\scshape PyBDSM} \citep{2015ascl.soft02007M}, with  {\tt rms\_box} = (boxsize, stepsize) $=(85,30)$\,pixels,  to create a $3\sigma$ island threshold map  from an initial image made without a mask. From the resulting images we created a new sky model, again using {\scshape PyBDSM} (with {\tt rms\_box}$=(150,40)$\,pixels, {\tt thresh\_pix}$=5\sigma$ and {\tt thresh\_isl}$=3\sigma$), and included all Gaussians in the model. In each additional iteration (first phase-only self-calibration, then two iterations of phase and amplitude self-calibration), solutions were obtained relative to the original un-calibrated data. We made no attempt to improve the resolution in each self-calibration cycle, as the direction-dependent effects become significantly worse at higher resolution. A part of the final self-calibrated image for the $10$-SB block at $149$\,MHz is shown in Fig.~\ref{fig:selfcalimage}. The noise level achieved is $\approx 1$\,{\mJybeam}. While deeper images at a similar resolution can be made with more bandwidth, this is sufficient for the purpose of initial calibration prior to the direction-dependent calibration.  

All the $10$-SB blocks were then corrected for the station beam in the phase centre before further processing. We then used the final image of the self-calibration cycle at $148-150$\,MHz to make a Gaussian input sky model with {\scshape PyBDSM} to (self)-calibrate the remaining $10$-SB blocks. All the components in this model were assumed to have a spectral index of $-0.8$. For all the bands we performed a single phase and amplitude calibration against this model { on $8$\,s solution intervals}. {The amplitude self-calibration calibration is done here to clean up some artefacts around the brightest, most dominant sources. The purpose of this step is to provide the best possible sky model before direction-dependent calibration. These solutions are not applied in the direction-dependent calibration; thus, if any frequency dependent amplitude errors are introduced, they  are not fixed in the data.} 

After calibration, images of each band were made with the Common Astronomy Software Applications  \citep[{\scshape Casa};][]{2007ASPC..376..127M} version 4.2.1 using W-projection \citep{2008ISTSP...2..647C,2005ASPC..347...86C} to handle the non-coplanar effects; {\scshape Casa} does not allow for the full beam correction across the field of view but it does allow a much larger field to be imaged than with {\scshape AWImager}. Because the beam correction was not performed in imaging, the resulting images are \emph{apparent} sky images. The imaging was carried out {in each  $10$-SB block} in two iterations. The first image was made at `medium resolution' using a $uv$-cut of $7$\,k$\lambda$ and Briggs weighting ({\tt robust=-0.25}) to limit the resolution to $\approx29$\,{\arcsec}, with $7.5$\,{\arcsec} pixels; the field of view imaged was $\approx10${\degree}. Next, in each $10$-SB dataset we subtracted the {\scshape clean} components with {these} calibration gain solutions applied and re-imaged the subtracted data using a `low resolution' of $\approx2${\arcmin} with a $uv$-cut of $2$\,k$\lambda$  and Briggs weighting ({\tt robust=-0.25}), $25$\,{\arcsec} pixels,  and a field of view of $\approx30${\degree}. This field of view extends to the second sidelobe of the station beams, allowing us to image and subtract these sources. The low-resolution image also picks up extended low-surface-brightness emission in the field not {\scshape clean}ed in the medium-resolution image. Both medium- and low-resolution images were created with {\scshape clean} masks automatically generated from images made without {\scshape clean} masks using {\scshape PyBDSM} with {\tt rms\_box}~$=(50,12)$ and {\tt rms\_box}~$=(60,12)$ respectively. The {\scshape clean} components in the low-resolution image were then subtracted in the same way as the medium-resolution components and a combined list of {\scshape clean} components for the medium- and low-resolution images was created.

The resulting products are $20$ sets of $2$-MHz residual datasets between $130$ and $169$\,MHz -- i.e.\ with all sources out to the second sidelobes subtracted using the gain solutions, but with the residual data itself not corrected for the gain solutions (in this way any errors in the self-calibration solutions can be corrected later). In addition, there are $20$ corresponding {\scshape clean}-component \emph{apparent} sky models of the sources that were subtracted. These products serve as the input for the direction-dependent calibration scheme, which is described in the following section.

\subsection{Directional calibration -- `Facet' scheme}
Significant artefacts remain in the self-calibrated images, even at $20$\,{\arcsec} resolution, and the rms noise of  {1-3}\,{\mJybeam} is a factor of $3-5$ higher than what is expected with these imaging parameters. Both of these issues result from the direction-dependent effects of the station beams and ionosphere.  {The variation in noise as a function of frequency is largely due to the variation in sensitivity of the HBA tiles and to variations in RFI across the LOFAR frequency band.} To correct for the artefacts, improve the noise and make high resolution images we follow the direction-dependent calibration, or `facet', scheme, of \citetalias[][]{2016ApJS..223....2V}, in which calibration is performed iteratively in discrete directions and imaging is carried out within \emph{mutually exclusive} facets around each calibration direction. The key concern is keeping the number of degrees of freedom in the calibration small with respect to the number of measured visibilities, in order to facilitate solving for DDEs in tens of directions. The `facet' scheme has the following underlying assumptions:
\begin{enumerate}
 \item the only calibration errors are a result of ionosphere and beam errors;
 \item the station beams vary slowly with time and frequency;
 \item differential Faraday rotation is negligible, so that XX and YY phases are affected identically by the ionosphere;
 \item the phase frequency dependence is phase $\propto v^{-1}$ as a result of ionosphere only; and
 \item DDEs vary slowly across the field of view.
\end{enumerate}
For a more detailed description and discussion of the underlying problems and assumptions see \citetalias[][]{2016ApJS..223....2V} and references therein. The following sub-sections describe our implementation of the `facet' scheme on the Bo\"otes field data. 

\subsubsection{Facetting the sky}
\begin{figure*}
 \centering
\includegraphics[width=0.9\textwidth, trim=5cm 2.1cm 3cm 1.8cm, clip]{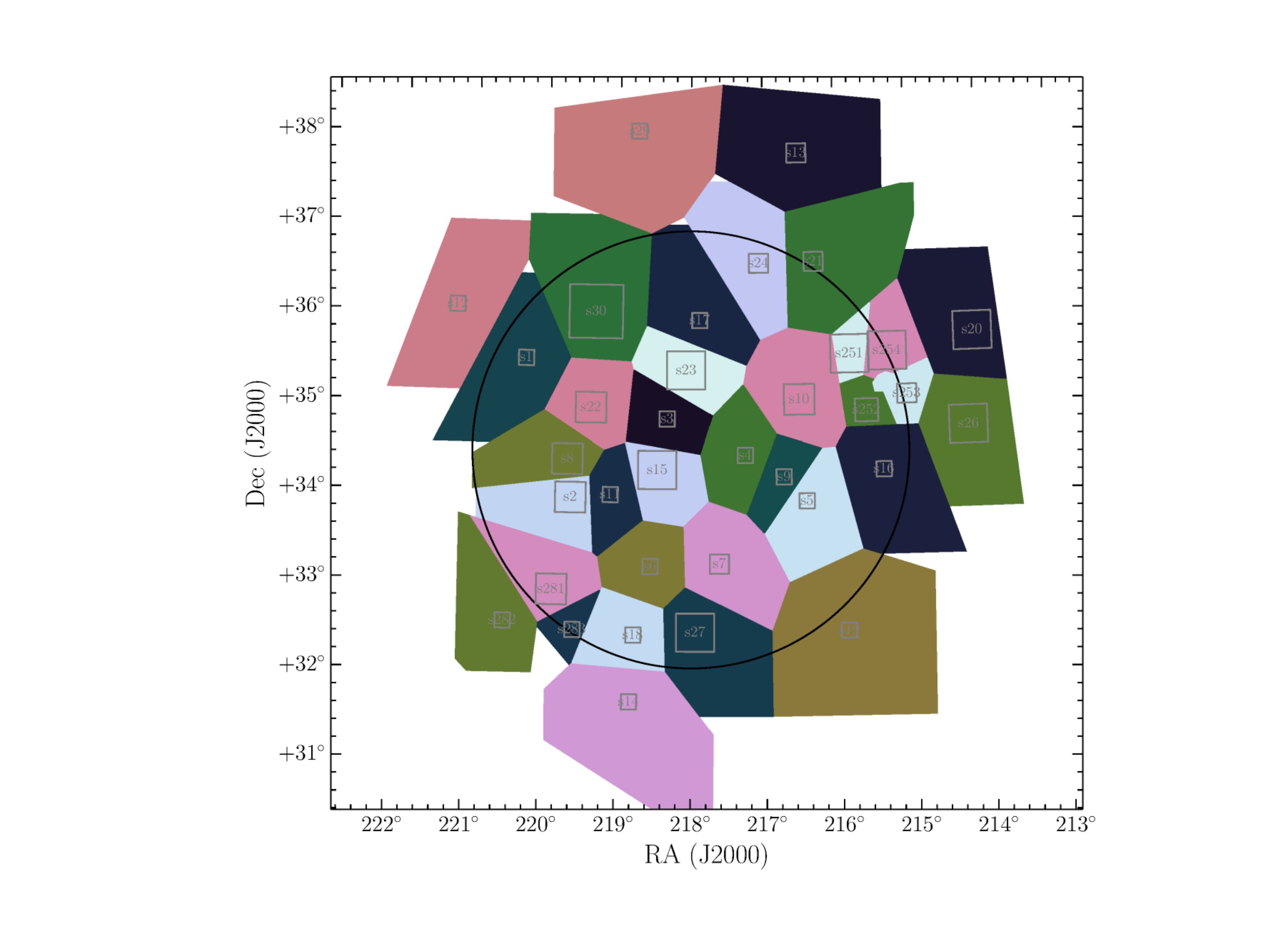}
\caption{Facet coverage of the Bo\"otes field. The grey boxes show the positions of the calibrator directions and the size of these boxes show the area used for self-calibration, i.e.\ larger boxes include more sources. The coloured polygons show the Voronoi tessellation of the image plane based on these calibrator positions. The maximum size of the facets is limited to $50${\arcmin} radius from the calibrator direction ($2048$ pixels at $1.5$\,{\arcsec}\,pixel$^{-1}$) resulting in some incomplete coverage particularly outside the FWHM. The black circle has a radius of $2.44${\degree},  at the approximate $40$~per~cent power point of the average  primary beam. }
\label{fig:facets}
\end{figure*}

Initially we identified $28$ calibration directions or groups consisting of single bright sources or closely (few arcminute) separated sources with a combined total flux density $\gtrsim 0.3$\,Jy. The bright source positions or centres of the groups define the facet directions that were used to tile the sky using Voronoi tessellation, ensuring that each point on the sky lies within the facet of the nearest calibrator source.  The assumption here is that the calibration solutions for the calibrator group applies to the full facet. The typical facet size is a few tens of arcminutes in diameter.  Around and beyond the HPBW of the station beam, the assumption that DDEs, in particular the beam, vary slowly across the field of view  {becomes less valid, and smaller facets are required to capture the changes}. We found that two of the facets initially defined were too large and showed worsening calibration artefacts away from the calibrator source. These two facets were subdivided into smaller facets each having new calibrator groups. The final set of facets is shown in Fig.~\ref{fig:facets}. {These facets are described by {\scshape Casa} regions and image masks. The image masks are constructed from the regions such that they are centred on the calibrator group for that direction and, when regridded to a single image centred on the pointing centre, will not overlap with any other facet.} The following steps were then performed for each direction sequentially, starting with the brightest calibrator sources.

\subsubsection{Directional self-calibration}
\label{sect:startdir}
The {\scshape clean} components of all the sources within the calibrator group were added back in each $10$-SB dataset (with the direction-independent calibration solutions with which they had been initially subtracted). These datasets were then phase-rotated to the direction of the calibration group and averaged in frequency (but not in time) to $1$ channel per $2$-MHz dataset. The frequency averaging greatly speeds up calibration without any bandwidth-smearing effects in the small (few arcminute) calibrator group images.

A self-calibration cycle with four iterations was then performed. The imaging at each iteration was carried out using {\scshape Casa} with the full $130-169$\,MHz bandwidth and multi-frequency synthesis (MFS) {\scshape clean} with a second-order frequency term \citep[i.e. including spectral index and curvature, {\tt nterms=2}][]{1990MNRAS.246..490C}, with the automated masking described in Section~\ref{sect:selfcaltarget}. {The masks made in this way exclude any negative bowls around sources, and can, based on visual inspection of the images and masks, include {\scshape Casa} regions specified by the user to account for extended sources with complicated sidelobes.} The imaging resulted in a {\scshape clean}-component model with both flux and spectral index. Multi-scale {\scshape clean} \citep[MS-MFS;][]{2011A&A...532A..71R,2008ISTSP...2..793C} was used only in the case of a few complex extended sources. We used a Briggs robust parameter of $-0.25$, a pixel size of $1.5$\,{\arcsec}, and imposed a $uv$-minimum of $80$\,$\lambda$ to achieve a resolution of approximately $5.6 \times 7.4$\,{\arcsec}. Using a more negative robust weighting allows for higher resolution images after the direction-dependant effects have been accounted for.

In the first two self-calibration cycles we solved for a single Stokes I phase-offset and TEC term per station in groups of $5$ $10$-SB datasets, i.e.\ within $10$\,MHz bands. {Across the full $40$\,MHz bandwidth, this gave a total of $8$ parameters (in $4$ frequency bands a phase and TEC solution) per station per solution interval}. The solutions were computed on timescales of $8$\,s where possible, but this was increased to $16$\,s for fainter calibration groups and even to $24$\,s for the faintest calibration groups where the signal-to-noise was lower.

{In the third and fourth iterations of self-calibration, we solved first for the short timescale phase-offset and TEC terms, pre-applied these `fast phase' solutions and subsequently solved for phases and amplitudes, i.e. XX and YY complex gains, independently per $10$-SB dataset on timescales between $5$ and $30$\,min, depending on the flux density of the calibrator group.  This additional `slow gain' calibration } yielded an additional $4$ parameters per $10$-SB block and primarily takes out the slowly-varying complex station beams. As expected the phase component of these solutions is small because the fast phase component has been taken out.  In the final self-calibration cycle, the amplitudes were normalised to unity across the full frequency range to prevent changes to the flux-scale. The normalisation corrections were typically smaller than a few percent. {Solutions were obtained for every time step and outliers were removed by filtering. The final solutions were filtered in time twice by passing them through a sliding window median filter of width $10$ solution intervals where outliers greater than $10$ (first pass) and $4$ (second pass) times $1.4826$ times the median distance from the median were replaced with the median value}. 

\begin{figure*}
 \centering
\includegraphics[width=0.33\textwidth]{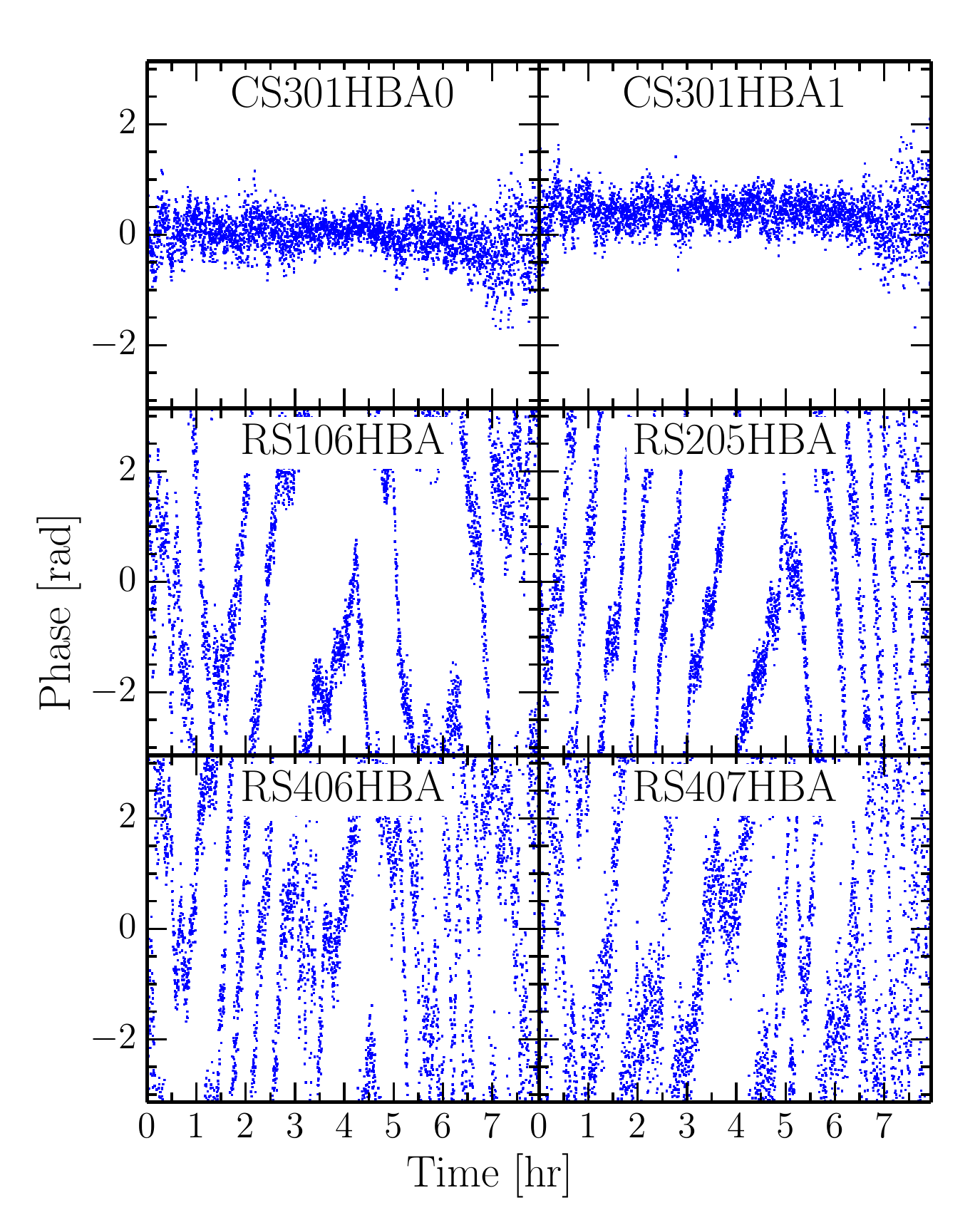}
\includegraphics[width=0.33\textwidth]{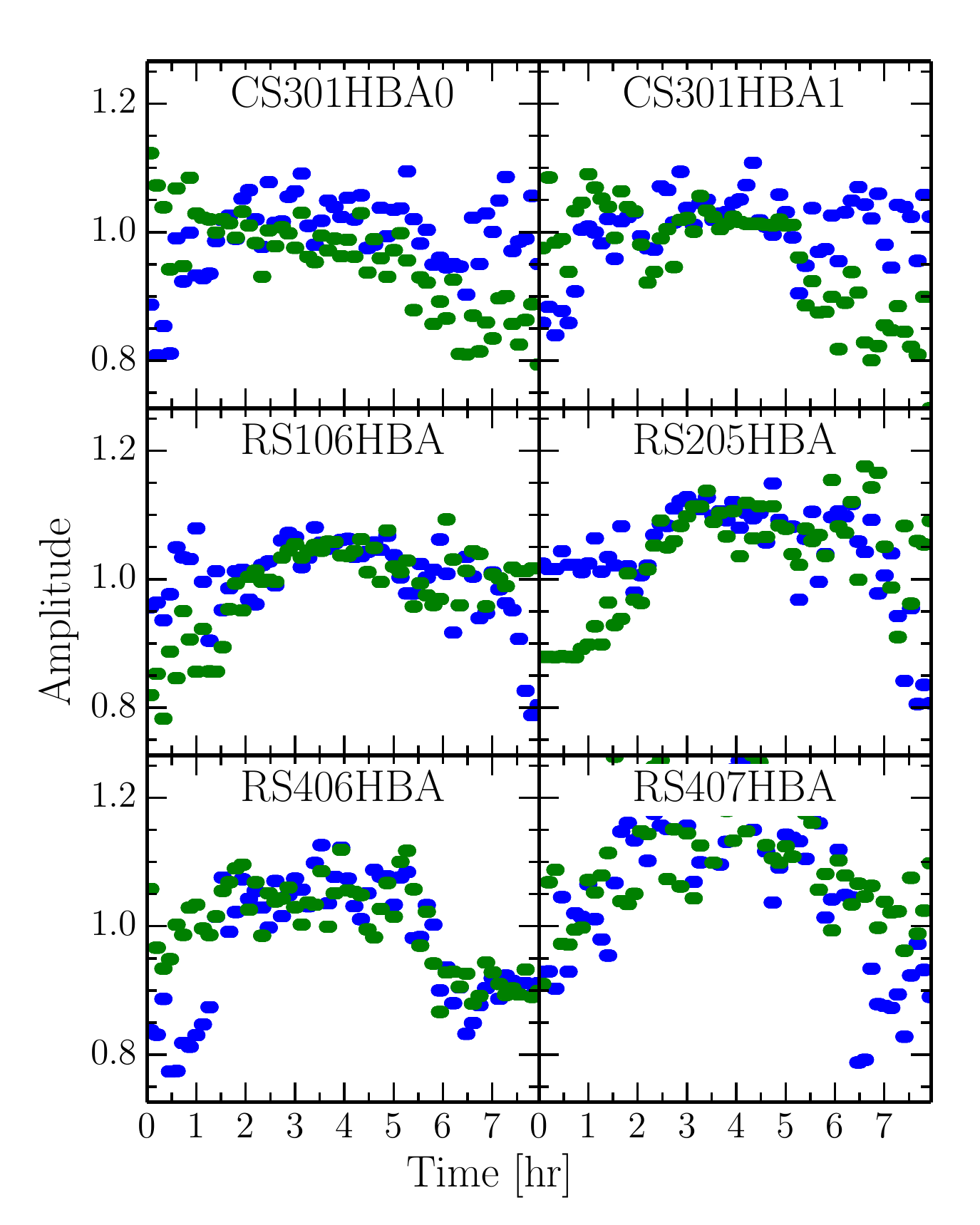}
\includegraphics[width=0.33\textwidth]{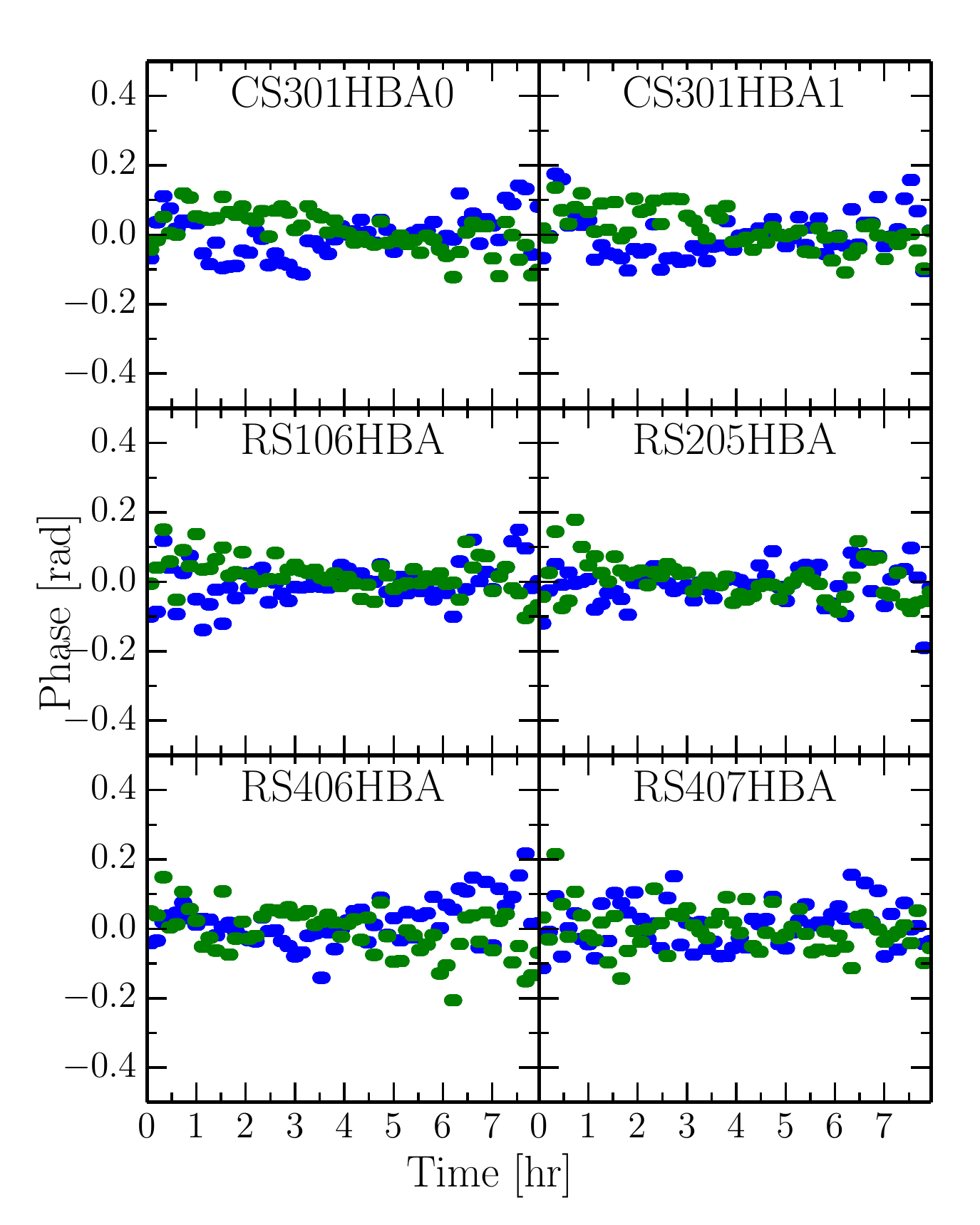} \\
\caption{Example good DDE solutions for a few selected stations obtained for a single direction s3 (the images corresponding to this direction are shown in  Fig.~\ref{fig:selfcalstages}). Left: The effective Stokes I phase corrections,  evaluated at an arbitrary frequency of $150$\,MHz. The solutions are obtained on a timescale of $8$\,s using $10$\,MHz
of bandwidth. Centre and Right: the additional XX and YY amplitude (centre) and phase (right) solutions for the $160-162$\,MHz SB block obtained on a timescale of $10$\,min after application of the short-timescale phase offsets and TEC solutions. In all cases  phases are plotted with respect to core station CS001HBA0.}
\label{fig:selfcalsols}
\end{figure*}

\begin{figure*}
 \centering
\includegraphics[width=0.33\textwidth]{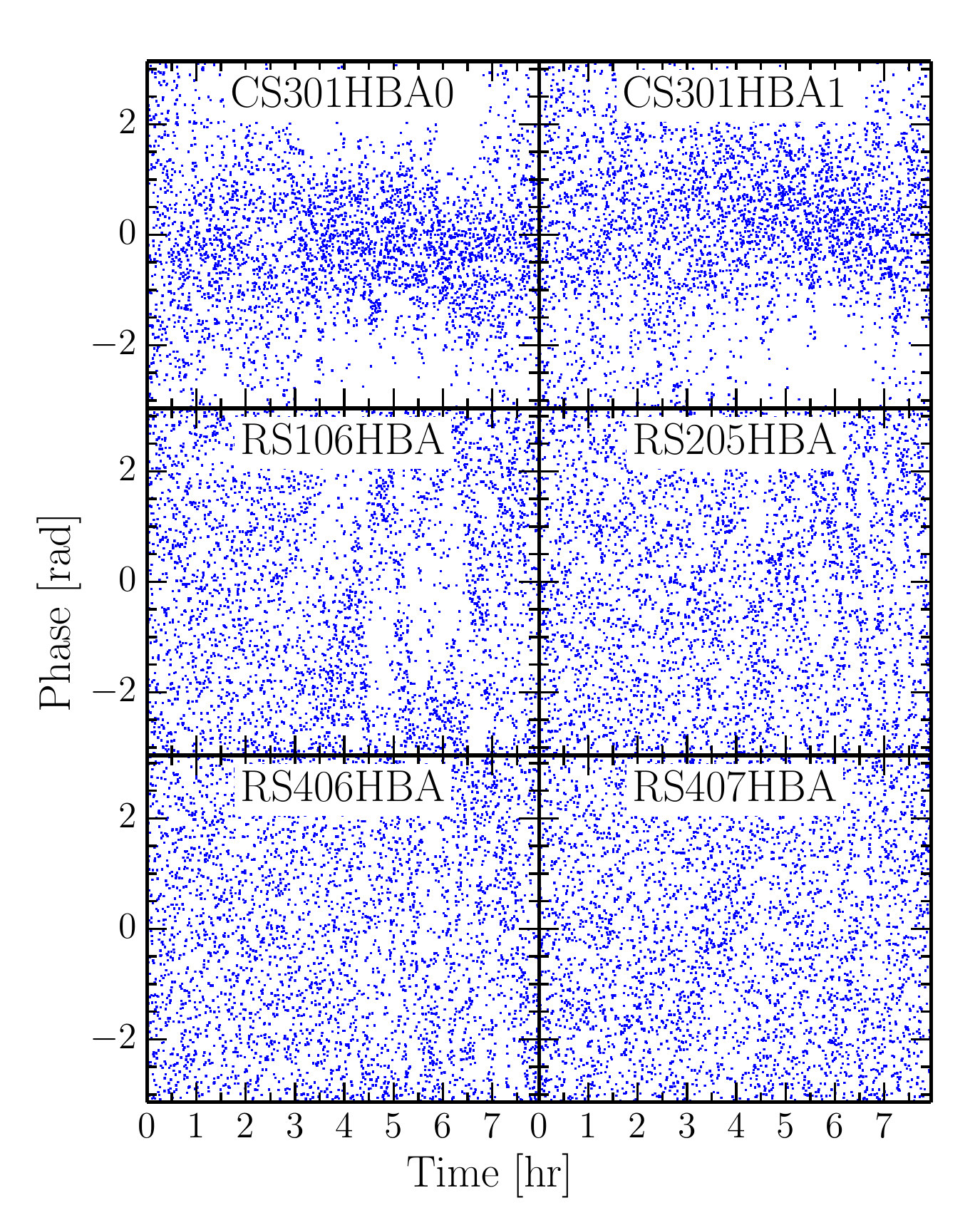}
\includegraphics[width=0.33\textwidth]{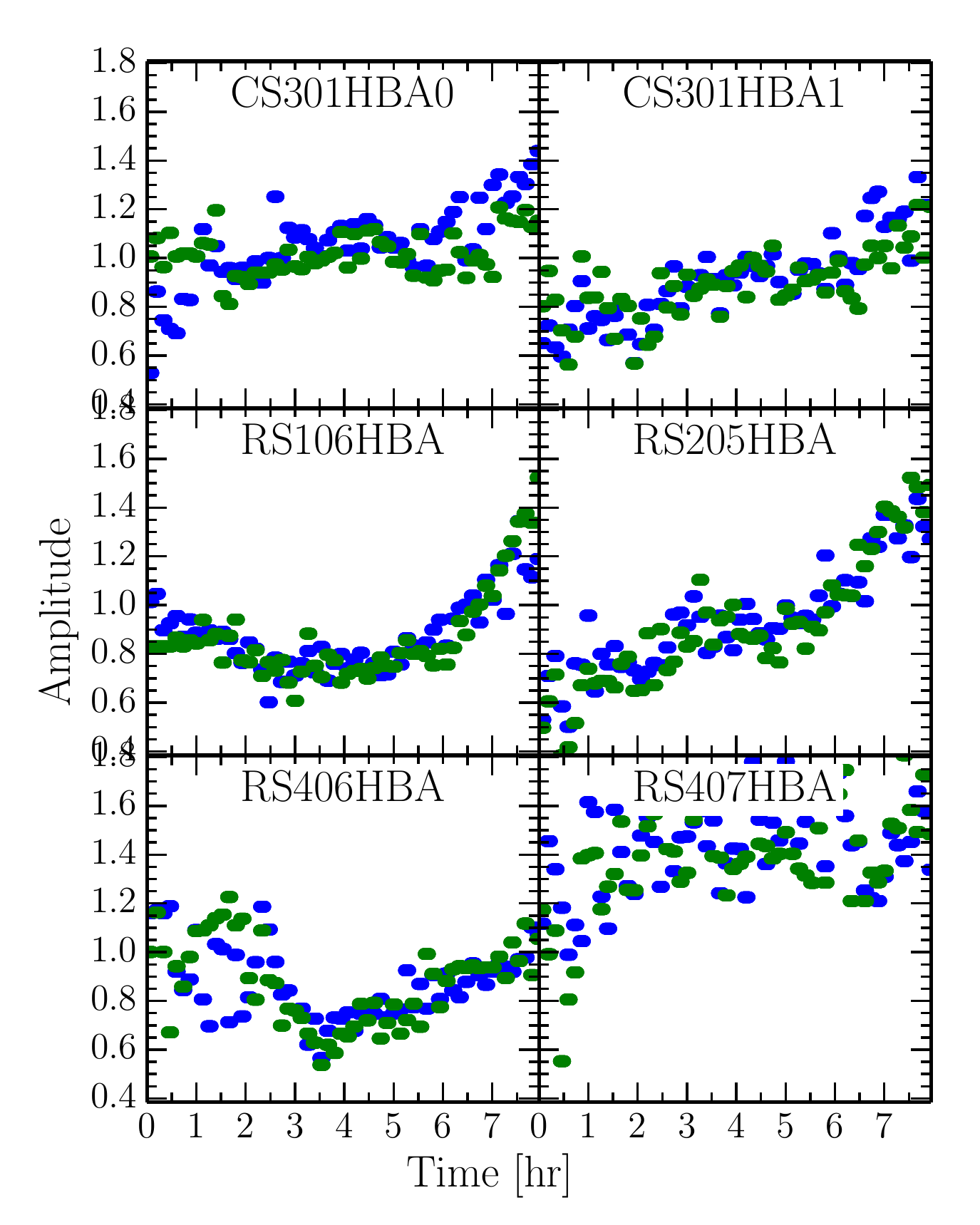}
\includegraphics[width=0.33\textwidth]{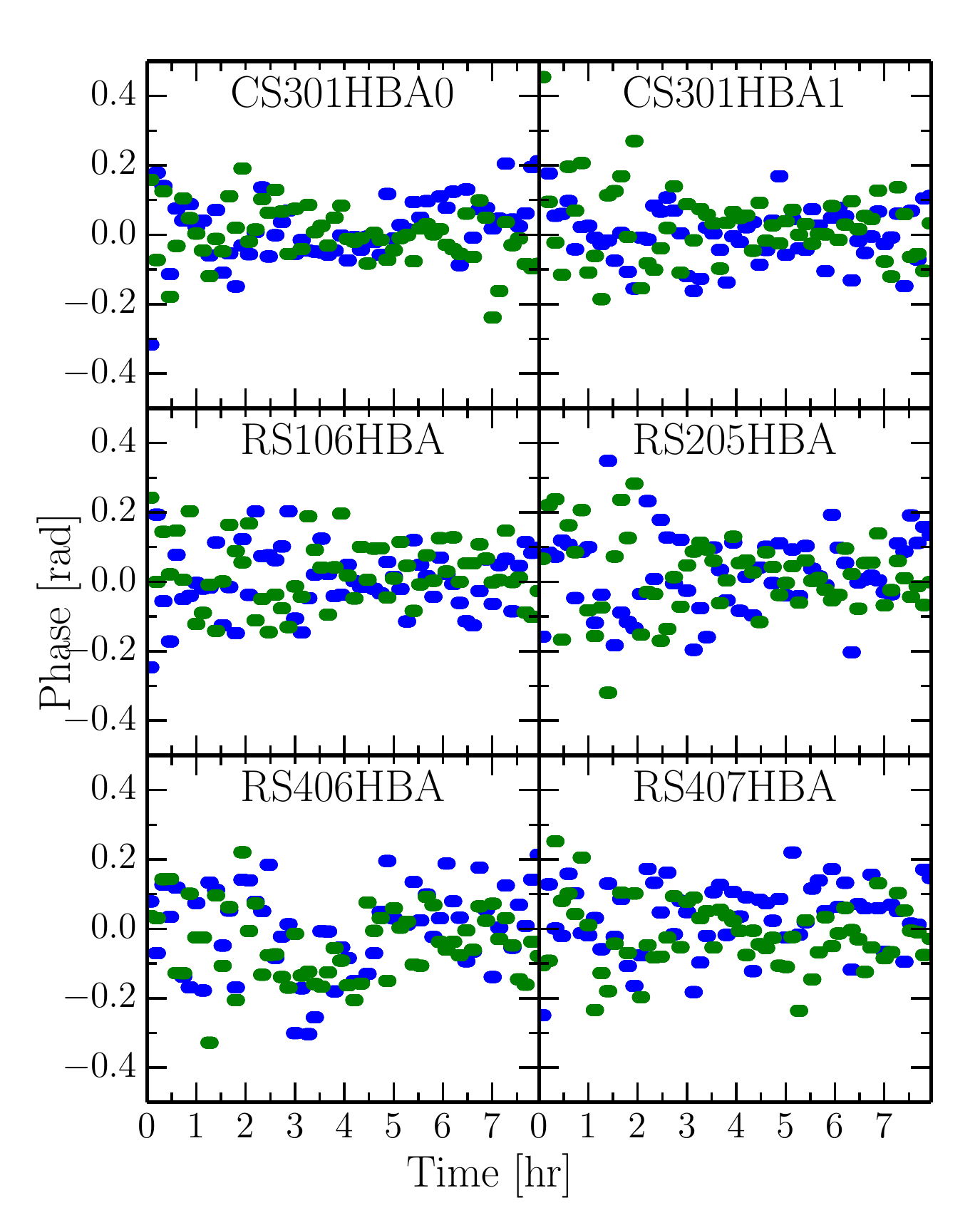} \\
\caption{{Example bad DDE solutions for a few selected stations obtained for a single direction s12. Left: The effective Stokes I phase corrections,  evaluated at an arbitrary frequency of $150$\,MHz. The solutions are obtained on a timescale of $8$\,s using $10$\,MHz
of bandwidth. Centre and Right: the additional XX and YY amplitude (centre) and phase (right) solutions for the $160-162$\,MHz SB block obtained on a timescale of $10$\,min after application of the short-timescale phase offsets and TEC solutions. In all cases  phases are plotted with respect to core station CS001HBA0.}}
\label{fig:selfcalsols_bad}
\end{figure*}

\begin{figure}
 \centering
\includegraphics[width=0.48\textwidth]{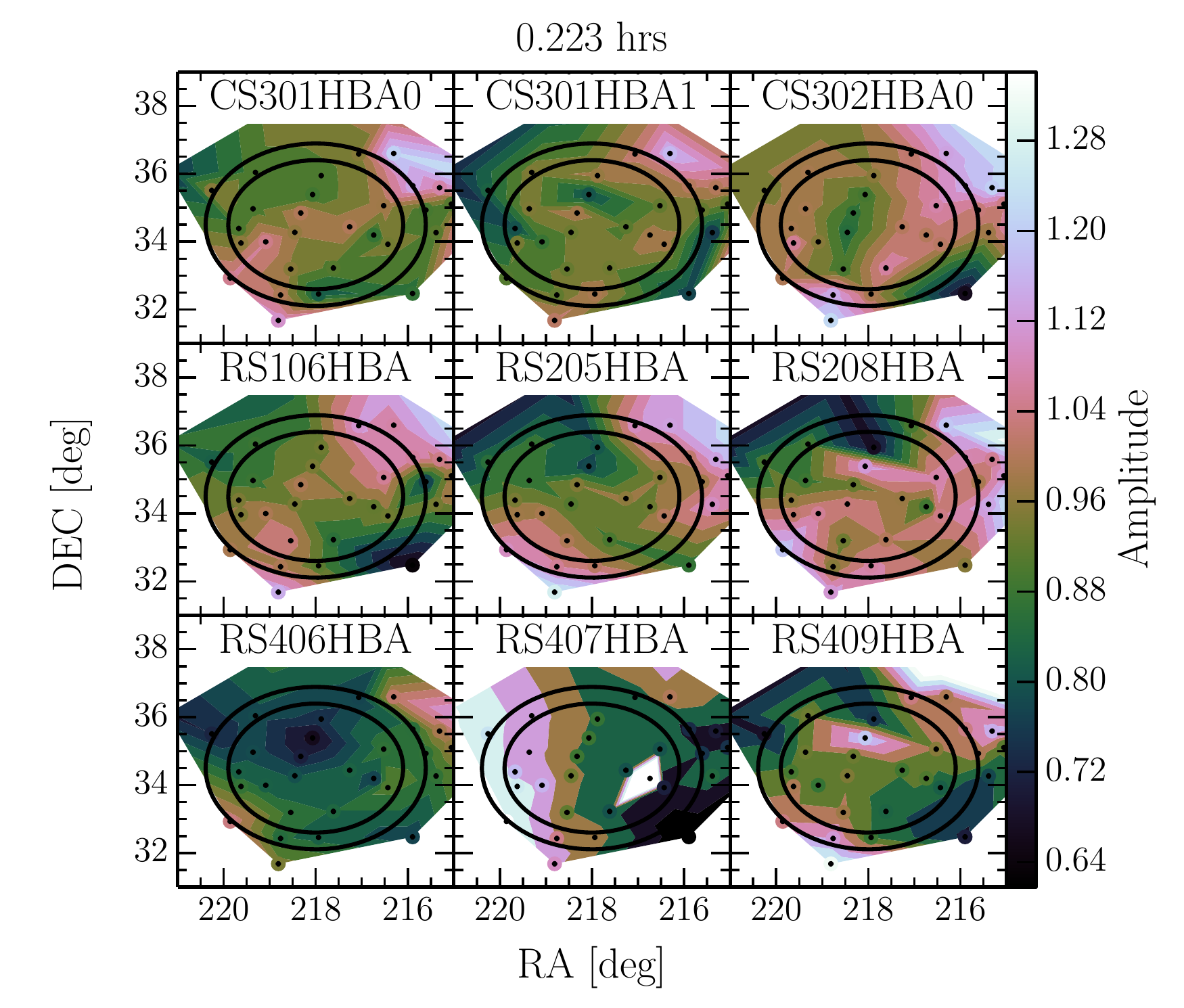}\\
\includegraphics[width=0.48\textwidth]{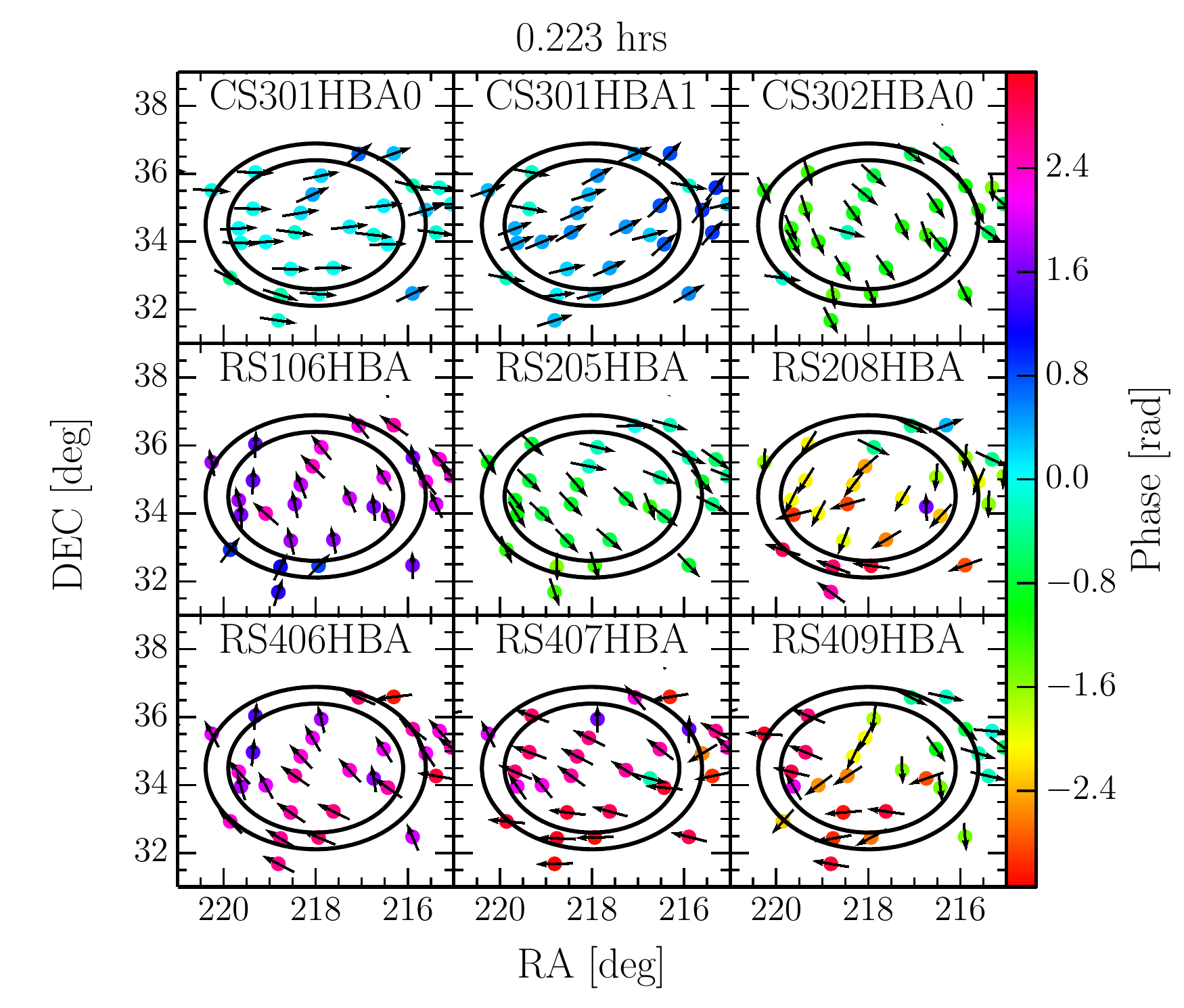}\\
\caption{Example good DDE solutions for a few selected stations obtained for all directions for a given timestep. The points show the facet centres. Top: XX amplitude solutions for the $160-162$\,MHz SB block obtained within a particular $10$\,min time interval. Bottom: The effective Stokes I phase corrections  evaluated at an arbitrary frequency of $150$\,MHz and plotted as an angle. In both plots, the two outer and inner circles show, respectively, the $30$ and $50$~per~cent power points of the average station beam.  The solutions are obtained on a timescale of $8$\,s using $10$\,MHz
of bandwidth. Phases are again plotted with respect to core station CS001HBA0.}
\label{fig:selfcalsols1}
\end{figure}

\begin{figure}
 \centering
\includegraphics[width=0.45\textwidth]{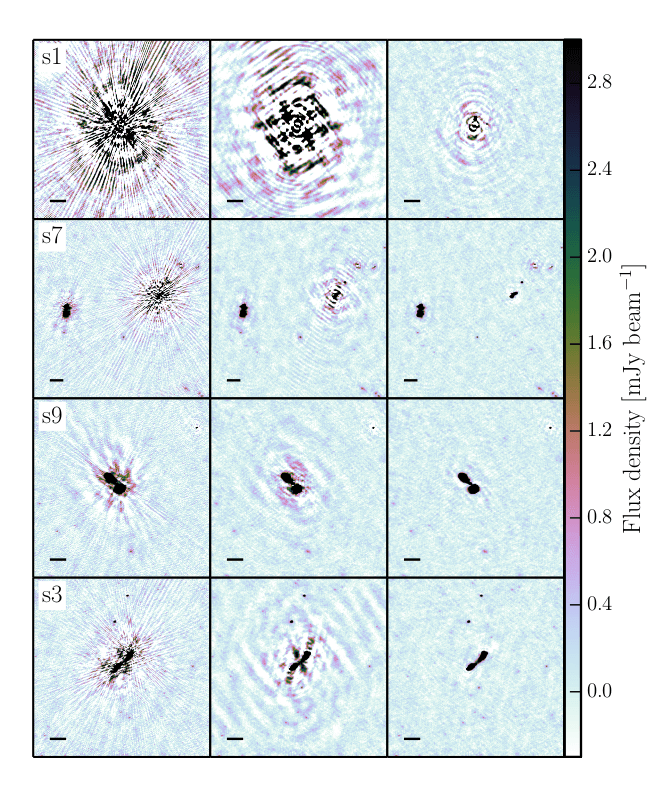}\\
\caption{Images showing the improvements during the DDE calibration for a few example directions. All images are made using the full dataset ({$130-169$\,MHz}, {\tt nterms}$=2$, {\tt robust}$=-0.25$) and have a resolution of $5.6 \times 7.4$\,{\arcsec}. Note that at this resolution many of the bright DDE calibrator sources are resolved. The leftmost column shows the initial images made with only the direction-independent self-calibration solutions. The centre column displays the improvements after two iterations of {fast phase (TEC and phase-offset)-only DDE calibration}. The right column  shows the improvement after two further iterations of {fast phase (TEC+phase offset) and slow phase and amplitude  (XX and YY gain) DDE calibration}. For all four directions shown, the TEC+phase offsets were solved for on $8$\,s time intervals. The XX and YY gains were solved for on  $10$\,min timescales. The scalebar in each image is $1${\arcmin}.}
\label{fig:selfcalstages}
\end{figure}

Example good and bad solutions for a small sample of stations are shown in Figs.~\ref{fig:selfcalsols} and ~\ref{fig:selfcalsols_bad} for a single direction. {We visually inspected these solutions and images at each self-calibration step for each direction. While the bad phase solutions do appear almost decorrelated there is still clear improvement in the image quality for these directions. We accept the bad solutions since we require solutions for each facet direction in order to fully cover the field of view.} Fig.~\ref{fig:selfcalsols1} shows example solutions for all directions as a snapshot in time. {We note that there is consistency in both the phase and amplitude solutions for the discrete directions across the field of view where the solutions for each direction have been obtained independently. This shows that the solutions do make physical sense. We have not yet attempted to spatially filter  or smooth the solutions, which would reduce outliers such as the one clearly visible in the amplitudes for station RS407HBA in Fig.~\ref{fig:selfcalsols1}. Indeed, an additional improvement in this method may be to interpolate the solutions between calibrator sources}. When viewed as a movie in time, trends can be seen `moving across' the field of view, in particular in the fast phases, which is consistent with ionospheric phase disturbances propagating through the field of view. The significant improvement made in the self-calibration cycles is demonstrated by the calibrator images shown in Fig.~\ref{fig:selfcalstages}. It is clear that both the phase \emph{and} amplitude calibration are currently required: the phase distortions resulting from ionospheric effects will always have to be corrected for, but future improvements in the LOFAR beam models may eliminate the need for the amplitude calibration. 

{The total number of parameters solved for across all $34$ facets is about $0.9$\,million, approximately $250$ times less than the $\sim 220$\,million visibilities. This is well within the requirement that the number of degrees of freedom be significantly less than the available visibilities. For comparison, solving for all $8$ Jones terms on short timescales in $34$ directions would result in over $20$ times as many parameters.}

\subsubsection{Facet imaging and subtraction}
\label{sect:facet_im}
After obtaining the direction-dependent calibration solutions for the given direction, the remaining sources within the facet were added back (with the direction-independent solutions with which they had been subtracted). Assuming the solutions for the calibrator group apply to the full facet, we applied those solutions to the facet data at the original $2$\,channels per SB resolution, which allows the $1/\nu$ dependence of the TEC term to be applied on a channel-to-channel basis. At this point the corrected facet data were averaged  $5$ times in frequency and $3$ times in time (to $0.5$\,MHz per channel and $24$\,s) to avoid excessive bandwidth and time smearing within the facet image. Each facet was then imaged with MS-MFS {\scshape clean} with {\tt nterms=2} over the full $130-169$\,MHz bandwidth, with a Briggs robust parameter of $-0.25$, $1.5$\,{\arcsec} pixels and a $uv$-minimum of $80$\,$\lambda$. Note that since all facets have different phase centres, their $uv$ coverage differs slightly and so the restoring beams are slightly different. {The facet masks were used to {\scshape clean} only within the facet. Sources outside the facet boundary appear only as residuals because they were not added back in the $uv$ data.}  As in Section~\ref{sect:selfcaltarget} we do not use {\scshape AWImager} mainly due to its limitations in imaging beyond the HPBW of the station beam.

Each facet image provided an updated sky model that was then subtracted from the full-resolution data with the corresponding direction-dependent solutions, thereby improving the residual data to which the subsequent facets were added. This process (from Section~\ref{sect:startdir}) was repeated until all facets had been calibrated and imaged. The order in which the facets were handled is determined by the severity of the calibration artefacts in the direction-independent images, which roughly corresponds to the brightness of the calibration groups, so that the directions with the worst artefacts were corrected first and did not influence later directions. 

After all the facets were calibrated in this way, we re-imaged all the facets {without doing any additional frequency or time averaging (i.e. leaving the data at $0.1$\,MHz per channel and $8$\,s)}. This step removes the artefacts present in the given facet  resulting from bright sources in neighbouring facets which had only been calibrated after  the given facet. It reduces the rms by a few per cent. This was achieved by adding back the facet sky model to the residual data,  applying the relevant directional-dependent solutions, and imaging with the same parameters (using the full $130-169$\,MHz bandwidth with {\tt nterms=2}, {\tt robust}$=-0.25$, a pixel-size of $1.5$\,{\arcsec}, and a $uv$-minimum of $80$\,$\lambda$). At this point we applied a common restoring beam of $5.6 \times 7.4$\,{\arcsec} to all the facets; this resolution was chosen as the smallest beam containing all the fitted restoring beams from the individual facets. {The facet templates were constructed such that, when regridded to a single image centred on the pointing centre, the facet images do not overlap by a single pixel. Thus, a single `mosaic' image was constucted by regridding the facet images, replacing the pixels outside the facet boundary with zeroes and summing the images. In this way there are no clear facet boundaries in the final image. Only four sources lie on the facet boundaries - i.e. they have {\scshape clean} components in two facets. For two of these sources that are found in the GMRT catalogue, their total flux is within the errors of the GMRT fluxes, so we infer that the total flux of sources on the boundaries is conserved. }

The resulting products are (i) high-resolution images of the facets, combined in a single image (`mosaic') covering the full field of view, (ii)  {short-timescale phase corrections for the variation of the ionosphere in each facet direction},  and (iii) long-timescale phase and amplitude corrections for the station beams {in each facet direction}.

\subsubsection{`Primary beam' correction}
\label{sect:pbcor}
{\scshape Casa} was used to image the individual facets so the images are in `apparent' flux units, with the station beam taken out only in the phase centre. In order to measure real sky fluxes we performed a primary-beam correction  to take into account the LOFAR station beam response across the field of view. We used {\scshape AWImager} to produce an average primary beam map from all the $20$ $10$-SB datasets, using the same Briggs weighting ({\tt robust}$=-0.25$), but with much larger pixels ($5.5$\,{\arcsec}), and imaging out to where the average station beam power drops to $20$~per~cent (approximately $6.2${\degree} in diameter). We corrected the combined `mosaic' image by dividing by the regridded {\scshape AWImager} average primary beam map\footnote{The actual average primary beam map used is the square root of the {\scshape AWImager} output.}. We imposed a `primary beam' cut where the average primary beam power drops below $40$~per~cent{\footnote{This is quite a liberal choice, but it fully includes the optical-infrared coverage of the Bo\"otes field.}}, at an approximate radius of $2.44${\degree}, which results in an image covering a total area of $\approx19$\,deg$^2$.

\section{Final Image and Catalogue}
\label{sect:imcat}

\begin{figure*}
 \centering
\includegraphics[width=\textwidth, trim=0.2cm 0.6cm 0.2cm 0.2cm, clip]{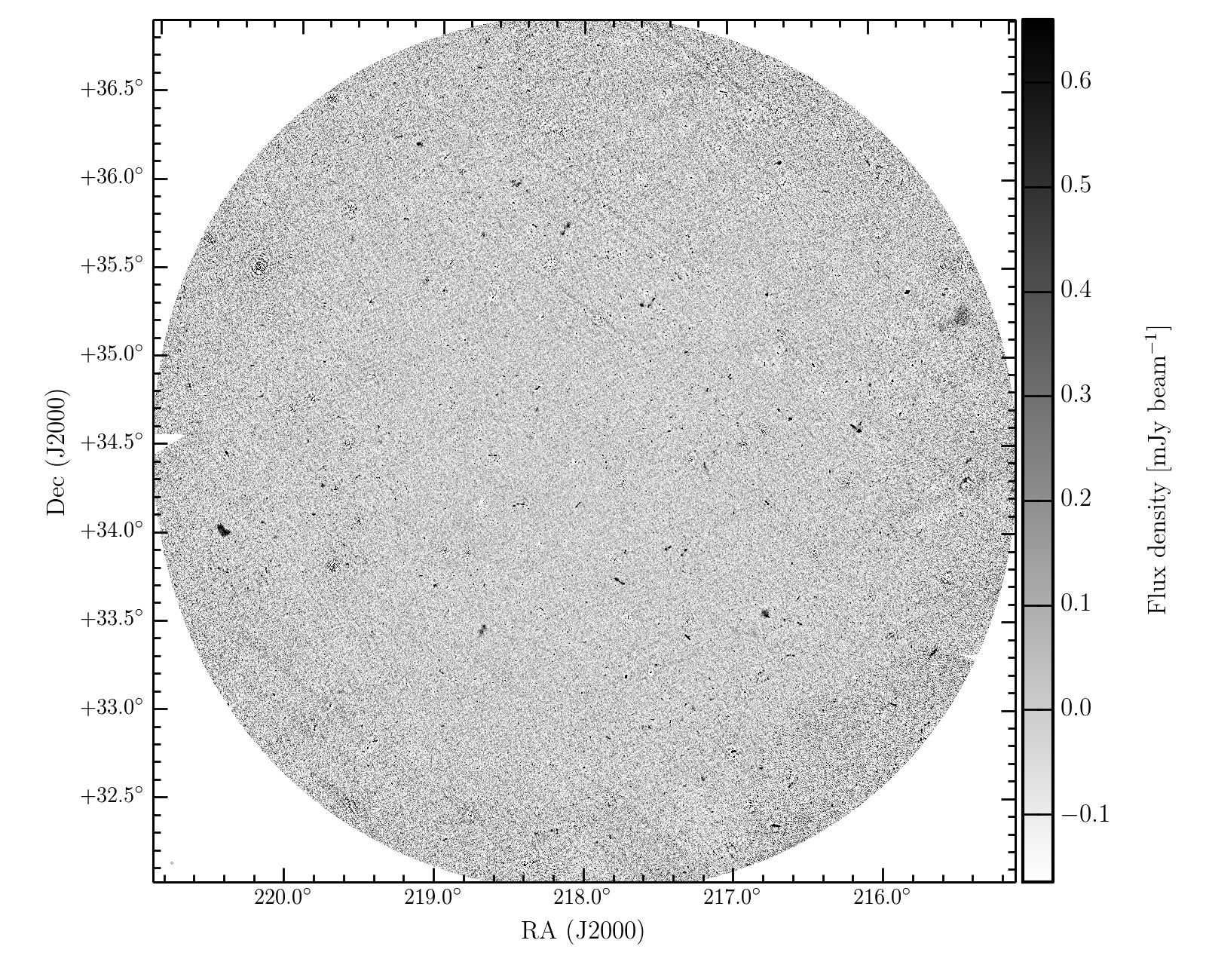}
\caption{Greyscale map showing the entire mosaic. The image covers $\approx19$\,deg$^2$. The greyscale shows the flux density from $-1.5\sigma_{\rm cen}$ to $6\sigma_{\rm cen}$ where  $\sigma_{\rm cen} = 110$\,{\muJybeam} is the approximate rms in the mosaic centre.}
\label{fig:mosaic}
\end{figure*}

\begin{figure*}
 \centering
\includegraphics[width=0.52\textwidth]{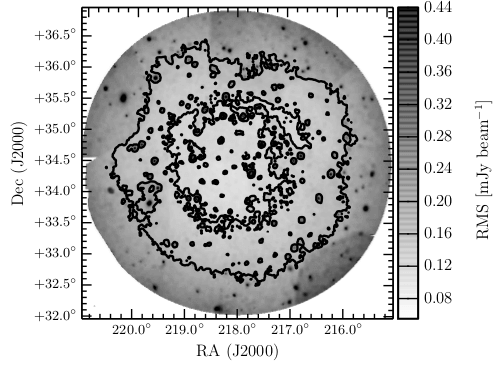}
\includegraphics[width=0.47\textwidth]{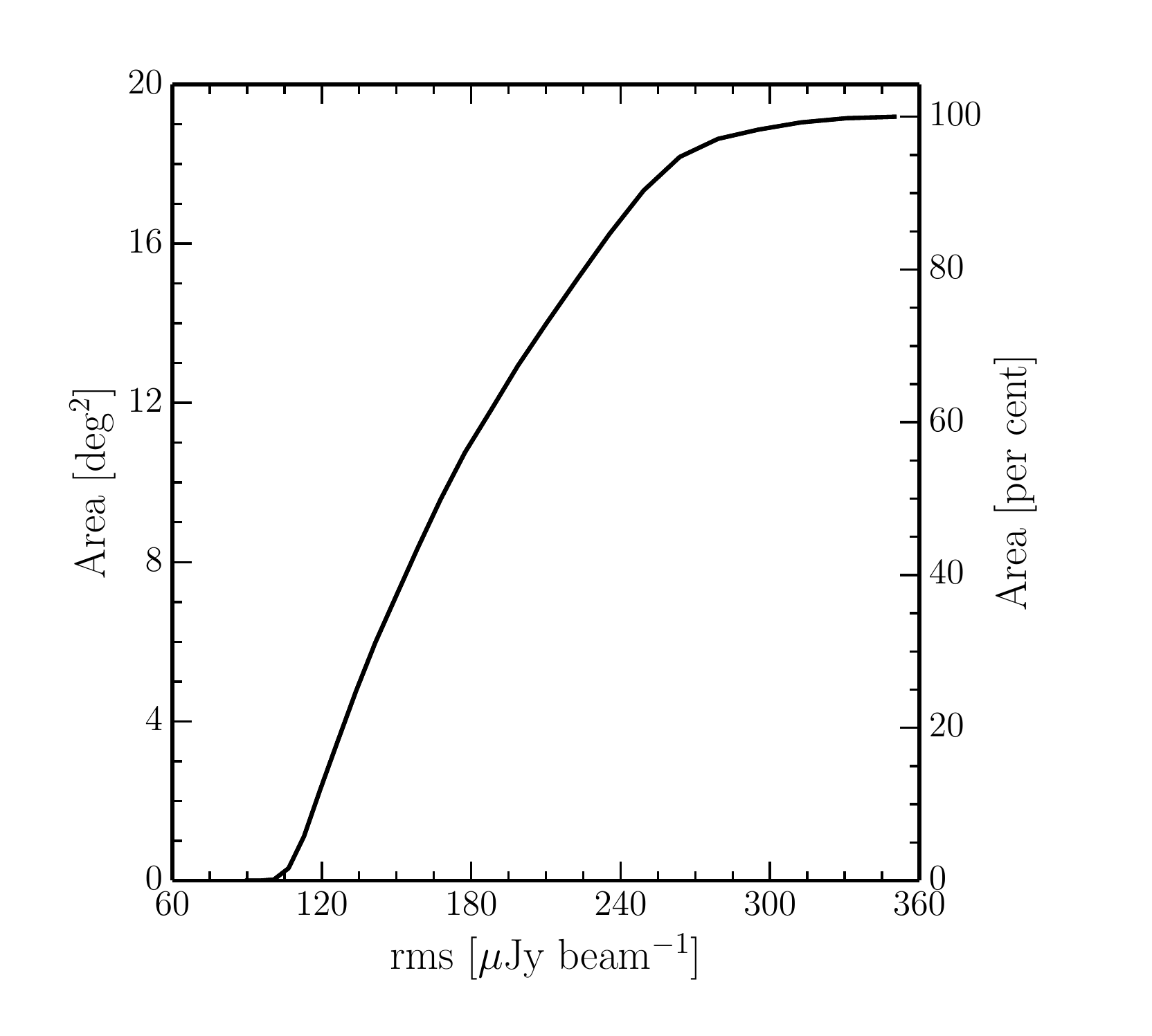}
\caption{Left:  Greyscale map showing the local rms noise measured in the mosaic image. The greyscale shows the rms noise from $0.5\sigma_{\rm cen}$ to $4\sigma_{\rm cen}$, where $\sigma_{\rm cen} = 110$\,{\muJybeam} is the approximate rms in the mosaic centre. The contours are plotted at $125$\,{\muJybeam} and $175$\,{\muJybeam}.  Peaks in the local noise coincide with the locations of bright sources. Right: Cumulative area of the map with a measured rms noise level below the given value.}
\label{fig:mosrms}
\end{figure*}

\begin{figure*}
 \centering
\includegraphics[width=\textwidth, trim=0.2cm 0.6cm 0.2cm 0.2cm, clip]{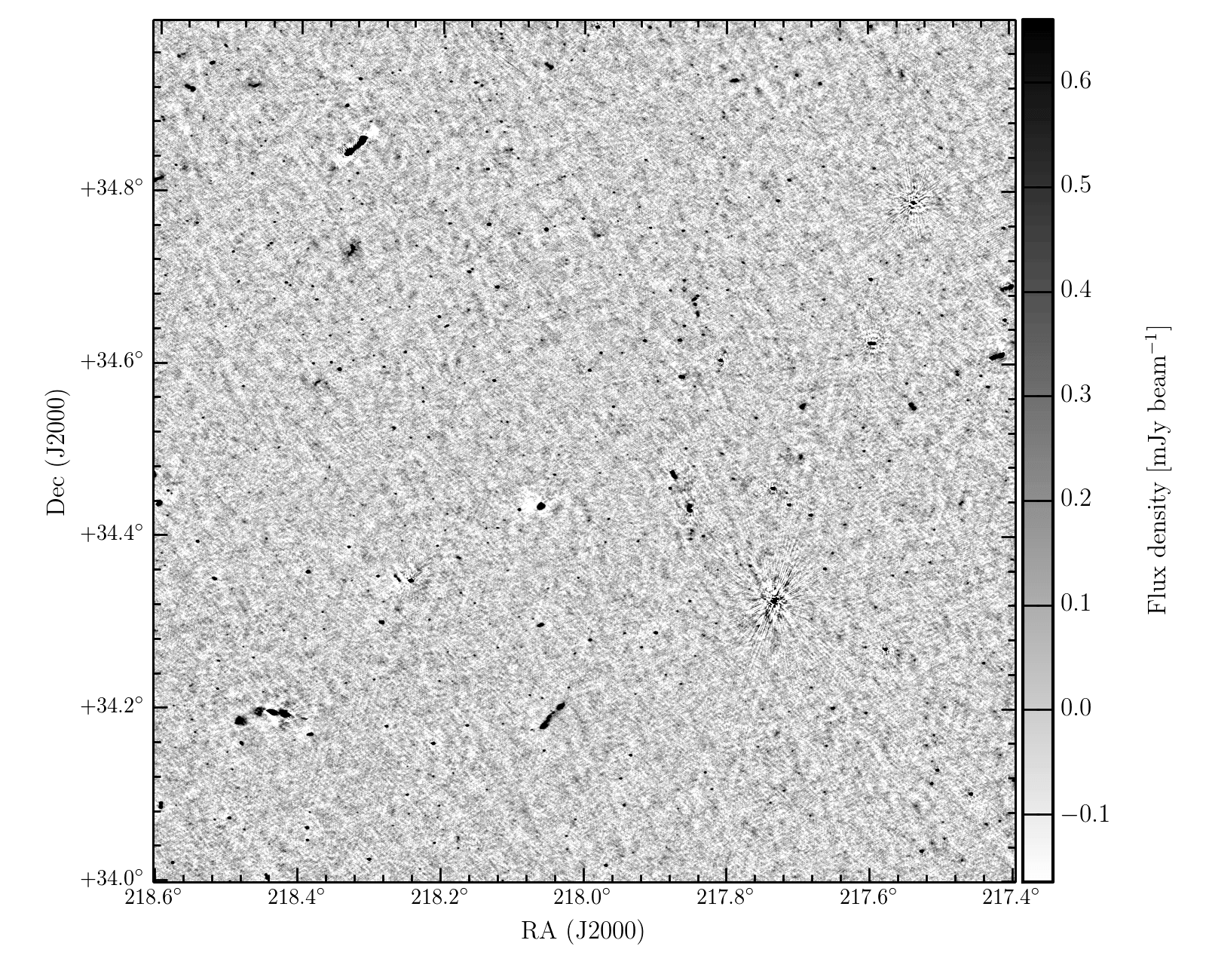}
\caption{Zoom-in of the central part of the mosaic. The image covers $0.25$\,deg$^2$. The greyscale shows the flux density from  $-1.5\sigma_{\rm cen}$ to $6\sigma_{\rm cen}$ where  $\sigma_{\rm cen} = 110$\,{\muJybeam} is the approximate rms in the mosaic centre.}
\label{fig:zoominnice}
\end{figure*}

\begin{figure*}
 \centering
\includegraphics[width=\textwidth, trim=0.2cm 0.6cm 0.2cm 0.2cm, clip]{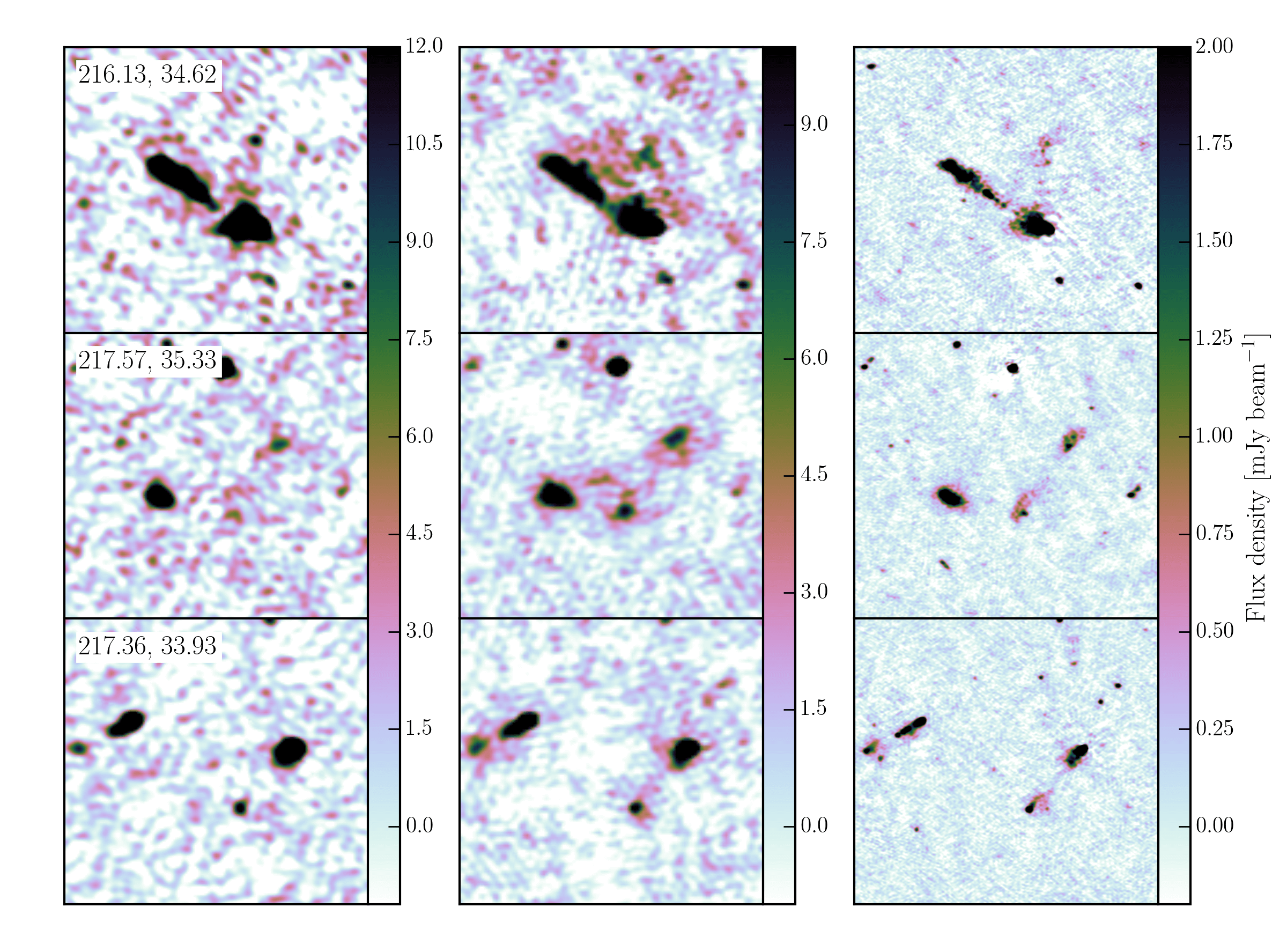}
\caption{Images showing a comparison between a few random sources in the GMRT image (\textit{left} column), the LOFAR $148-150$\,MHz direction-independent self-calibration image (\textit{centre} column, see Section~\ref{sect:selfcaltarget}) and the final $130-169$\,MHz direction-dependent calibrated image (\textit{right} column). The noise in the three images is respectively $\approx3$, $\approx1$ and $\approx0.15$\,{\mJybeam} and the resolution is $25$, $20$ and $5.6 \times 7.4$\,{\arcsec} respectively. Each image is $15${\arcmin} on a side. The image centre J2000 coordinates (right ascension and declination) are shown in degrees in the top left of each row. {Note that they are not all plotted on the same colourscale.}}
\label{fig:zoomincompare}
\end{figure*}

The combined `mosaic' image at $5.6 \times 7.4$\,{\arcsec} resolution is shown in  Fig.~\ref{fig:mosaic}. The central rms noise level is relatively smooth and $\lesssim125$\,{\muJybeam}, and $50$ per cent of the map is at a noise level below $180$\,{\muJybeam} (see also Fig.\ref{fig:mosrms}). {There is a small amount of `striping' (see e.g. the Northern part of the image) indicative of some low-level residual RFI. This is localised and at the level of $<2\sigma$, but should be addressed before deeper images are made. } A small portion of the image covering the inner $0.25$\,deg$^2$ is shown in Fig.~\ref{fig:zoominnice} to illustrate the resolution and quality of the map. There remain some phase artefacts  around the brightest sources (see for example the source in the lower right of the image in Fig.~\ref{fig:zoominnice}), which have not been entirely removed during the facet calibration. {While these are localised around the bright sources and have little impact on the majority of the map, they do affect some nearby sources.} Fig.~\ref{fig:zoomincompare} shows a comparison between the $153$\,{MHz} GMRT image, the LOFAR $148-150$\,MHz direction-independent self-calibration image (see Section~\ref{sect:selfcaltarget}) and the final $130-169$\,MHz direction-dependent calibrated image for  three arbitrary positions. This serves to illustrate the significant improvement in both noise and resolution achieved with the new LOFAR observations over the existing GMRT data, which has an rms noise level of $\approx3$\,{\mJybeam} and resolution of $25$\,{\arcsec}.

\subsection{Source Detection and Characterisation}
\label{sect:sources}

We compiled a source catalogue using {\scshape PyBDSM}  to detect and characterise sources. We ran {\scshape PyBDSM} on the final `mosaic' image, using the pre-beam-corrected image as the detection image and the primary beam-corrected image as the extraction image. The rms map was determined with a sliding box {\tt rms\_box}~$=(160,50)$\,pixels (i.e.\ a box size of $160$\,pixels every $50$\,pixels), with a smaller box {\tt rms\_box}~$=(60,15)$\,pixels in the regions around bright sources (defined as having peaks exceeding $150\sigma${, where $\sigma$ is the sigma-clipped rms across the entire field}). Using a smaller box near bright sources accounts for the increase in local rms as a result of calibration aretefacts. For source extraction we used {\tt thresh\_pix}$=5\sigma$ and {\tt thresh\_isl}$=3\sigma$ (i.e.\ the limit at which flux is included in the source for fitting). Fig.~\ref{fig:mosrms} illustrates the variation in rms noise thus determined across the combined `mosaic' image.

{\scshape PyBDSM} fits each island with one or more Gaussians, which are subsequently grouped into sources. We used the {\tt group\_tol} parameter with a value of $10.0$ to allow larger sources to be formed. This parameter controls how Gaussians within the same island are grouped into sources; the value we chose is a compromise between selecting all Gaussians in a single island as a single source, thus merging too many distinct nearby sources, and selecting them as separate sources, thus separating the radio lobes belonging to the same radio source. Sources are classified as `S' for single sources and `M' for multiple-Gaussian sources. {\scshape PyBDSM} reports the fitted Gaussian  parameters as well as the deconvolved sizes, computed assuming the image restoring beam. Uncertainties on the fitted parameters are computed following \cite{1997PASP..109..166C}.  The total number of sources detected by {\scshape PyBDSM} in all the facets is  $6\,349$ comprised of $10\,771$ Gaussian components of which $3\,010$ were single-component sources. We allowed {\scshape PyBDSM} to include sources that were poorly fitted by Gaussians; these $197$ sources are included in the catalogue with the integrated flux density being the total flux density within the source island and flagged as having poor Gaussian fits (`Flag\_badfit') and have no associated errors. {Additionally, based on visual inspection, {$71$} sources  were flagged as artefacts near bright sources, or detections on the edge of the image, or otherwise bad (`Flag\_artefact', `Flag\_edge', `Flag\_bad'). These flags are included in the final catalogue presented in Section~\ref{sect:catalog}.}

\subsection{Resolved Sources}
\label{sect:sources_ext}

In the absence of noise, resolved sources can easily be identified based on the ratio of the integrated flux density to the peak flux density, $S_{\rm int}/S_{\rm peak} > 1$. However, since the uncertainties on $S_{\rm int}$ and $S_{\rm peak}$ are  correlated, the $S_{\rm int}/S_{\rm peak}$ distribution is skewed, particularly at low signal-to-noise. We note that the scatter that we observe is large and skewed towards values $>1$. This is in some part due to the effects of bandwidth- and time-smearing, which both reduce the peak flux densities of sources as a function of distance from the phase centre. Given the averaged channel, time and  imaging resolution, we estimate \citep[using the equations given by][{for the reduction in peak flux}]{1989ASPC....6..247B} the combined effect of bandwidth and time smearing due to the averaging in the full dataset to be of the order of $74$~per~cent (i.e.\ measured peak flux densities at  $74$~per~cent of the expected value) at the edge of our field at $2.4${\degree}. {Our estimate for the correction factor for each source due to the combined effect is included in the source catalogue (at each source position). }

\begin{figure*}
 \centering
\includegraphics[width=0.45\textwidth]{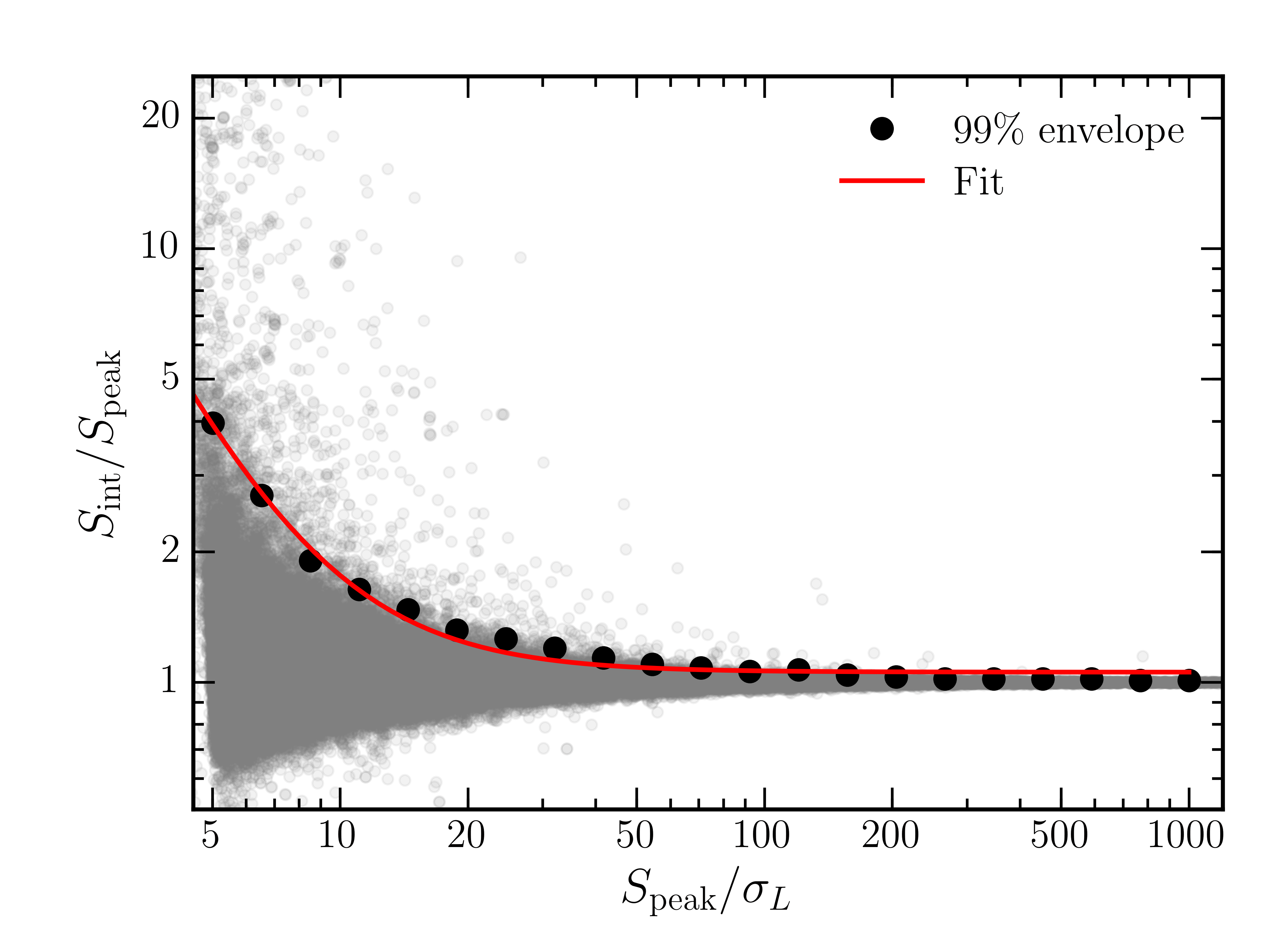}
\includegraphics[width=0.45\textwidth]{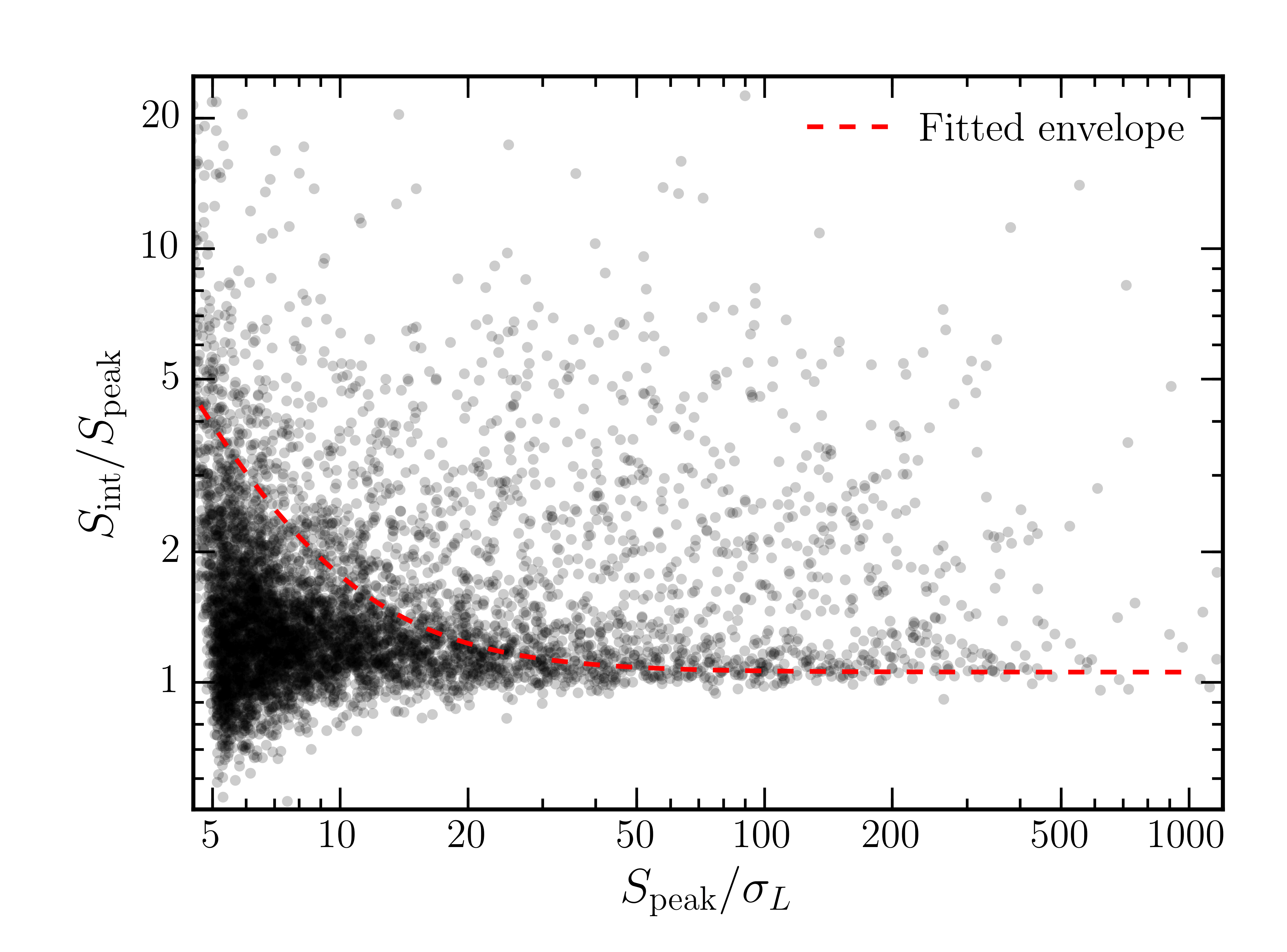}
\caption{\textit{Left} The simulated ratio of integrated to peak flux density as a function of signal-to-noise ratio for sources from the $20$ Monte-Carlo simulations. For $20$ logarithmic bins in signal-to-noise ratio, the black points show the threshold below which $99$~per~cent of the sources lie in that bin.   The red line shows a fit to this upper envelope. \textit{Right} The measured ratio of integrated to {(smearing-corrected)} peak flux density as a function of signal-to-noise ratio {for all sources in the catalogue}. The red line shows the fitted $99$~per~cent envelope. }
\label{fig:resolved}
\end{figure*}

The effect of noise on the total-to-peak flux density ratios as a function of signal-to-noise can be determined by running complete simulations in which artificial sources are injected into the real data and imaged and detected in the same way as the observed data. To determine an upper envelope of this distribution, we performed a Monte-Carlo simulation in the image plane in which we generated $20$ random fields containing $\sim20 000$ randomly positioned point sources with peak flux densities between $0.1\sigma$ and $20\sigma$, where $\sigma$ was taken to be $110${\,\muJybeam}. The source flux densities were drawn randomly from the real source count distribution, using a power law slope. Sources were injected in the residual mosaic map produced after source detection with {\scshape PyBDSM}.  Source detection was performed in the same manner described in Section~\ref{sect:sources}, thus only $\sim 5,000$ sources in each field satisfy the detection criterion of peak flux density $>5\sigma$. The  $S_{\rm int}/S_{\rm peak}$ distribution produced from the Monte-Carlo simulation is plotted in the left panel of Fig.~\ref{fig:resolved}, {after removing artefacts and false sources detected in the residual mosaic map}. To determine the $99$ per cent envelope, a curve was fit to the 99th percentile of $20$ logarithmic bins across signal-to-noise ratio. {The fitted envelope is characterised by by a function of the form $1.06+74.6/SNR^{2.02}$.} 

{The flux-density ratio as a function of signal-to-noise for our catalogued sources is shown in the right panel of Fig.~\ref{fig:resolved}, where we have used the smearing-corrected peak flux densities.  We therefore consider the $1,748$ sources above the fitted envelope from the point source simulation as resolved. These are flagged as resolved  in the final catalogue presented in Section~\ref{sect:catalog}. Note that not all the {\scshape PyBDSM} sources with  multiple Gaussian components are resolved by this criterion. Conversely, not all single-component sources are unresolved.}

{Additionally, by visual inspection of the images, we have identified $54$ large extended sources that appear as separate sources within the {\scshape PyBDSM} catalogue because the emission from the lobes of these giant radio galaxies is not contiguous. These sources span sizes (largest angular size, LAS) of $\approx20$\,{\arcsec} to $\approx250$\,{\arcsec}. We have merged these components into single sources in the final catalogue, by taking the flux-weighted mean positions, summing their total flux densities, and retaining the maximum peak flux density value. these merged sources are flagged in the catalouge (`Flag\_merged'). All but one of these sources meet the envelope criterion above and so are also classified as `resolved`. All the sources with LAS$>45$\,{\arcsec} (including these merged sources and those already identified by {\scshape PyBDSM}),  are presented in Fig.~\ref{fig:app:extendedsources} in the Appendix.}

Many diffuse extended sources are also clearly visible in the facet images. These sources are not detected by {\scshape PyBDSM}  as their peak flux densities fall below the detection threshold. We identifed four very clear large diffuse sources -- see  Fig.~\ref{fig:app:diffusesources} in the Appendix. A full study of diffuse emission is deferred to a subsequent paper, as this will require re-imaging of the facets with optimized parameters.

\subsection{Flux Density Uncertainties}
\label{sect:errors}
In this section we evaluate the uncertainties in the measured LOFAR flux densities. We make corrections to the catalogue we present in Section~\ref{sect:catalog}, to account for systematic effects.

\subsubsection{Systematic errors}
\label{sect:systerrors}
Given the uncertainties in the low-frequency flux density scale \citepalias[e.g.][]{2012MNRAS.423L..30S} and the LOFAR station beam models, we may expect some systematic errors in the measured LOFAR flux densities. In order to  determine any systematic offsets {and place the final catalogue onto the \citetalias{2012MNRAS.423L..30S} flux scale}, we have  compared the LOFAR flux densities to those of the GMRT image of the Bo\"otes field at $153$\,MHz \citep{2013A&A...549A..55W}. {The veracity of the GMRT flux densities was evaluated both by comparing flux densities measured in the overlap areas of the $7$ individual pointings and through comparison with NVSS \citep[the NRAO VLA Sky Survey;][]{1998AJ....115.1693C}. The assumed uncertainty on the GMRT flux density scale is $20$~per~cent. For the comparison we selected} only sources detected at high signal-to-noise ($S_{\rm peak}/\sigma > 10$) in both maps. {We have further limited the selection to \emph{isolated} LOFAR sources, i.e. those with no other LOFAR source within $25$\,{\arcsec} (the size of the GMRT beam), to ensure one-to-one matches. Finally, we consider only \emph{small} LOFAR sources, with measured sizes of less than $50$\,{\arcsec}, to rule out resolution effects, i.e. sources being resolved out by the poorer short baseline coverage of the GMRT.} {This yielded a sample of $420$ objects. Limiting the selection only to those unresolved by LOFAR (and hence the GMRT) would give only $129$ sources, so for the purposes of robust statistics we opt to select the larger \emph{small} source sample over unresolved sources only.} {Before the comparison, the GMRT flux densities were multiplied by $1.078$ to place them on the  \citetalias{2012MNRAS.423L..30S} flux scale, based on the calibration model used}. For this sub-sample of sources we determined the ratio of integrated flux densities between LOFAR and the GMRT $f_S = S_{\mathrm{LOFAR}}/S_{\mathrm{GMRT}}$. {We measured a median ratio of $\langle f_S \rangle= 1.10$ with a standard deviation of $\sigma_{f_S} = 0.20$.} 

\begin{figure}
 \centering
\includegraphics[width=0.45\textwidth]{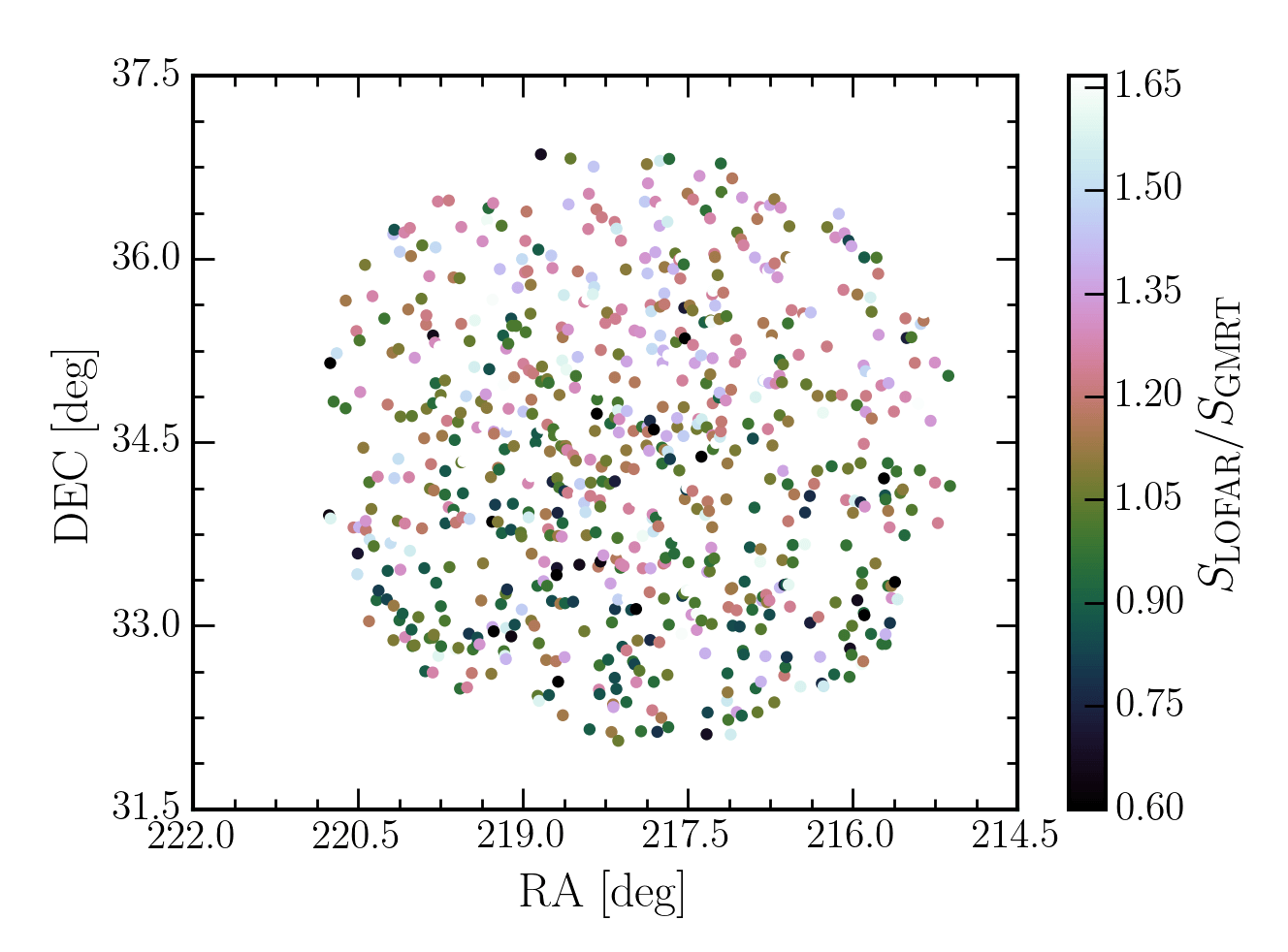}
\caption{Map of the measured ratios of integrated flux densities  for high signal-to-noise{, small and isolated} LOFAR sources with respect to the GMRT. The colourscale shows the the flux density ratio. }
\label{fig:flux_comparemap}
\end{figure}

The flux density ratio showed no significant variation with distance from the phase centre. However we noticed a small variation with position on the sky,  plotted  in  Fig.~\ref{fig:flux_comparemap}. The variation is such that the LOFAR flux densities are slightly higher towards the North-North-West (upper right of the map) and lower towards the South-South-East (lower left of the map). We note that the trend is consistent across different facets so is a result of some global effect and not a result of discrete calibration in facets. Moreover, the trend is similar, but noisier, if we consider LOFAR sources extracted from the self-calibrated-only image from the single $10$-SB block at $148$--$150$\,MHz. This further suggests that the flux density offset is not introduced by the direction dependent calibration procedure.

\begin{figure}
 \centering
\includegraphics[width=0.45\textwidth]{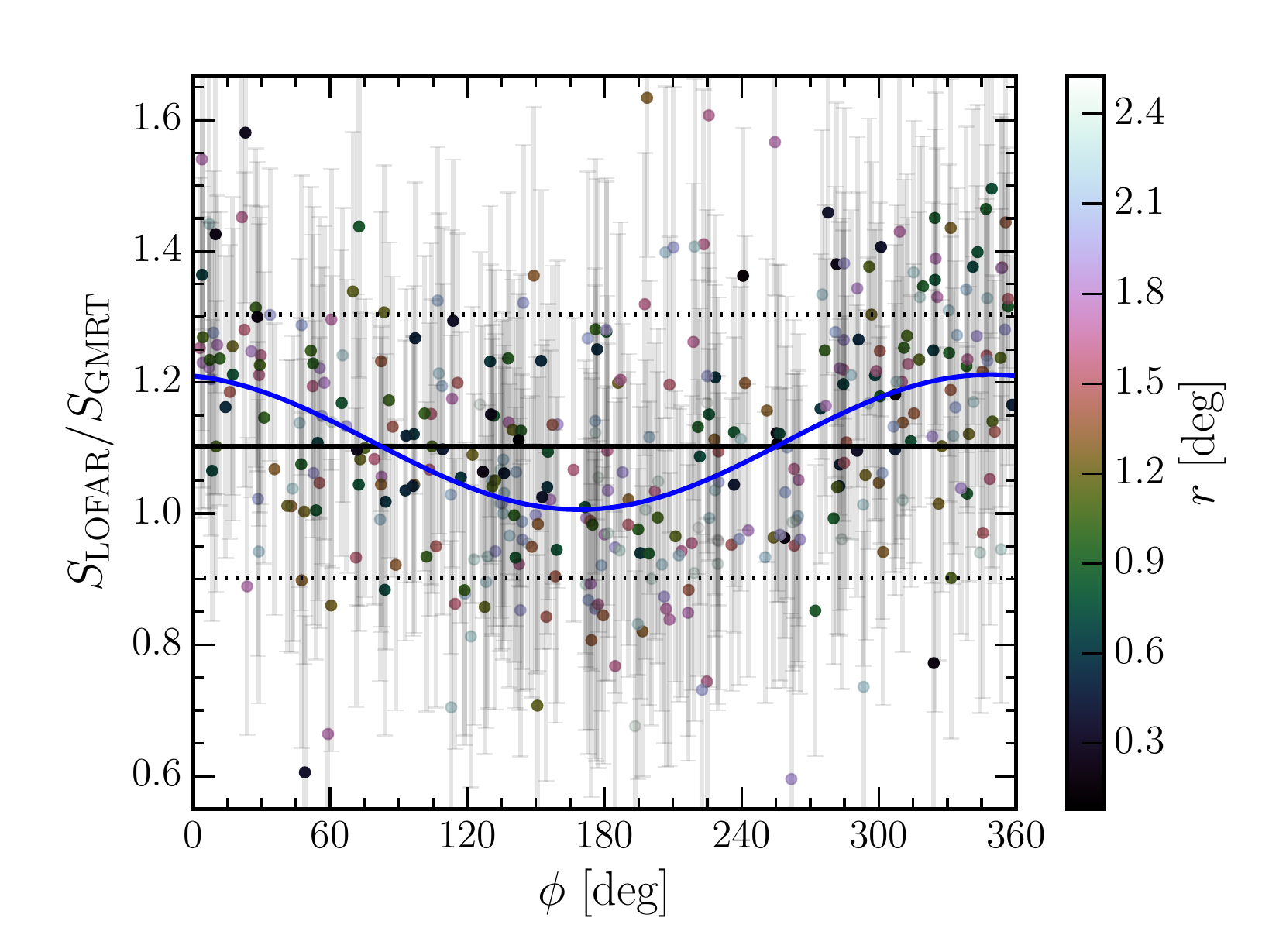}
\caption{Measured ratios of integrated flux densities  for high signal-to-noise{, small and isolated} LOFAR sources with respect to the GMRT as a function of source position angle with respect to the phase centre (plotted as points). The colourscale shows the distance of each source to the pointing centre. The blue curve shows a fitted sinusoid which we used for correcting the flux densities of the LOFAR catalogued sources. The solid and dashed horizontal lines show the measured median and standard deviation of the flux density ratio.}
\label{fig:flux_compare}
\end{figure}

The flux density ratio is plotted as  a function of the position angle between the source and the phase centre in Fig.~\ref{fig:flux_compare}, to which we fitted a  sinusoid of the form $f_S = 1.11+0.10\sin(101^{\circ}+\phi)${.\footnote{The trend is robust to the sample selection used. We note that the smaller \emph{unresolved} sample yields a quantitatively similar result with a slightly lower normalisation $f_S = 1.05+0.10\sin(90.21^{\circ}+\phi)$.}} The median ratio was $\langle f_S \rangle = 1.00$ with a standard deviation of $\sigma_{f_S} = 0.17$ after applying this sinusoid function to correct the LOFAR flux densities. This correction assumes that the GMRT flux densities are `correct'{ and, in particular, have no variation on the sky.}  {We emphasise that after making this correction to the LOFAR flux densities, they are on the  \citetalias{2012MNRAS.423L..30S} flux scale.}

To {validate} this correction we performed a similar comparison with sources in the deep $1.4$~GHz WSRT catalogue of the Bo\"{o}tes Field \citep{2002AJ....123.1784D}, {again considering only the isolated, small and bright sources}. Scaling the higher frequency flux densities to $150$\,MHz with a spectral index of $-0.8$ we recovered a similar, albeit noisier, trend. {Given the large uncertainty in spectral index scaling, we neglect the flux scale differences of only $~2-3$~per~cent between our  LOFAR flux densities and the WSRT catalogue, which was tied to the NVSS scale \citep{1977A&A....61...99B}. Moreover, in this case we are only investigating spatial variations, not any absolute scaling errors.} This suggests that it is likely not to be GMRT pointing errors causing systematic trends in the GMRT flux densities.   We speculate that the observed trend in the flux density errors may be the result of an incorrect primary-beam correction, which itself may be caused by errors in the LOFAR station beam model used in \textsc{AWImager}, or a map projection error. 

\subsubsection{Flux scale accuracy}
\begin{figure}
 \centering
\includegraphics[width=0.45\textwidth]{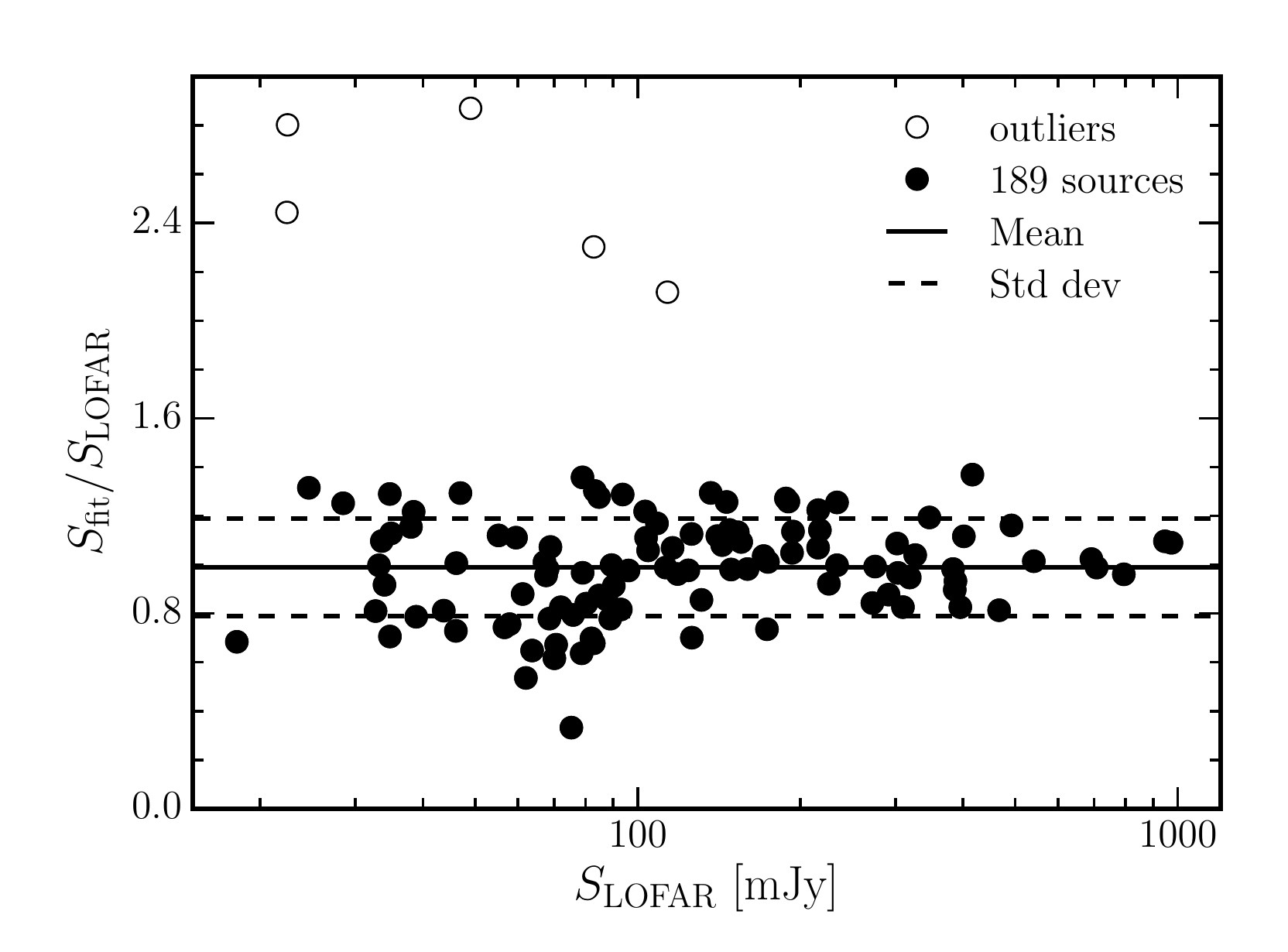}
\caption{{Comparison of LOFR measured integrated flux densities with those predicted from WENSS and NVSS at higher frequencies. The mean ratio of  $1.01  \pm 0.10$ shows that the LOFAR flux densities are on the same flux scale.  Five outliers, which are consistent with having spectra which turn over or are inverted, are excluded.}}
\label{fig:flux_scale_compare}
\end{figure}

To investigate the overall reliability of the fluxscale, we have compared $189$ small, isolated within 1\,{\arcmin}, and high signal-to-noise sources that are detected at higher frequencies both in NVSS at $1.4$\,GHz and WENSS at $325$\,MHz \citep[the Westerbork Northern Sky Survey;][]{1997A&AS..124..259R}. For these sources we computed the spectral index between the two higher frequencies and predicted the LOFAR flux density. The  WENSS fluxes were first scaled by a factor of $0.9$ to put them on the same flux scale \citepalias{2012MNRAS.423L..30S}. The predicted to observed flux density ratio is plotted in Fig.~\ref{fig:flux_scale_compare} for this sample.  We calculate a mean flux density ratio of   $1.01  \pm 0.10$ with standard deviation of $\sigma = 0.20$. We thus conclude that the corrected LOFAR flux densities are consistent with being on the \citetalias{2012MNRAS.423L..30S} flux scale. Moreover,  the scatter includes source-to-source variations in spectral shape and uncertainties in the spectral index fits, so we can conclude that the flux scale is accurate to better than $20$~per~cent. Five outliers, at greater than $4\sigma$, were excluded from the statistics. Based on inspection of the spectra of these outliers, considering also VLSS (at $74$\,MHz) and LOFAR LBA (at $62$\,MHz) detections and upper limits, they are consistent with having turnovers below $\approx200$\,MHz or inverted spectra.    

{\subsubsection{{\scshape clean} bias}}
{The {\scshape clean}ing process can redistribute flux from real sources on to noise peaks, resulting in a ``{\scshape clean}  bias'' \citep{1995ApJ...450..559B,1998AJ....115.1693C}. The effect is worse for observations with poor $uv$ coverage due to increased sidelobe levels. Despite our good $uv$ coverage, and {\scshape clean}  masking, we have checked for the presence of {\scshape clean}  bias in our images. We injected point sources with flux densities drawn from the distribution of real sources at random positions into the residual $uv$ data for a representative sample of facets. The simulated data were then imaged and {\scshape clean}ed in exactly the same manner as the real data, including automated {\scshape clean}  masking. We compared the measured peak flux density  with the input flux density and measure a difference consistent with zero. We therefore conclude that there is no significant {\scshape clean}  bias.}

\begin{figure}
 \centering
\includegraphics[width=0.45\textwidth]{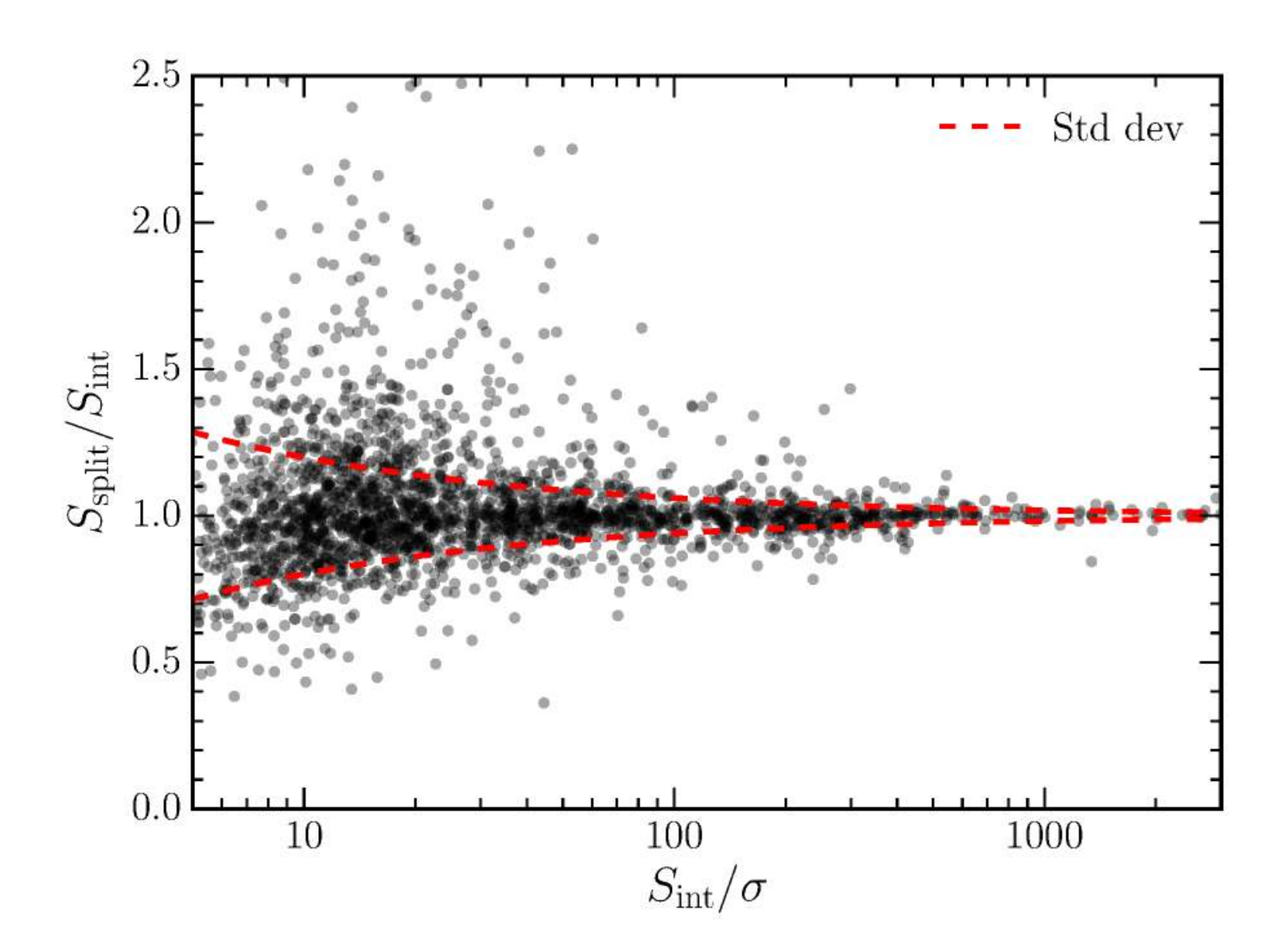}
\caption{{Integrated flux densities of sources extracted from images made using subsets of the data over shorter time chunks ($2$\,hrs each) compared to their fluxes in the final densities in the final LOFAR image. The red lines show the signal-to-noise-dependent standard deviation.}}
\label{fig:fluxes_errors}
\end{figure}

{\subsubsection{Flux Density Uncertainty}}
{The uncertainties of the flux density measurements themselves has been investigated by through a jackknife test, whereby we have split the visibility data into four time segments of $2$\,hrs each. Due to the computational expense, this was only done for a representative sample of seven facets. The $uv$ data for the subsamples were then imaged and sources extracted in a way identical to the original data. Fig.~\ref{fig:fluxes_errors}  shows a comparison of the measured total flux densities in the split images compared to the deep images as a function of the signal-to-noise ratio of each source in the split image. From this we determined the signal-to-noise-dependent standard deviation  of $0.66/SNR^{0.52}$. }

{\subsection{Astrometric Uncertainties}}
\label{sect:asterrors}

\begin{figure*}
 \centering
\includegraphics[width=0.45\textwidth]{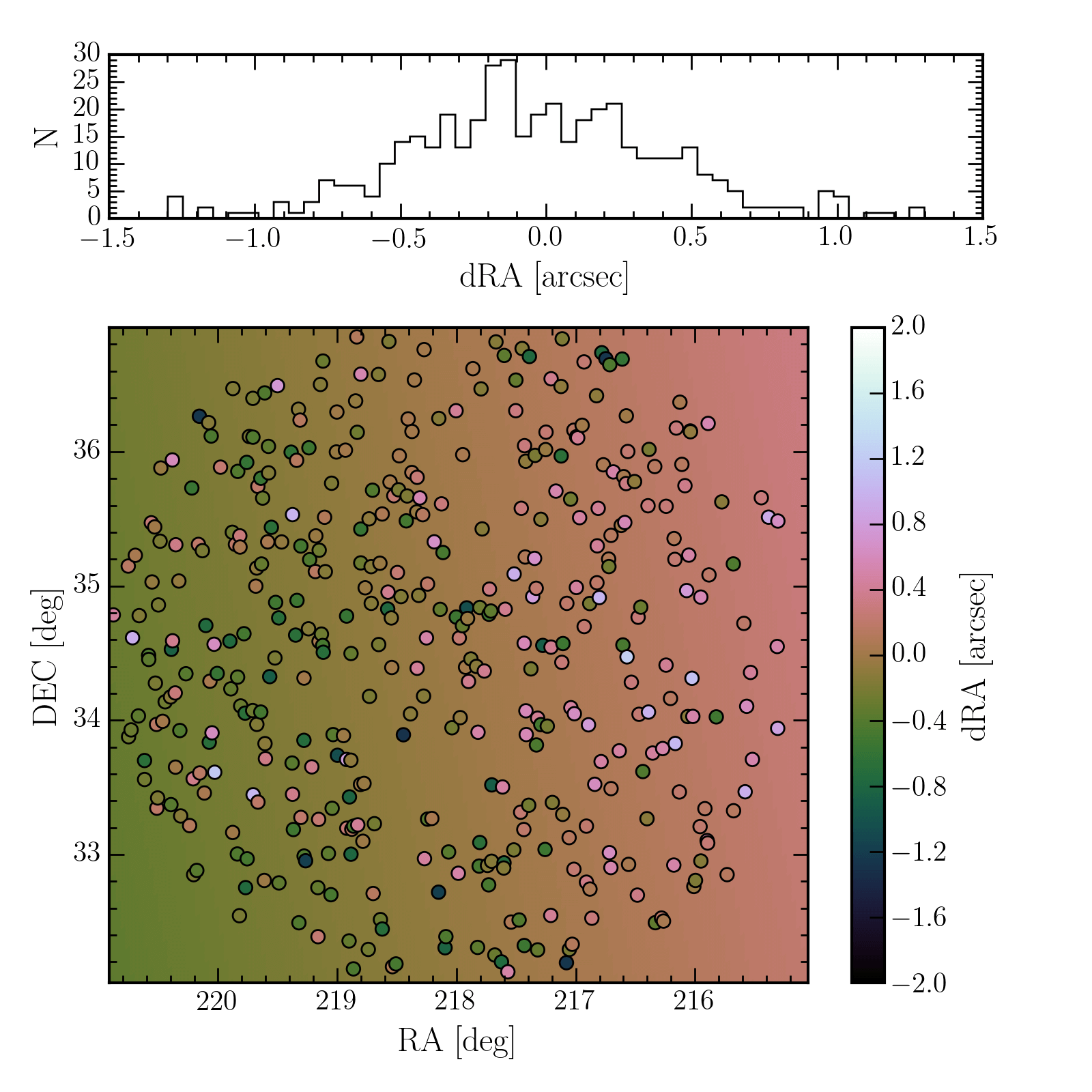}
\includegraphics[width=0.45\textwidth]{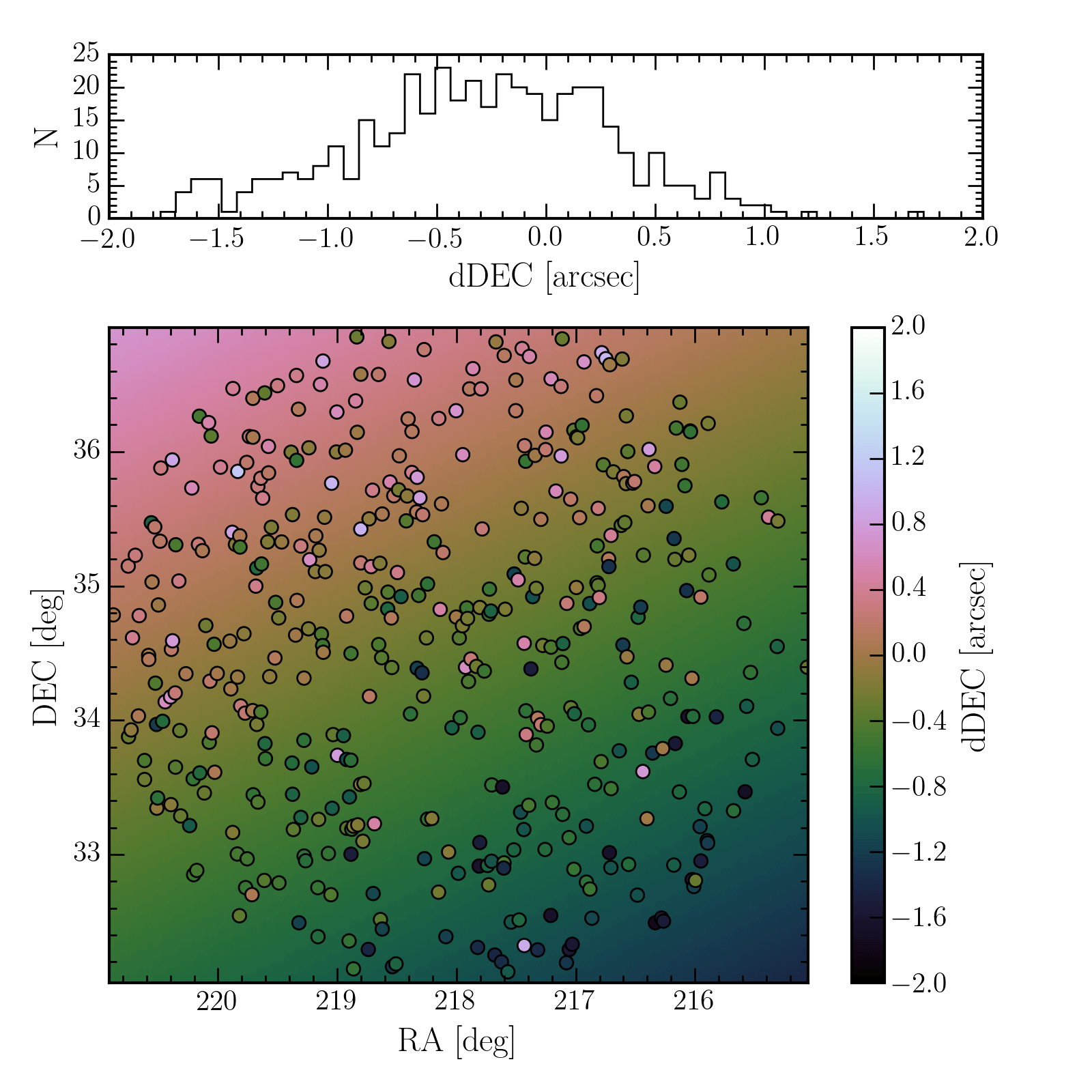}\\
\caption{Measured offsets in right ascension (left) and declination (right) for high signal-to-noise sources of LOFAR with respect to FIRST (plotted as points). The colourscale shows the value of the given offset. The top panels show the distribution of the offsets. The offsets, particularly in declination, show a variation across the field of view. The background colourscale shows a fitted plane used for correcting the LOFAR catalogued sources.}
\label{fig:pos_compare}
\end{figure*}

It is possible that errors in the phase calibration can introduce uncertainties in the source positions. Here we evaluate any systematic offsets in the measured source positions and determine their uncertainties.  We used the $1.4$~GHz FIRST catalogue \citep{1995ApJ...450..559B} to assess the LOFAR positional accuracy. The uncertainty on the FIRST positions is given by $\sigma = s\left( 1/SNR+ 0.05 \right)$\,{\arcsec}, where $s$ is the source size. FIRST counterparts were identified within $2\,{\arcsec}$ of our LOFAR sources with high signal-to-noise ratios, i.e.\ $S_{\rm peak}/\sigma_{\rm local} > 10$. This yielded $968$ matches, of which $313$ were fitted with a single Gaussian in the LOFAR image, making them more likely to be point sources. For the  high signal-to-noise point source sample, we measured a small mean offset in right ascension of 
$\Delta \alpha = \alpha_{\mathrm{LOFAR}} - \alpha_{\mathrm{FIRST}} = -0.037 \pm 0.001 \,{\arcsec}$, with rms $\sigma_\alpha = 0.44\,{\arcsec}$ and a somewhat larger mean offset in declination  of 
$\Delta \delta = \delta_{\mathrm{LOFAR}} - \delta_{\mathrm{FIRST}} = -0.301  \pm 0.001 \,{\arcsec}$ with rms $\sigma_\alpha = 0.59\,{\arcsec}$ . 
The offset is  negligible and we note that it is within the size of the LOFAR image pixels ($1.5$\,{\arcsec}). However, closer inspection of the offsets showed that the offset in declination varied systematically across the full $5{\degree}$ field of view, and to a lesser extent for the offsets in right ascension. A correction for this offset has been made by fitting a plane, $\Delta = a\alpha +b\delta +c$, where $\Delta$ is in {\arcsec}, to the offset values and applying the fitted offsets to all sources in the catalogue. While this could be expressed as a rotation and therefore a sinusoidal correction could be made, as we have done for the flux densities, we find that the positional offests are asymmetric with respect to the pointing centre. The fitted planes were 
$\Delta \alpha =-0.10(\alpha-218{\degree}) +0.02(\delta-34.5{\degree}) -0.01 $ and  
$\Delta\delta = 0.13(\alpha-218{\degree}) +0.29(\delta+34.5{\degree}) -0.34$. 
These offsets and functional corrections are shown in Fig.~\ref{fig:pos_compare}. After applying this correction we measured 
$\Delta \alpha = \alpha_{\mathrm{LOFAR}} - \alpha_{\mathrm{FIRST}} = -0.018 \pm 0.002 \,{\arcsec}$ with rms deviation $\sigma_{\Delta\alpha} = 0.42\,{\arcsec}$, and  
$\Delta \delta = \delta_{\mathrm{LOFAR}} - \delta_{\mathrm{FIRST}} = -0.008 \pm 0.001 \,{\arcsec}$, with $\sigma_{\Delta\delta} = 0.31\,{\arcsec}$. We note that since the initial phase calibration of the LOFAR data was performed against the GMRT model sky, these positional offsets may originate from the GMRT model. We did evaluate the GMRT positions in a similar way and found no significant variations, but this may be due to the much larger positional uncertainty and the $25$\,{\arcsec} synthesised beam of the GMRT map.  We also consider LOFAR sources extracted from the self-calibrated-only image from the single $10$-SB block at $148$--$150$\,MHz compared to FIRST and find a similar trend. This  suggests that the position offsets are not  introduced by the direction dependent calibration procedure.  We hypothesise that a map projection or mosaic construction error could be the cause of these offsets, but note that it is a small correction.

The scatter in the offsets between the LOFAR and FIRST positions is a combination of noise-independent  calibration errors, $\epsilon$, in both the LOFAR and FIRST data as well as a noise-dependent error, $\sigma$, from position determination via Gaussian-fitting:
\[ \sigma^2 = \epsilon_{\mathrm{LOFAR}}^2 + \epsilon_{\mathrm{FIRST}}^2  + \sigma_{\mathrm{LOFAR}}^2 + \sigma_{\mathrm{FIRST}}^2\]
To separate the noise-dependent and -independent uncertainties we select from the above sample only the FIRST sources with position errors of less than $0.6\,{\arcsec}$ and measure an rms scatter of $(\sigma_{\alpha}, \sigma_{\delta})_{\mathrm{LOFAR}} = (0.37\,{\arcsec}, 0.35\,{\arcsec})$  between the corrected LOFAR and FIRST source positions for this very high signal-to-noise sub-sample of $89$ sources. 
From \cite{1995ApJ...450..559B}, the FIRST calibration errors are $(\epsilon_{\alpha}, \epsilon_{\delta})_{\mathrm{FIRST}} = (0.05\,{\arcsec}, 0.05\,{\arcsec})$.  The noise-dependent fit errors for both the LOFAR and FIRST can be assumed to be small so we determine the LOFAR calibration errors to be $(\epsilon_{\alpha}, \epsilon_{\delta})_{\mathrm{LOFAR}} = (0.37\,{\arcsec}, 0.35\,{\arcsec})$. This  scatter may contain a small contribution resulting from any spectral variation between $150$ and $1400$\,MHz on scales smaller than the resolution of the surveys ($\approx5$\,{\arcsec}). 

\subsection{Completeness}
\label{sect:completeness1}

\begin{figure*}
 \centering
\includegraphics[width=0.48\textwidth, trim=0cm 0cm 0cm 0.825cm, clip]{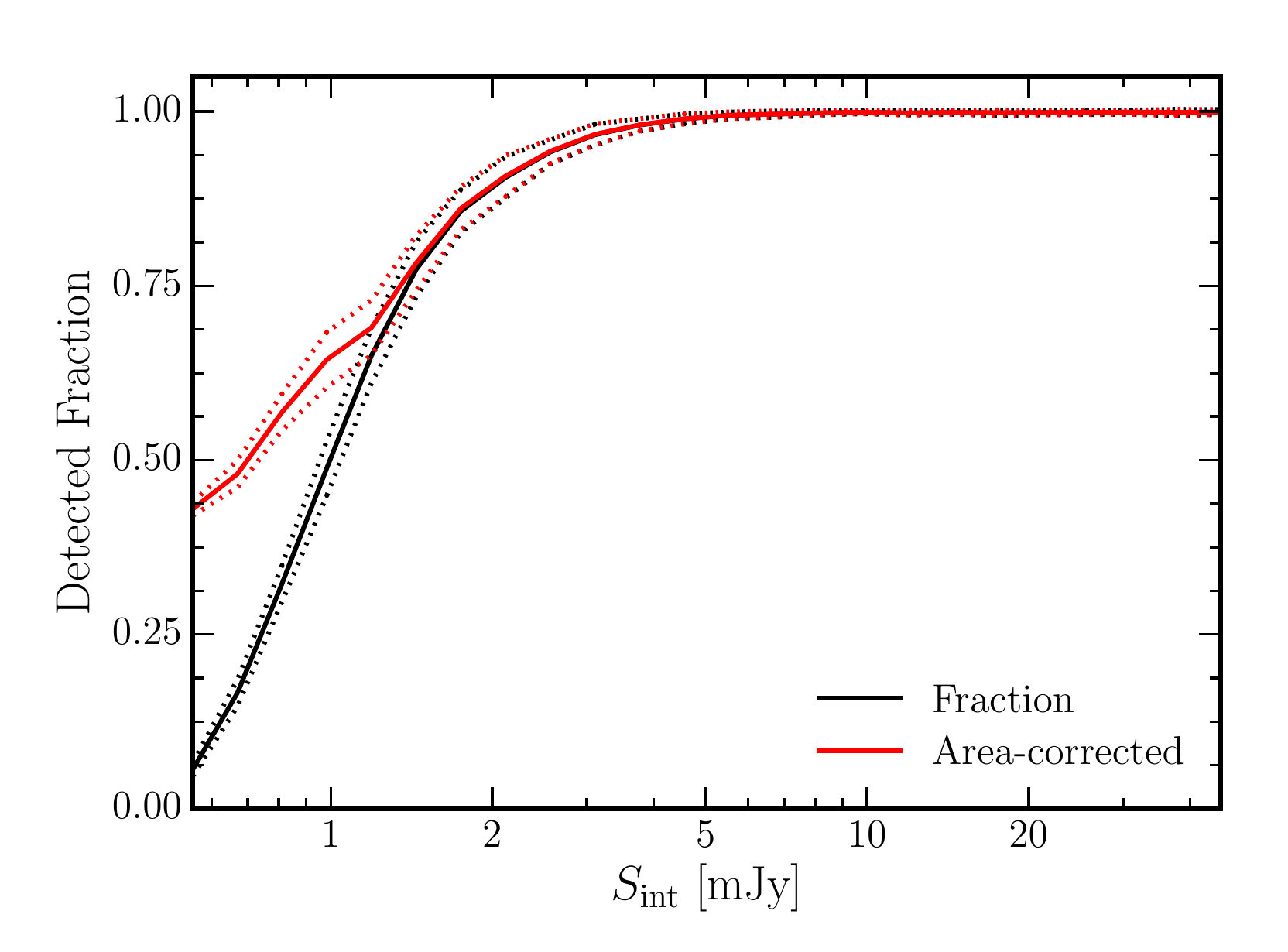}
\includegraphics[width=0.48\textwidth, trim=0cm 0cm 0cm 0.825cm, clip]{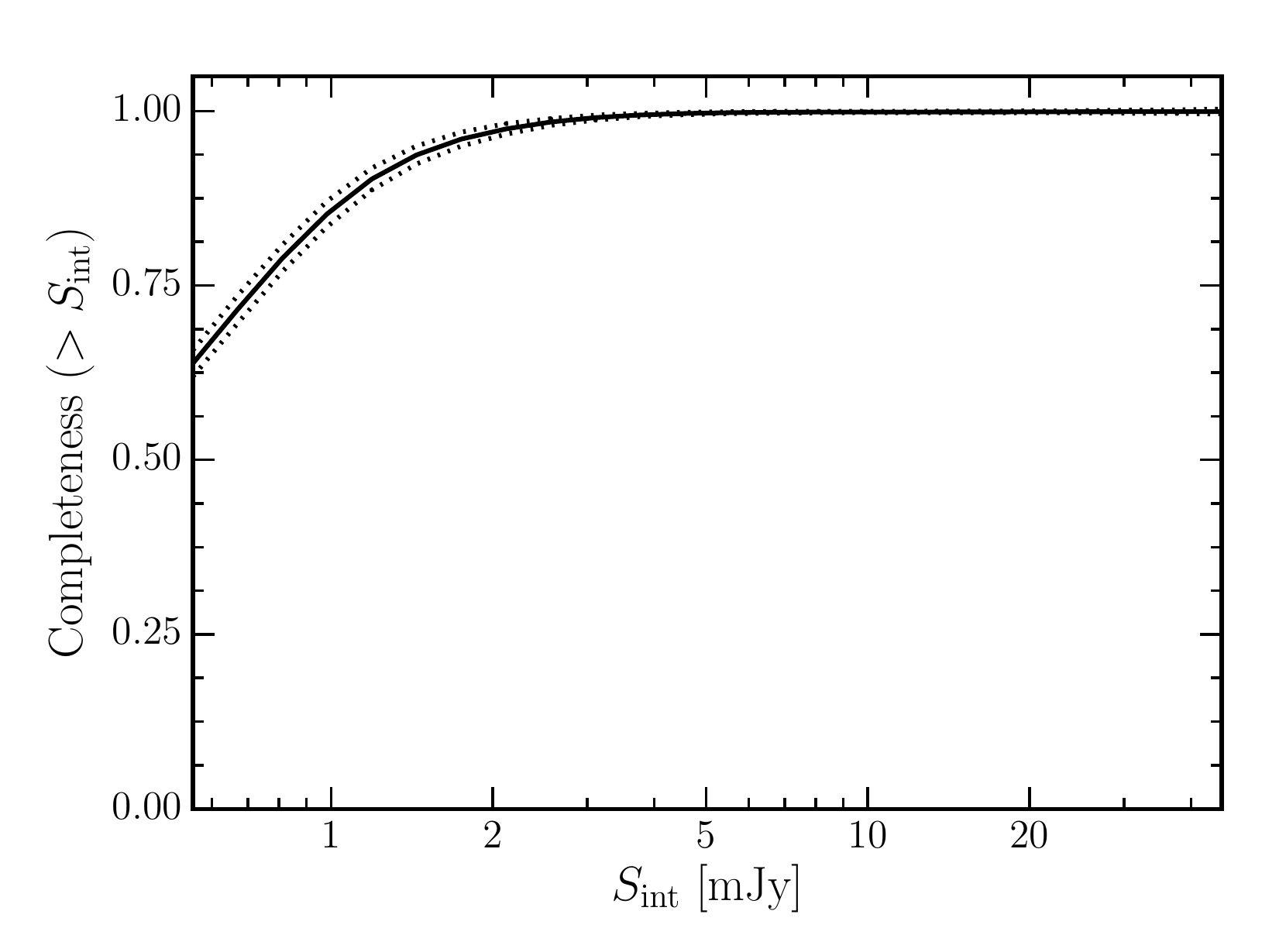}
\caption{{Monte-Carlo completeness simulations: \textit{Left} Fraction of sources detected as a function of {total flux density}. The red curves show the detected fraction after taking out the effect of the visibility area. \textit{Right} Estimated completeness of the whole catalogue as a function of integrated flux density limit. The dotted lines show the $1\,\sigma$ uncertainty derived from Poissonian errors on the source counts}.}
\label{fig:completeness}
\end{figure*}

To quantify the completeness of the catalogue, we performed another Monte-Carlo simulation in which we added simulated sources to the residual image (cf. Section~\ref{sect:sources_ext}). However, in this case approximately $10$~per~cent of the artificial sources inserted into the noise map were extended sources -- Gaussians with FWHM larger than the beamsize. This allows for a better estimate of the completeness in terms of integrated flux densities. The completeness of a catalogue represents the probability that all sources above a given flux density are detected. We have estimated this by plotting the fraction of detected sources in our simulation as a function of integrated flux density (left panel of Fig.~\ref{fig:completeness}), i.e. the fraction of input sources that have a catalogued flux density using the same detection parameters.  This detection fraction is largely driven by the variation in rms across the image, or visibility area. The number of detected sources as a fraction of sources that \emph{could} be detected, accounting for the visibility area, is also shown in the Figure. The completeness at a given flux density is determined by integrating the detected fraction upwards from a given flux density limit and is plotted as a function of integrated flux density in the right panel of Fig.~\ref{fig:completeness}. This shows the completeness of the \emph{full} catalogue. We thus estimate that the catalogue is 85~per~cent complete above a peak flux density of $1$\,mJy. 

\subsection{Reliability}
\label{sect:reliability}
The reliability of a source catalog  indicates the probability that all sources above a given flux density are real. It is defined as the fraction of all detected sources in the survey area above a certain total flux density limit that are real sources and are not accidental detections of background features or noise. To estimate the reliability, we extracted sources from the inverted residual mosaic image, assuming that negative image background features are statistically the same as positive ones. We grouped the detected negative `sources' by total flux density into 20 logarithmic flux density bins and compare these to the binned results of the regular (positive) source extraction as described in Section~\ref{sect:sources}. For convenience, we define the real number of sources to be the number of positive sources minus the number of negative sources.

\begin{figure*}
 \centering
\includegraphics[width=0.48\textwidth, trim=0cm 0cm 0cm 0.825cm, clip]{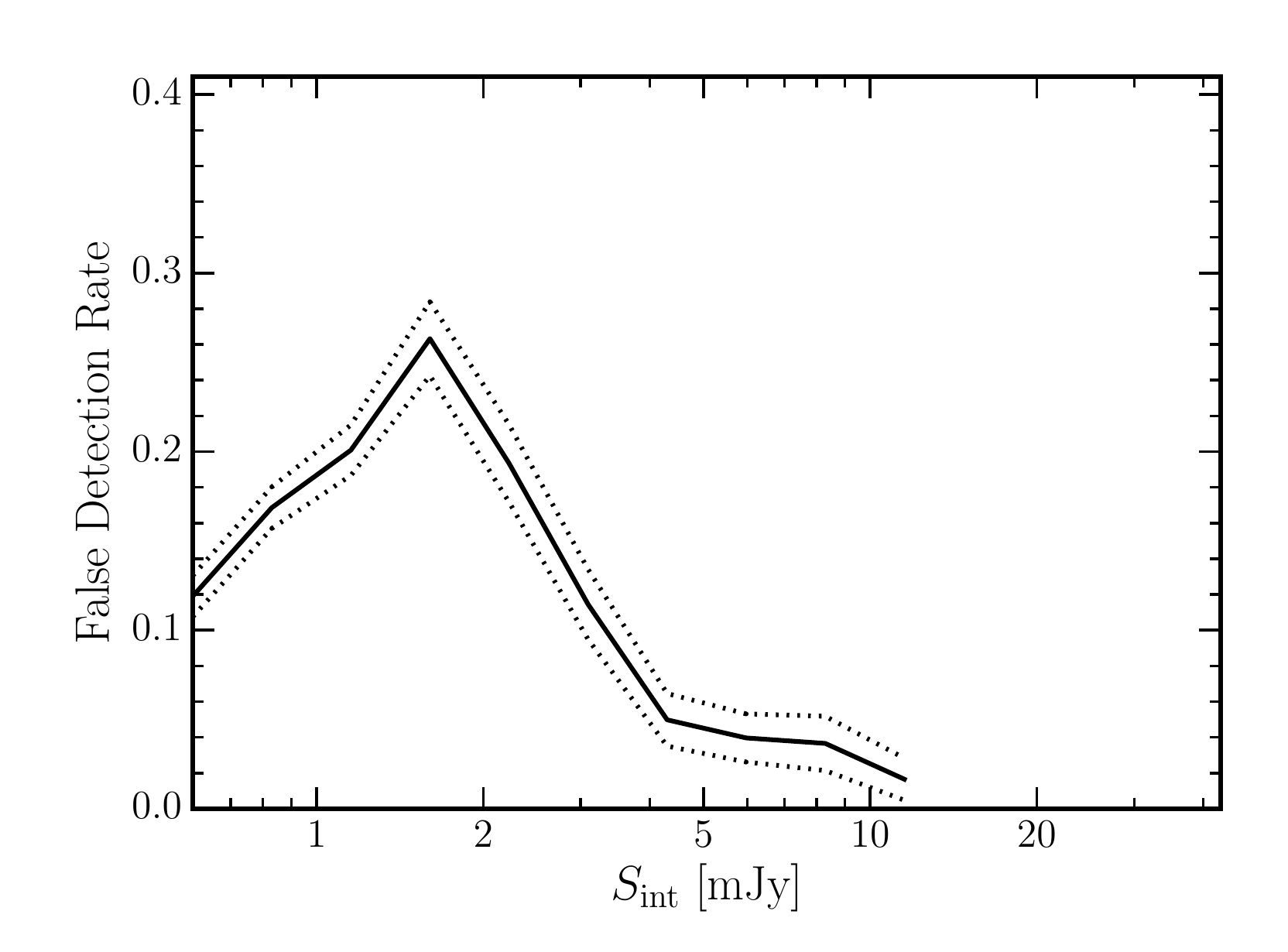}
\includegraphics[width=0.48\textwidth, trim=0cm 0cm 0cm 0.825cm, clip]{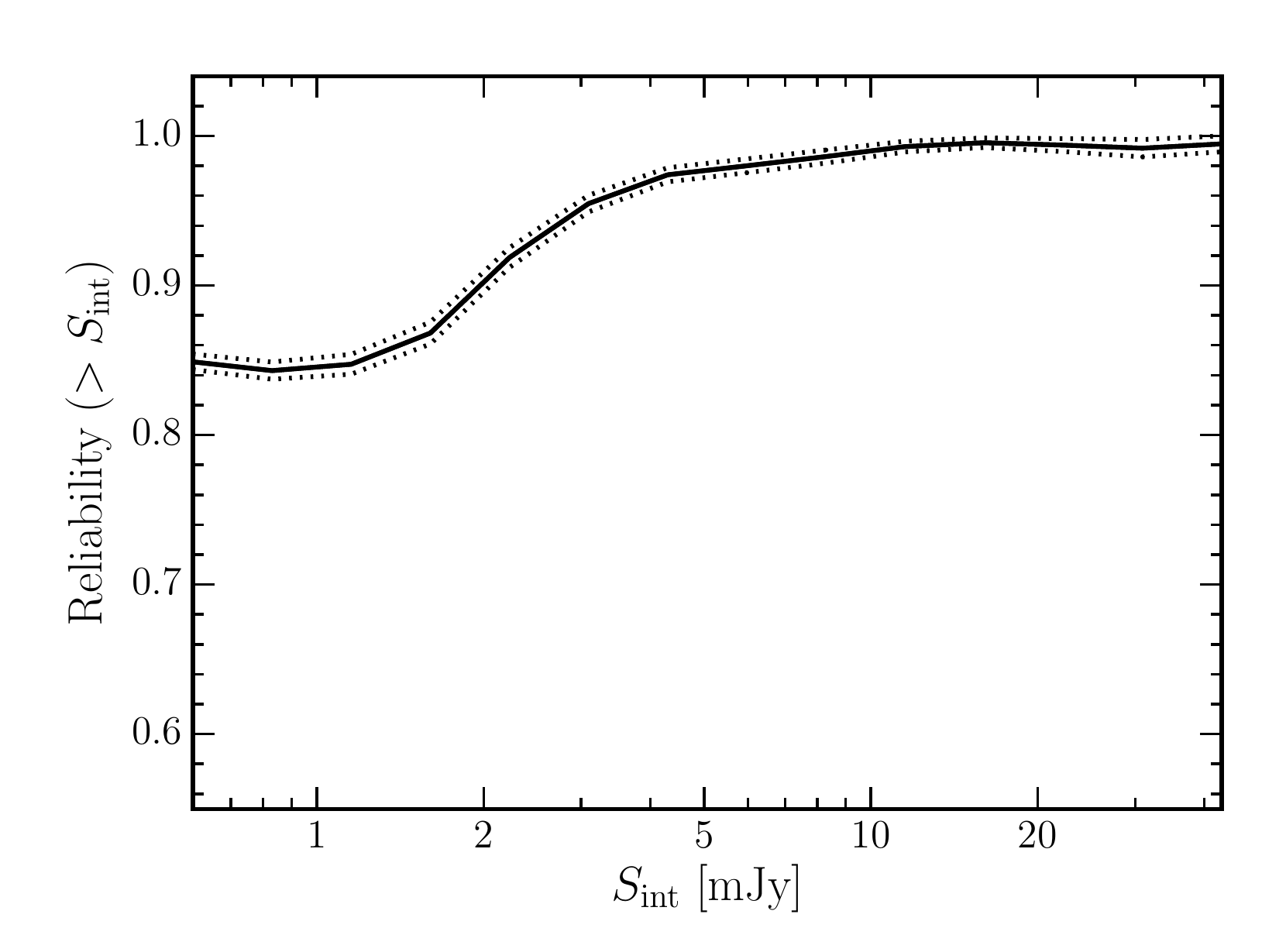}
\caption{{\textit{Left} False detection rate as a function of {total} flux density calculated from source detection on an inverted residual map. \textit{Right} Estimated reliability of the catalogue as a function of integrated flux density limit accounting for the varying sensitivity across the field of view.} {The dotted lines show the $1\,\sigma$ uncertainty derived from Poissonian errors on the source counts}.}
\label{fig:rely}
\end{figure*}

The left panel of Fig.~\ref{fig:rely} shows the false detection rate, $FDR$, determined from the number ratio of negative sources over positive sources per flux density bin. {The peak around $2$\,mJy is explained by the fact that the detection efficiency drops off below this flux density as shown in Section~\ref{sect:completeness1}}. The right panel shows the {integrated reliability curve, $R = 1-FDR(>S_{\rm int})$,} determined from the number ratio of real sources over positive sources above the total flux density limits that define the lower edges of the flux density bins. {Errors are calculated based on Possionian errors on the number of sources per flux density bin.}  For a $1$\,mJy total flux density threshold, the reliability is ${85}$~per~cent.

\subsection{Source Catalogue}
\label{sect:catalog}
{The final catalogue consists of $6\,272$ sources with flux densities between $0.4$\,{\mJy} and $5$\,Jy and is available as {a table in FITS format} as part of the online version of this article. {The catalogue is also available from the CDS}. {The astrometry in the catalogue has been corrected for the systematic offset described in Section~\ref{sect:asterrors}}. {Both the integrated and peak flux densities  in the catalogue have been corrected for the systematic offset (Section~\ref{sect:systerrors})}. Resolved sources are identified as described in Section~\ref{sect:sources_ext}. Errors given in the catalogue are the nominal fit errors. All flux densities and rms values are on the \citetalias{2012MNRAS.423L..30S} flux density scale. A sample of the catalogue is shown in Table~\ref{tab:cataloguesample}. 
Not included in this sample in Table~\ref{tab:cataloguesample}, the catalogue also contains a number of simple flags, based on visual inspection, where non-zero values indicate: \\
Column (14) -- Flag\_badfit, a bad Gaussian fit and so no parameters derived from the Gaussian fit,\\
Column (15) -- Flag\_edge, a source on the edge of the mosaic such that some of the flux is missing, \\
Column (16) -- Flag\_bad, a source has been successfully fit with Gaussians but visual inspection indicates a likely poor fit or other problem\\
Column (17) -- Flag\_artefact, a source is identified as an artefact (this flag has two values -- a `1' indicating a likely artefact and a `2' indicating an almost certain artefact), and\\
Column (18) -- Flag\_merged, a large source whose source components have been manually merged into single sources.}

\begin{table*}
 \centering
 \begin{center}
 \caption{Sample of the LOFAR $150$-MHz source catalogue.}
 \label{tab:cataloguesample}
\begin{tabular}{rccccr@{\,$\pm$\,}lr@{\,$\pm$\,}llcccccccc}
\hline
\multicolumn{1}{c}{Source ID} & \multicolumn{1}{c}{RA} & \multicolumn{1}{c}{$\sigma_{\rm RA}$} & \multicolumn{1}{c}{DEC}  & \multicolumn{1}{c}{$\sigma_{\rm DEC}$} & \multicolumn{2}{c}{$S_{\rm int}$}  & \multicolumn{2}{c}{$S_{\rm peak}$} & \multicolumn{1}{c}{$F_{\rm smear}$}  &  \multicolumn{1}{c}{rms}  &  \multicolumn{1}{c}{Gaussians}  &  \multicolumn{1}{c}{Resolved}\\
  & \multicolumn{1}{c}{[deg]} & \multicolumn{1}{c}{[${\arcsec}$]} & \multicolumn{1}{c}{[deg]} & \multicolumn{1}{c}{[${\arcsec}$]} & \multicolumn{2}{c}{[mJy]} & \multicolumn{2}{c}{[mJy\ beam$^{-1}$]} &  & \multicolumn{1}{c}{[mJy\ beam$^{-1}$]} &  \\
\multicolumn{1}{c}{(1)} & \multicolumn{1}{c}{(2)} & \multicolumn{1}{c}{(3)} & \multicolumn{1}{c}{(4)} & \multicolumn{1}{c}{(5)} & \multicolumn{2}{c}{(6--7)} & \multicolumn{2}{c}{(8--9)} & \multicolumn{1}{c}{(10)} & \multicolumn{1}{c}{(11)} & \multicolumn{1}{c}{(12)} & \multicolumn{1}{c}{(13)} \\
\hline
J142956.07+350244.8 & 217.48 &  0.16 & 35.05  &  0.10 &  3.46 &  0.18  &  2.51&  0.12 &  1.02  &  0.09  & 1 & R  \\
J144256.37+334516.9 & 220.73 &  1.12 & 33.75  &  0.76 &  3.46 &  0.31  &  1.24&  0.24 &  1.30  &  0.22  & 1 & U  \\
J143219.55+330127.6 & 218.08 &  1.30 & 33.02  &  0.37 &  3.45 &  0.17  &  1.05&  0.13 &  1.11  &  0.14  & 1 & R  \\
J142921.88+355821.7 & 217.34 &  0.19 & 35.97  &  0.16 &  3.45 &  0.26  &  2.51&  0.16 &  1.12  &  0.13  & 1 & U  \\
J142311.64+333503.5 & 215.80 &  0.76 & 33.58  &  0.63 &  3.45 &  0.26  &  1.27&  0.20 &  1.21  &  0.17  & 1 & U  \\
J143744.79+330715.9 & 219.44 &  0.69 & 33.12  &  0.92 &  3.45 &  0.22  &  1.15&  0.17 &  1.17  &  0.17  & 1 & R  \\
J143044.45+355716.5 & 217.69 &  0.32 & 35.95  &  0.19 &  3.44 &  0.27  &  2.10&  0.18 &  1.11  &  0.14  & 1 & R  \\
J143751.95+322342.2 & 219.47 &  0.19 & 32.40  &  0.25 &  3.44 &  0.64  &  2.50&  0.27 &  1.31  &  0.26  & 2 & U  \\
J143629.65+362949.9 & 219.12 &  0.16 & 36.50  &  0.20 &  3.44 &  0.49  &  2.31&  0.20 &  1.25  &  0.17  & 2 & U  \\
J144302.47+342335.7 & 220.76 &  0.66 & 34.39  &  0.31 &  3.44 &  0.32  &  1.76&  0.22 &  1.27  &  0.20  & 1 & U  \\
\hline
\multicolumn{11}{l}{The FITS catalogue columns are:} \\
\multicolumn{11}{l}{(1) -- IAU Source name} \\
\multicolumn{11}{l}{(2) and (3) -- flux-weighted right ascension (RA) and {uncertainty}} \\
\multicolumn{11}{l}{(4) and (5) -- flux-weighted declination (DEC) and {uncertainty}} \\
\multicolumn{11}{l}{(6--7) -- integrated source flux density and uncertainty} \\
\multicolumn{11}{l}{(8--9) -- peak flux density and uncertainty} \\
\multicolumn{11}{l}{(10) -- approximate correction factor to the peak flux density to account for bandwidth- and time-smearing} \\
\multicolumn{11}{l}{(11) -- the local rms noise used for the source detection} \\
\multicolumn{11}{l}{(12) -- number of Gaussian components} \\
\multicolumn{11}{l}{(13) -- a flag indicating the resolved parameterisation of the source. `U' refers to unresolved sources and `R' to resolved sources.} \\
 \end{tabular}
 \end{center}
\end{table*}

\section{Results}
\label{sect:results}
In this section, we report two results based on the LOFAR catalogue: the spectral indices between $150$ and $1400$\,MHz, and the $150$\,MHz faint source counts. Further analysis of these data will be presented in future publications.

\subsection{{Spectral Index Distributions}}
\label{sect:spec_ind}
\begin{figure}
 \centering
\includegraphics[width=0.45\textwidth]{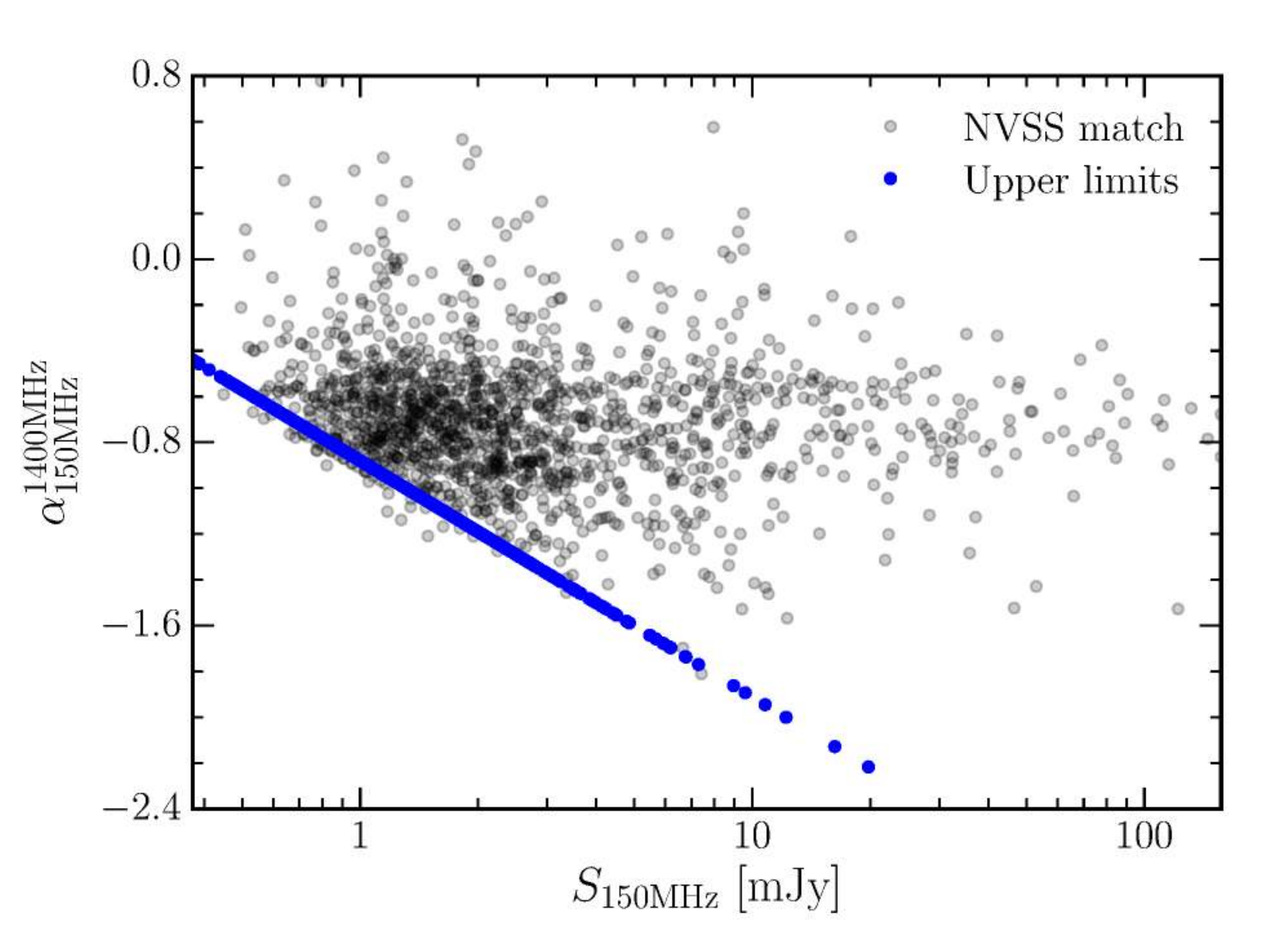}
\caption{Spectral index, $\alpha^{1400}_{150}$, distribution of sources matched between $1.4$~GHz and $150$~MHz (points). The  difference in resolution is  $13 \times 27\,{\arcsec}$ (WSRT) and $5.6 \times7.4\,{\arcsec}$ (LOFAR). The blue points show upper limits to the spectral index for LOFAR sources which do not have a higher frequency counterpart. The horizontal line shows the median spectral index of $-0.79$ determined using LOFAR sources with flux densities greater than $2$\,{\mJy}.}
\label{fig:specind_wsrt}
\end{figure}

We use the deep WSRT $1.4$~GHz data covering the Bo\"{o}tes Field \citep{2002AJ....123.1784D} to calculate spectral indices between $150$\,MHz and $1.4$\,GHz, $\alpha^{1400}_{150}$, the distribution of which is shown in Fig.~\ref{fig:specind_wsrt}. The WSRT  map has a resolution of $13 \times 27\,{\arcsec}$, so some sources appear as separate sources in the LOFAR map but are identified as single sources in the WSRT image. To exclude erroneous spectral indices derived for such sources, we limit this selection to sources that are not identified as extended in either the LOFAR or WSRT catalogues and that do not have multiple matches within a $30$\,{\arcsec} search radius.

Using LOFAR sources with flux densities greater than $2$\,{\mJy}, we find a median spectral index between $1400$ and $150$~MHz of $-0.79 \pm 0.01$ and scatter of $\sigma=0.30$ which is consistent with previously reported values: $-0.87 \pm 0.01$, \citep{2013A&A...549A..55W},  $-0.79$ \citep{2011A&A...535A..38I}, $-0.78$ \citep{2010MNRAS.405..436I},  $-0.82$ \citep{2009MNRAS.392.1403S}, and $-0.85$ \citep{2007ASPC..380..237I}. A detailed spectral-index analysis using the other available radio data  is deferred to later works. {Analysis of the in-band LOFAR spectral indicies is also deferred to later work after the current LOFAR gain transfer problems have been solved.}

\subsection{Source Counts}
\label{sect:source_counts}

We used the LOFAR catalogue to compute the $150$\,MHz source counts down to $\approx1$\,{\mJy}. This is at least an order of magnitude deeper than previously studied at these low frequencies \citep[e.g.][]{1990MNRAS.246..110M,2011A&A...535A..38I,2013A&A...549A..55W}. The source counts are computed using the integrated flux densities, but sources are detected based on their \emph{measured} peak flux density over the local noise level. Thus, the completeness of the source counts depends both on the variation of the noise in the image and on the relation between integrated and peak flux densities. The latter is dependant both on systematic effects (e.g.\ smearing) and the intrinsic relation between integrated and peak flux densities of radio sources due to their intrinsic sizes. In the following paragraphs we discuss these effects and how we correct for them in deriving the source counts.

\subsubsection{Visibility Area}
Due to the large variation in rms across the single pointing image (see Fig.~\ref{fig:mosrms}),  sources of different flux densities are not uniformly detected across the image, i.e.\ faint sources can only be detected in a smaller area in the inner part of the image. Moreover, smearing causes a reduction in peak flux density while conserving the integrated flux density, and the amount of smearing depends on the distance to the phase-centre. We have noted already the effect of bandwidth- and time-smearing (see Section~\ref{sect:sources_ext}) and use the equations given by \cite{1989ASPC....6..247B} to calculate an approximate correction to the peak flux density of each source based on its position in the map. The maximal correction is at most $S_{\rm peak}^{\rm meas} \approx0.74 S_{\rm peak}^{\rm corr}$, so sources with \emph{corrected} peak flux densities $> 6.7\sigma$ will have effective measured peak flux densities above the $5\sigma$ detection threshold. We therefore select only sources based on this threshold for deriving the source counts. To correct for the varying rms, we  weight each source by the reciprocal of the area in which it can be detected, its visibility area, \citep[e.g.][]{1985ApJ...289..494W}, based on its smearing-corrected peak flux density value. This also accounts for the varying detection area within a given flux density bin. 

\subsubsection{Completeness and Reliability}
We consider also a correction for the completeness of the catalogue (see Section~\ref{sect:completeness1}). As the visibility area is considered separately, we use the red curves in Fig.~\ref{fig:completeness} to determine a correction factor to account for the fraction of sources missed in each flux density bin. {Additionally, we make a correction for the reliability by applying the $FDR$ derived in Section~\ref{sect:reliability}, which acts in the opposite direction to the completeness correction.}

\subsubsection{Systematic Effects}
Another effect that could potentially influence both the peak and integrated flux densities is {\scshape clean} bias, which could bias both downwards at the lowest flux densities, thus leading to low source counts. However, we {have shown (Section~\ref{sect:errors} that} this is negligible because the use of masks in the imaging and good $uv$-coverage.  In general, noise can scatter sources into adjacent bins, again most noticeably at low flux densities.  A positive bias is introduced by the enhancement of weak sources by random noise peaks \citep[Eddington bias;][]{1913MNRAS..73..359E}. Both of these effects could be quantified by simulations, but our source counts are not corrected for them, due to the computational expense of running the full required simulation.

\subsubsection{Resolution Bias}
\label{sect:resbias}
\begin{figure}
 \centering
\includegraphics[width=0.45\textwidth]{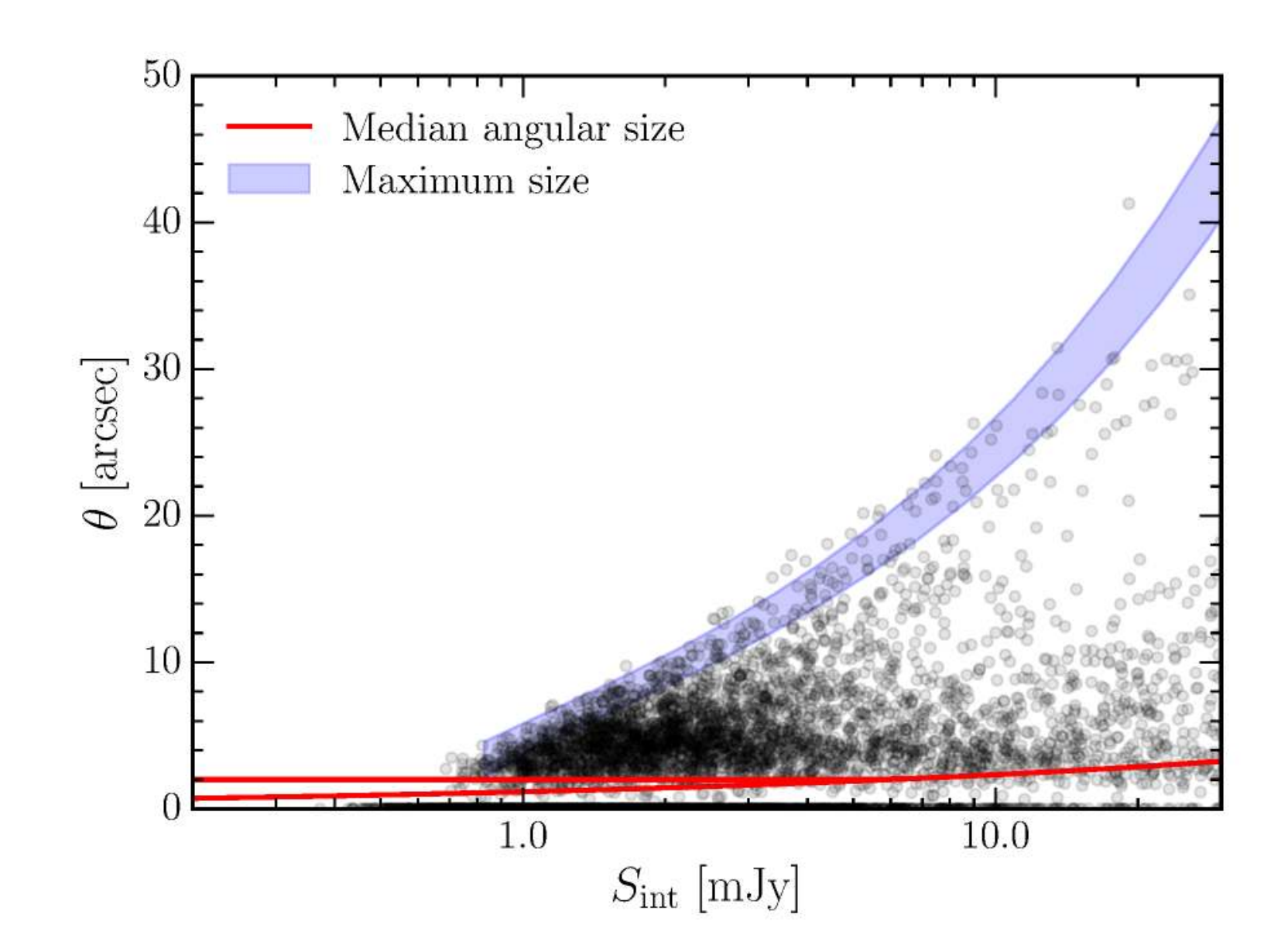}
\caption{Angular size, $\theta$ (geometric mean), for the LOFAR sources as a function of integrated flux density. The blue shaded region shows the range of maximum size ($\theta_{\rm max}$) a source of a given integrated flux density can have before dropping below the peak-flux detection threshold (the range reflects the range of rms noise in the LOFAR map). The red lines show the two functions used for the median angular size ($\theta_{\rm med}$) as a function of integrated flux density.}
\label{fig:sizeflux}
\end{figure}

\begin{figure}
 \centering
\includegraphics[width=0.45\textwidth]{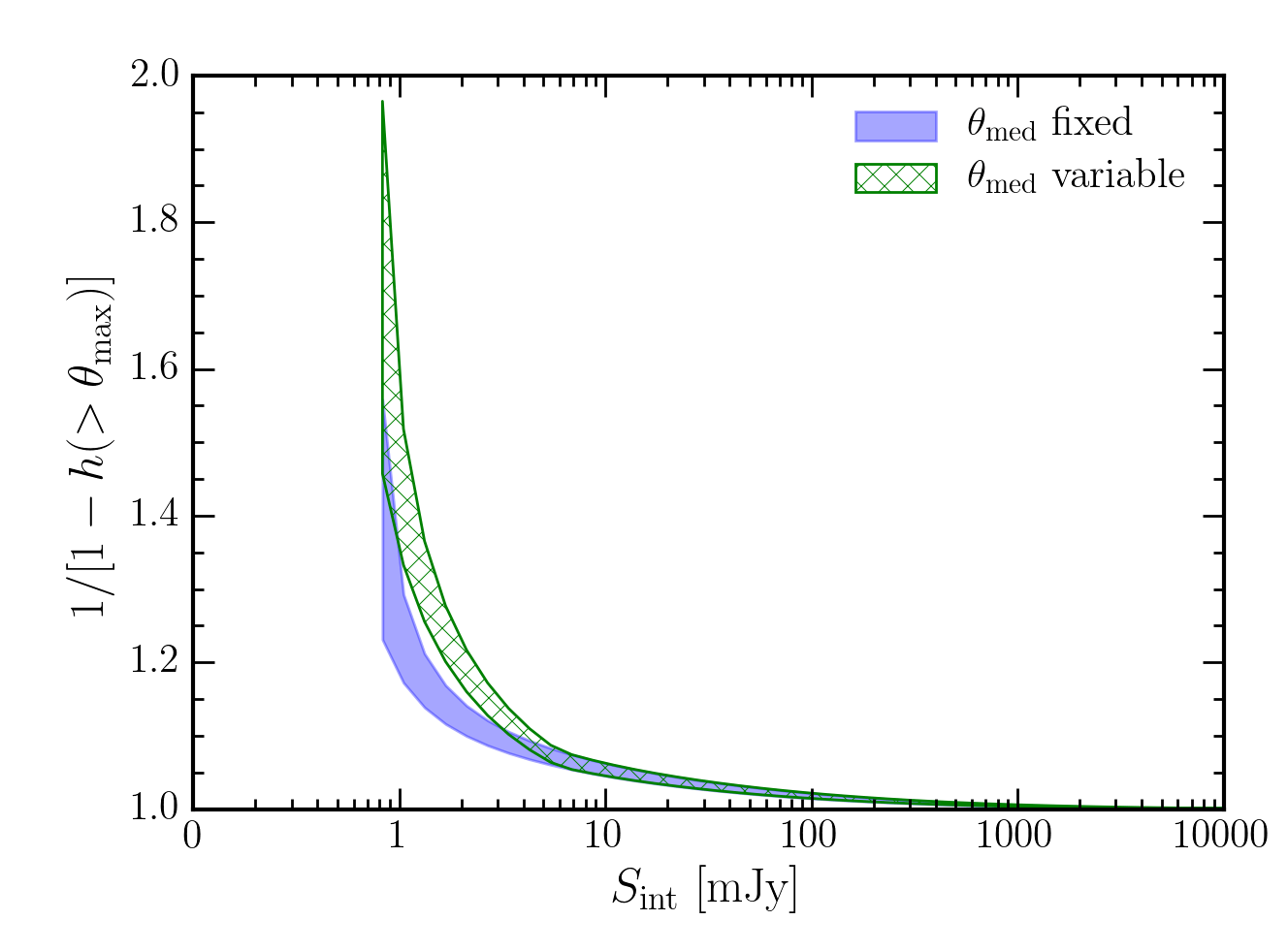}
\caption{Resolution bias correction $1/[1-h(>\theta_{\rm max})]$ for the fraction of sources with angular size larger than $\theta_{\rm max}$ at a given integrated flux density. For the faintest sources two curves are shown: the blue curve shows  $\theta_{\rm med}$ as a function of $S_{\rm int}$ and the green curve shows the result assuming $\theta_{\rm med}=2$\,{\arcsec} at these flux densities (the range of each reflects the range of rms noise in the LOFAR map).}
\label{fig:rescor}
\end{figure}

A resolved source of a given integrated flux density will be missed by the peak-flux-density selection more easily than a point source of the same integrated flux density. This incompleteness is called the resolution bias and to make a correction for it requires some knowledge of the true angular size distribution of radio sources. We have estimated a correction for the resolution bias following \cite{2001A&A...365..392P}. First we calculate the approximate maximum size $\theta_{\rm max}$ a source could have for a given integrated flux density before it drops below the peak-flux detection threshold. Fig.~\ref{fig:sizeflux} shows the angular size of the LOFAR sources. We use the relation
\[ \frac{S_{\rm int}}{S_{\rm peak}} = \frac{\theta_{\rm min}\theta_{\rm max}}{b_{\rm min}b_{\rm max}}, \]
where $b_{\rm min}$ and $b_{\rm max}$ are the synthesized beam axes and $\theta_{\rm min}$ and $\theta_{\rm max}$ are the source sizes, to estimate the maximum size a source of a given integrated flux density can have before dropping below the peak-flux detection threshold. Given this $\theta_{\rm max}$ we estimate the fraction of sources with angular sizes larger than this limit using the assumed true angular size distribution proposed by \cite{1990ASPC...10..389W}: $h(>\theta_{\rm max}) = \exp[-\ln 2 (\theta_{\rm max}/\theta_{\rm med})^{0.62}]$ with $\theta_{\rm med} = 2 S_{1.4\mbox{\,GHz}}^{0.30}$\,{\arcsec} (with $S$ in {\mJy}; we have scaled the $1.4$\,GHz flux densities to $150$\,MHz with a spectral index of $-0.8$). We have also calculated the correction using $\theta_{\rm med} = 2\,{\arcsec}$ for sources with $S_{1.4\mbox{\,GHz}}<1$\,{\mJy} \citep[see][]{1993ApJ...405..498W,2000ApJ...533..611R}.  The resolution bias correction $c = 1/[1-h(>\theta_{\rm max})]$  is plotted in Fig.~\ref{fig:rescor} for the two different assumed distributions. In correcting the source counts we use an average of the two functions. We use the uncertainty in the forms of $\theta_{\rm med}$  and in $\theta_{\rm max}$ to estimate the uncertainty in the resolution bias correction. We further include a  overall $10$~per~cent uncertainty following \cite{1990ASPC...10..389W}. While we have used the extrapolated \cite{1990ASPC...10..389W} size distribution from $1.4$\,GHz to correct the source counts presented here, we note that the observed size distribution (see Appendix~\ref{sect:ap:sizes}) suggests that the low frequency emission is more extended. Thus the real resolution bias correction factor is likely to be somewhat larger, particularly in the lowest flux density bins, and may explain the turndown in source counts (see Fig.~\ref{fig:srccounts}). A full study of the true low frequency angular size distrubution of radio sources is beyond the scope of this paper.

\subsubsection{Complex Sources}
The source counts need to be corrected for multi-component sources, i.e. cases where the radio-lobes are detected as two separate sources. The flux densities of physically related components should be summed together, instead of counted as separate sources. We use the method described in \citet{2012MNRAS.427.1830W} and \citet{1998MNRAS.300..257M} to identify the double and compact source populations. This is done by considering the separation of the nearest neighbour of each component and the summed flux of the component and its neighbour. Pairs of sources are regarded as single sources if the ratio of their flux densities is between $0.25$ and $4$, and their separation is less than a critical value dependent on their total flux density, given by
\[ \theta_{\mathrm{crit}}  = 100 \left[ \frac{S}{10} \right]^{0.5},  \]
where $S$ is in mJy and $\theta$ is in arcsec. Approximately  $460$ sources in the sample used for calculating the source counts, or $8.5$~per~cent,  are considered to be a part of double or multiple sources for the source count calculation.

\subsubsection{The Low-Frequency mJy Source Counts}
\begin{figure*}
 \centering
\includegraphics[width=\textwidth, trim=0cm 2cm 0cm 1.8cm, clip]{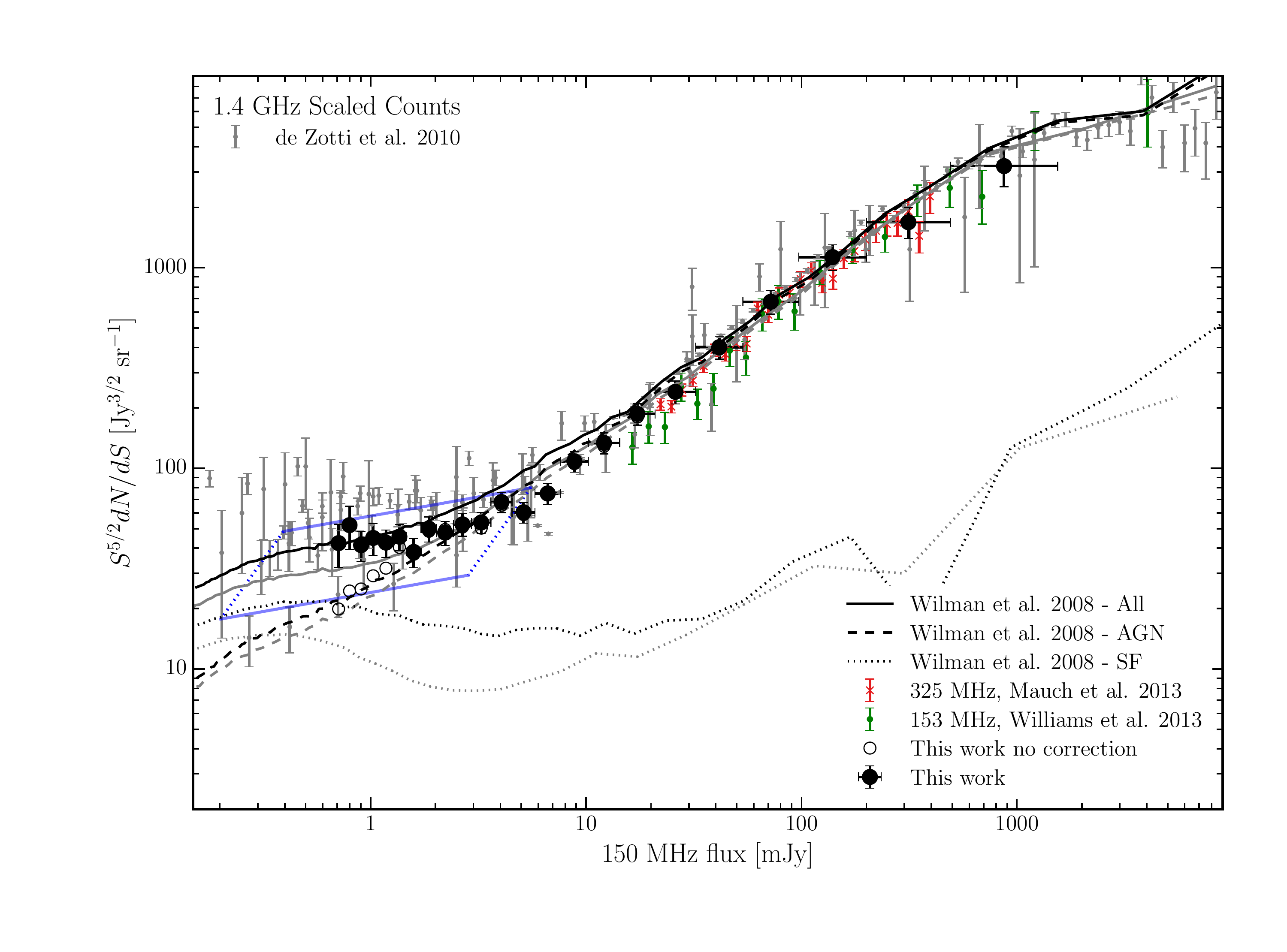}
\caption{Euclidean-normalized differential source counts for the LOFAR $150$-MHz catalogue (filled black circles) between $1$~mJy and $5$~Jy. Open black circles show the sources counts without the resolution bias correction. For comparison we plot the SKADS  model counts \citep{2008MNRAS.388.1335W} separated into AGN and SF for $151$\,MHz (in grey) and scaled from $610$\,MHz (in black). We show also the measured $150$\,MHz source counts \citep{2013A&A...549A..55W} (in green) and scaled from $325$\,MHz \citep{2013MNRAS.435..650M} (in red). Additionally, we have included several source counts determinations from $1.4$\,GHz from the compilation of \citet{2010A&ARv..18....1D} scaled down to $150$\,MHz assuming a spectral index of $-0.8$ (grey points).  The blue lines at the low flux density end show how the higher frequency counts would scale if a flatter spectral index of $-0.5$ was used, i.e. the tail of the source counts would lie further down and to the left in the plot. }
\label{fig:srccounts}
\end{figure*}

The Euclidean-normalized differential source counts are shown in Fig.~\ref{fig:srccounts}. Uncertainties on the final normalised source counts are propagated from the errors on the correction factors and the Poisson errors \citep{1986ApJ...303..336G} on the raw counts per bin. {The source counts are tabulated in Table~\ref{tab:src_counts}. }

\begin{table*}
 \begin{center}
 \caption{{Euclidean-normalized differential source counts for the LOFAR $150$-MHz catalogue.}}
 \label{tab:src_counts}
\begin{tabular}{cccccccccc}
\hline
\multicolumn{1}{c}{$S$ Range} & \multicolumn{1}{c}{$S_c$} & \multicolumn{1}{c}{Raw Counts} & \multicolumn{1}{c}{Area} & \multicolumn{1}{c}{$A(S_c)$} & \multicolumn{1}{c}{$\langle W \rangle$}&\multicolumn{1}{c}{FDR} &\multicolumn{1}{c}{Completeness} &\multicolumn{1}{c}{Resolution}  & \multicolumn{1}{c}{Normalised counts}\\
\multicolumn{1}{c}{[mJy]} & \multicolumn{1}{c}{[mJy]} & \multicolumn{1}{c}{ } & \multicolumn{1}{c}{[deg$^2$]} & \multicolumn{1}{c}{[deg$^2$]} & \multicolumn{1}{c}{ }&\multicolumn{1}{c}{ } &\multicolumn{1}{c}{ } &\multicolumn{1}{c}{ }  & \multicolumn{1}{c}{ [Jy$^{3/2}$ sr$^{-1}$] }\\
\multicolumn{1}{c}{(1)} & \multicolumn{1}{c}{(2)} & \multicolumn{1}{c}{(3)} & \multicolumn{1}{c}{(4)} & \multicolumn{1}{c}{(5)} & \multicolumn{1}{c}{(6)}&\multicolumn{1}{c}{(7)}  & \multicolumn{1}{c}{(8)}  & \multicolumn{1}{c}{(9)}& \multicolumn{1}{c}{(10)}\\
\hline
$   0.67-   0.75$ & $   0.71$ & $174_{-13}^{+14}$ & $ 4.74- 7.19$ & $ 6.01$ & $4.07$& $0.8$ & $1.8$  & $1.5 \pm 0.7$ & $  42_{-  1}^{+   1}$ \\[+1.5pt]
$   0.75-   0.85$ & $   0.80$ & $264_{-16}^{+17}$ & $ 7.19- 9.94$ & $ 8.55$ & $2.90$& $0.8$ & $1.8$  & $1.5 \pm 0.7$ & $  52_{-  1}^{+   1}$ \\[+1.5pt]
$   0.85-   0.96$ & $   0.90$ & $283_{-17}^{+18}$ & $ 9.94-12.44$ & $11.27$ & $2.42$& $0.8$ & $1.6$  & $1.3 \pm 0.4$ & $  41_{-  1}^{+   1}$ \\[+1.5pt]
$   0.96-   1.10$ & $   1.03$ & $334_{-18}^{+19}$ & $12.44-15.03$ & $13.74$ & $2.06$& $0.8$ & $1.4$  & $1.3 \pm 0.4$ & $  45_{-  1}^{+   1}$ \\[+1.5pt]
$   1.10-   1.26$ & $   1.18$ & $377_{-19}^{+20}$ & $15.03-17.75$ & $16.46$ & $1.72$& $0.7$ & $1.4$  & $1.3 \pm 0.3$ & $  43_{-  1}^{+   1}$ \\[+1.5pt]
$   1.26-   1.47$ & $   1.36$ & $410_{-20}^{+21}$ & $17.75-18.86$ & $18.55$ & $1.71$& $0.7$ & $1.3$  & $1.2 \pm 0.2$ & $  46_{-  1}^{+   1}$ \\[+1.5pt]
$   1.47-   1.71$ & $   1.59$ & $397_{-20}^{+21}$ & $18.86-19.17$ & $19.08$ & $1.38$& $0.7$ & $1.2$  & $1.2 \pm 0.2$ & $  38_{-  1}^{+   1}$ \\[+1.5pt]
$   1.71-   2.03$ & $   1.86$ & $408_{-20}^{+21}$ & $19.17-19.24$ & $19.21$ & $1.46$& $0.8$ & $1.1$  & $1.2 \pm 0.2$ & $  50_{-  2}^{+   2}$ \\[+1.5pt]
$   2.03-   2.43$ & $   2.22$ & $420_{-20}^{+22}$ & $19.24-19.27$ & $19.25$ & $1.19$& $0.8$ & $1.1$  & $1.1 \pm 0.1$ & $  48_{-  2}^{+   2}$ \\[+1.5pt]
$   2.43-   2.94$ & $   2.67$ & $354_{-19}^{+20}$ & $19.27-19.29$ & $19.28$ & $1.18$& $0.9$& $1.0$  & $1.1 \pm 0.1$ & $  53_{-  2}^{+   2}$ \\[+1.5pt]
$   2.94-   3.62$ & $   3.26$ & $306_{-17}^{+19}$ & $19.29-19.30$ & $19.29$ & $1.08$& $1.0$ & $1.0$  & $1.1 \pm 0.1$ & $  54_{-  3}^{+   3}$ \\[+1.5pt]
$   3.62-   4.53$ & $   4.05$ & $312_{-18}^{+19}$ &  $  \ldots $  & $19.30$ & $1.08$& $1.0$ & $1.0$ & $1.1 \pm 0.1$ & $  68_{-  3}^{+   4}$ \\[+1.5pt]
$   4.53-   5.80$ & $   5.13$ & $221_{-15}^{+16}$ &  $  \ldots $ & $19.30$ & $1.05$& $1.0$& $1.0$  & $1.1 \pm 0.1$ & $  60_{-  4}^{+   4}$ \\[+1.5pt]
$   5.80-   7.59$ & $   6.63$ & $211_{-15}^{+16}$ &  $  \ldots $  & $19.31$ & $1.03$& $1.0$ & $1.0$ & $1.1 \pm 0.1$ & $  75_{-  5}^{+   5}$ \\[+1.5pt]
$   7.59-  10.25$ & $   8.82$ & $214_{-15}^{+16}$ &  $  \ldots $  & $19.31$ & $1.05$& $1.0$ & $1.0$ & $1.0 \pm 0.1$ & $ 108_{-  7}^{+   7}$ \\[+1.5pt]
$  10.25-  14.32$ & $  12.11$ & $188_{-14}^{+15}$ &  $  \ldots $  & $19.31$ & $1.02$& $1.0$ & $1.0$ & $1.0 \pm 0.1$ & $ 134_{-  9}^{+  10}$ \\[+1.5pt]
$  14.32-  20.91$ & $  17.31$ & $175_{-13}^{+14}$ &  $  \ldots $  & $19.31$ & $1.01$& $1.0$ & $1.0$ & $1.0 \pm 0.1$ & $ 187_{- 13}^{+  14}$ \\[+1.5pt]
$  20.91-  32.26$ & $  25.97$ & $146_{-12}^{+13}$ &  $  \ldots $  & $19.31$ & $1.00$& $1.0$ & $1.0$ & $1.0 \pm 0.1$ & $ 240_{- 20}^{+  21}$ \\[+1.5pt]
$  32.26-  53.42$ & $  41.52$ & $138_{-12}^{+13}$ &  $  \ldots $  & $19.31$ & $1.01$& $1.0$ & $1.0$ & $1.0 \pm 0.1$ & $ 402_{- 33}^{+  36}$ \\[+1.5pt]
$  53.42-  97.07$ & $  72.01$ & $122_{-11}^{+12}$ &  $  \ldots $  & $19.31$ & $1.01$& $1.0$ & $1.0$ & $1.0 \pm 0.1$ & $ 676_{- 60}^{+  65}$ \\[+1.5pt]
$  97.07- 199.99$ & $ 139.33$ & $ 93_{-10}^{+11}$ &  $  \ldots $  & $19.31$ & $1.00$& $1.0$ & $1.0$ & $1.0 \pm 0.1$ & $1128_{-115}^{+ 128}$ \\[+1.5pt]
$ 199.99- 490.54$ & $ 313.21$ & $ 52_{- 7}^{+ 8}$ &  $  \ldots $  & $19.31$ & $1.00$& $1.0$ & $1.0$ & $1.0 \pm 0.1$ & $1683_{-231}^{+ 265}$ \\[+1.5pt]
$ 490.54-1544.46$ & $ 870.42$ & $ 28_{- 5}^{+ 6}$ &  $  \ldots $  & $19.31$ & $1.00$& $1.0$ & $1.0$ & $1.0 \pm 0.1$ & $3207_{-599}^{+ 725}$ \\[+1.5pt]
\hline
\multicolumn{9}{l}{(1) the flux density bins} \\
\multicolumn{9}{l}{(2) the central flux density of the bin}\\
\multicolumn{9}{l}{(3) the raw counts}\\
\multicolumn{9}{l}{(4) the effective detection areas for sources at the lower and upper limits of the flux density bin where they are different}\\
\multicolumn{9}{l}{(5) the effective area corresponding to the bin centre}\\
\multicolumn{9}{l}{(6) the mean weight of the sources in the bin}\\
\multicolumn{9}{l}{(7) the false detection rate  correction factor} \\
\multicolumn{9}{l}{(8) the completeness  correction factor} \\
\multicolumn{9}{l}{(9) the resolution bias correction factor} \\
\multicolumn{9}{l}{(10) the corrected normalised source counts} \\
 \end{tabular}
 \end{center}
\end{table*}

Model source counts have been derived by \cite{2008MNRAS.388.1335W} for the $151$~MHz and $610$\,MHz source populations predicted from the extrapolated radio luminosity functions of different radio sources in a $\Lambda$CDM framework. We show the source counts for both AGN and star-forming (SF) galaxies on Fig.~\ref{fig:srccounts}. The \cite{2008MNRAS.388.1335W} model catalogue has been corrected with their recommended default post-processing, which effectively reduces the source count slightly at low flux densities.  At low flux densities it is likely that the \cite{2008MNRAS.388.1335W} counts slightly overestimate the true counts due to double counting of hybrid AGN-SF galaxies. These models are based on low-frequency data at higher flux density limits and higher frequency data so some deviations are not unexpected; however, our observed counts do follow their model quite well. \citet{2013MNRAS.435..650M} suggest that the spectral curvature term used in the \citet{2008MNRAS.388.1335W} models mean that their  $151$\,MHz counts under-predict reality. For this reason we include the model counts at both frequencies.

Source counts below $1$\,{\mJy} at $1.4$\,GHz have been the subject of much debate. For comparison, in Fig.~\ref{fig:srccounts}, we have included the several source count determinations from $1.4$\,GHz scaled down to $150$\,MHz from the compilation of \citet{2010A&ARv..18....1D} \citep[including counts from][]{1972AJ.....77..405B,1997ApJ...475..479W,1999MNRAS.302..222C,1999MNRAS.305..297G,2000ApJ...533..611R,2003AJ....125..465H,2006ApJS..167..103F,2008ApJ...681.1129B,2008ApJS..179...71K,2008AJ....136.1889O,2008MNRAS.386.1695S}. This is  a representative comparison and not an exhaustive list of available source counts. In particular, there are even deeper models of higher frequency counts using statistical methods  \citep[e.g.][]{2014MNRAS.440.2791V,2015arXiv150302493Z}. The source counts are scaled assuming a spectral index of $-0.8$.   The blue lines at the low flux density end show how the higher frequency counts would scale if a flatter spectral index of $-0.5$ was used, i.e. the blue line drawn through the tail of the source counts would lie at lower $150$\,MHz fluxes (left in the plot) and at lower normalised count values (down in the plot) mostly due to the $S^{5/2}$ term in the normalised counts. We note that the flattening of the source counts at $5-7$\,{\mJy}, associated with the growing population of SF galaxies and faint radio-quiet AGN at lower flux densities \citep[see e.g.][]{2004NewAR..48.1173J,2006MNRAS.372..741S,2015MNRAS.452.1263P}, is clear and is the same as that seen at the higher frequencies. The further drop in the lowest flux density bins may be the result of some unaccounted for incompleteness in our sample or different resolution bias correction (see Section~\ref{sect:resbias} and Appendix~\ref{sect:ap:sizes}).

\section{Conclusion}
\label{sect:concl}
We have presented LOFAR High Band Antenna observations of the Bo\"otes field made as part of the LOFAR Surveys Key Science Project. These are the first wide area (covering $19$\,deg$^2$), deep (reaching $\approx120-150$\,{\muJybeam}), high resolution ($5.6 \times 7.4$\,{\arcsec}) images of one of the extragalactic deep fields  made at $130-169$\,MHz.  These observations are at least an order of magnitude deeper and $3-5$ times higher in resolution than previously obtained  at these frequencies. We have used a new calibration and imaging method to correct for the corrupting effects of the ionosphere and LOFAR digital beams.

The radio source catalogue presented here contains $6\,276$\ sources detected with peak flux densities exceeding $5\sigma$. We have quantified the positional and flux density accuracy of the LOFAR sources and used the source catalogue to derive spectral indices between $150$ and $1400$\,MHz, finding a median spectral index of $-0.79 \pm 0.01$. Finally, we have presented the deepest differential source counts at these low frequencies. These source counts follow quite well the model predictions of  \cite{2008MNRAS.388.1335W} and show the flattening at a few mJy as a result of the increasing contribution of SF galaxies.

\section*{Acknowledgements}

The authors thank the anonymous referee for useful comments which have improved this manuscript. WLW, HJR gratefully acknowledge support from the European Research Council under the European Union's Seventh Framework Programme (FP/2007-2013)/ERC Advanced Grant NEWCLUSTERS-321271. WLW, MJH acknowledge support from the UK Science and Technology Facilities Council [ST/M001008/1]. RJW is supported by a Clay Fellowship awarded by the Harvard-Smithsonian Center for Astrophysics. ADK acknowledges financial support from the Australian Research Council Centre of Excellence for All-sky Astrophysics (CAASTRO), through project number CE110001020. JZ gratefully acknowledges a South Africa National Research Foundation Square Kilometre Array Research Fellowship. RM gratefully acknowledges support from the European Research Council under the European Union's Seventh Framework Programme (FP/2007-2013)/ERC Advanced Grant RADIOLIFE-320745. This research made use of \textsc{astropy}, a community-developed core Python package for astronomy \citep{2013A&A...558A..33A} hosted at \url{http://www.astropy.org/}, of \textsc{APLpy}, an open-source astronomical plotting package for Python hosted at \url{http://aplpy.github.com/}, and of \textsc{topcat} \citep{2005ASPC..347...29T}.

LOFAR, the Low Frequency Array designed and constructed by ASTRON, has facilities in several countries, that are owned by various parties (each with their own funding sources), and that are collectively operated by the International LOFAR Telescope (ILT) foundation under a joint scientific policy. The Open University is incorporated by Royal Charter (RC 000391), an exempt charity in England \& Wales and a charity registered in Scotland (SC 038302). The Open University is authorized and regulated by the Financial Conduct Authority.

\bibliographystyle{mnras}
\bibliography{thisbibfile}

\bsp

\appendix
\section{Postage stamp images}
\begin{figure*}
\centering
\includegraphics[width=0.245\textwidth, trim=0.5cm 0.5cm 0.5cm 0.cm, clip]{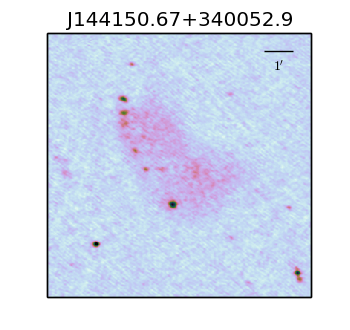}
\includegraphics[width=0.245\textwidth, trim=0.5cm 0.5cm 0.5cm 0.cm, clip]{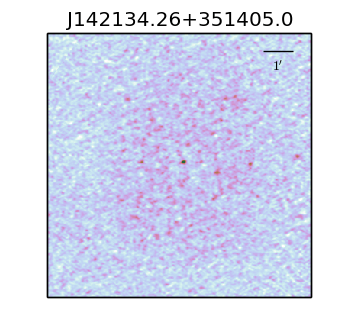}
\includegraphics[width=0.245\textwidth, trim=0.5cm 0.5cm 0.5cm 0.cm, clip]{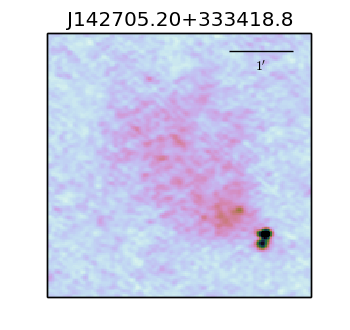}
\includegraphics[width=0.245\textwidth, trim=0.5cm 0.5cm 0.5cm 0.cm, clip]{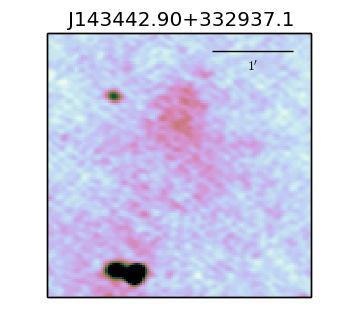}
\\
 \caption{Postage stamps of large diffuse sources identified by eye. The greyscale shows the flux density from  $-3\sigma_{\rm local}$ to $15\sigma_{\rm local}$ where  $\sigma_{\rm local}$ is the local $rms$ noise.  The scalebar in each image is $1${\arcmin}.}
 \label{fig:app:diffusesources}
\end{figure*}

The largest (LAS$>45$\,{\arcsec}) extended sources, including those merged into single sources, are shown in Fig.~\ref{fig:app:extendedsources}, and the four  very clear large diffuse sources are shown in Fig.~\ref{fig:app:diffusesources}.

%

\input{ext_srcs_v4_extendedimages.tex}

\section{Source size distribution}
\label{sect:ap:sizes}

%

\begin{figure*}
 \centering
\includegraphics[width=0.4\textwidth, trim=0cm 0cm 0cm 0cm, clip]{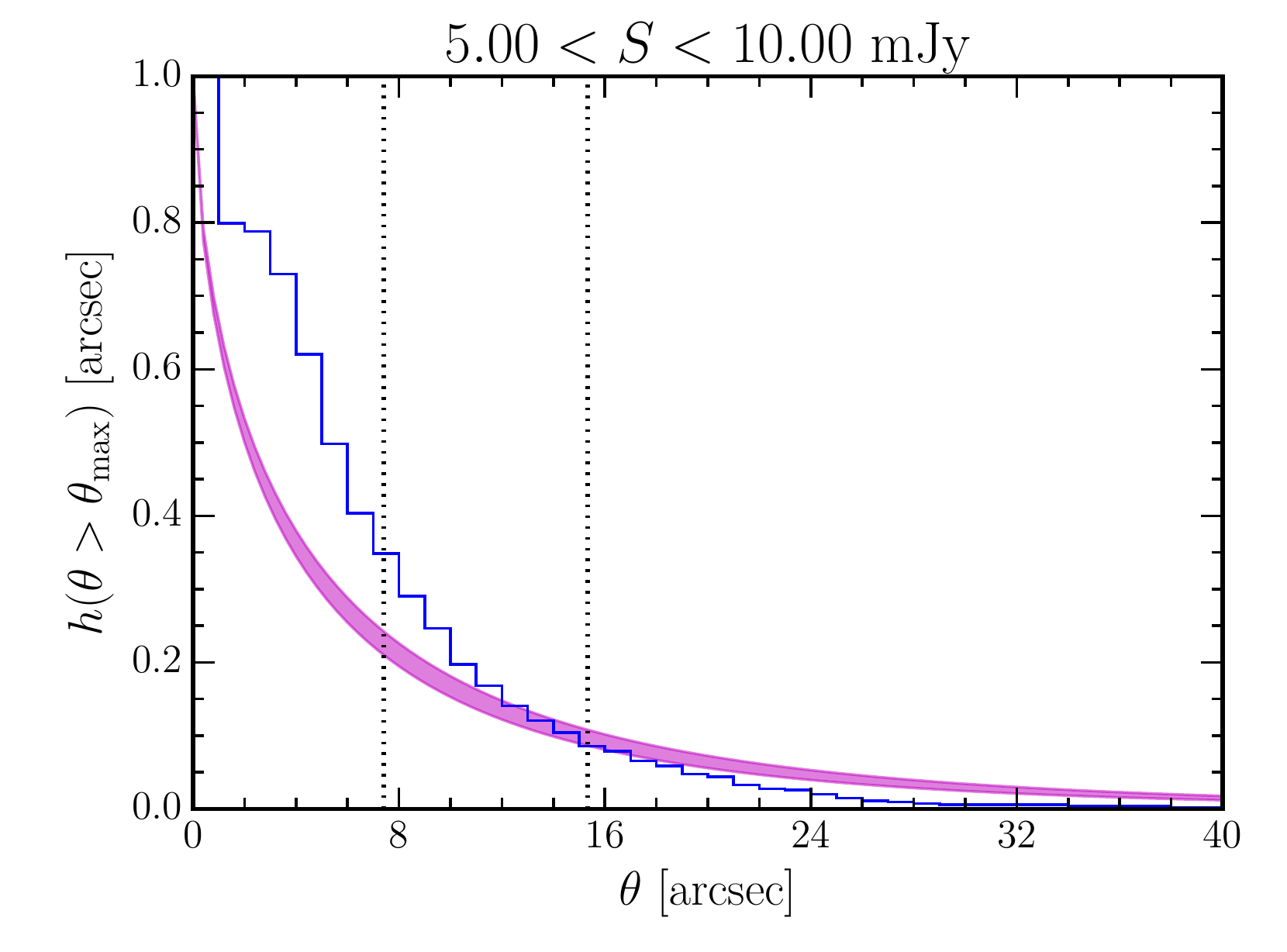}
\includegraphics[width=0.4\textwidth, trim=0cm 0cm 0cm 0cm, clip]{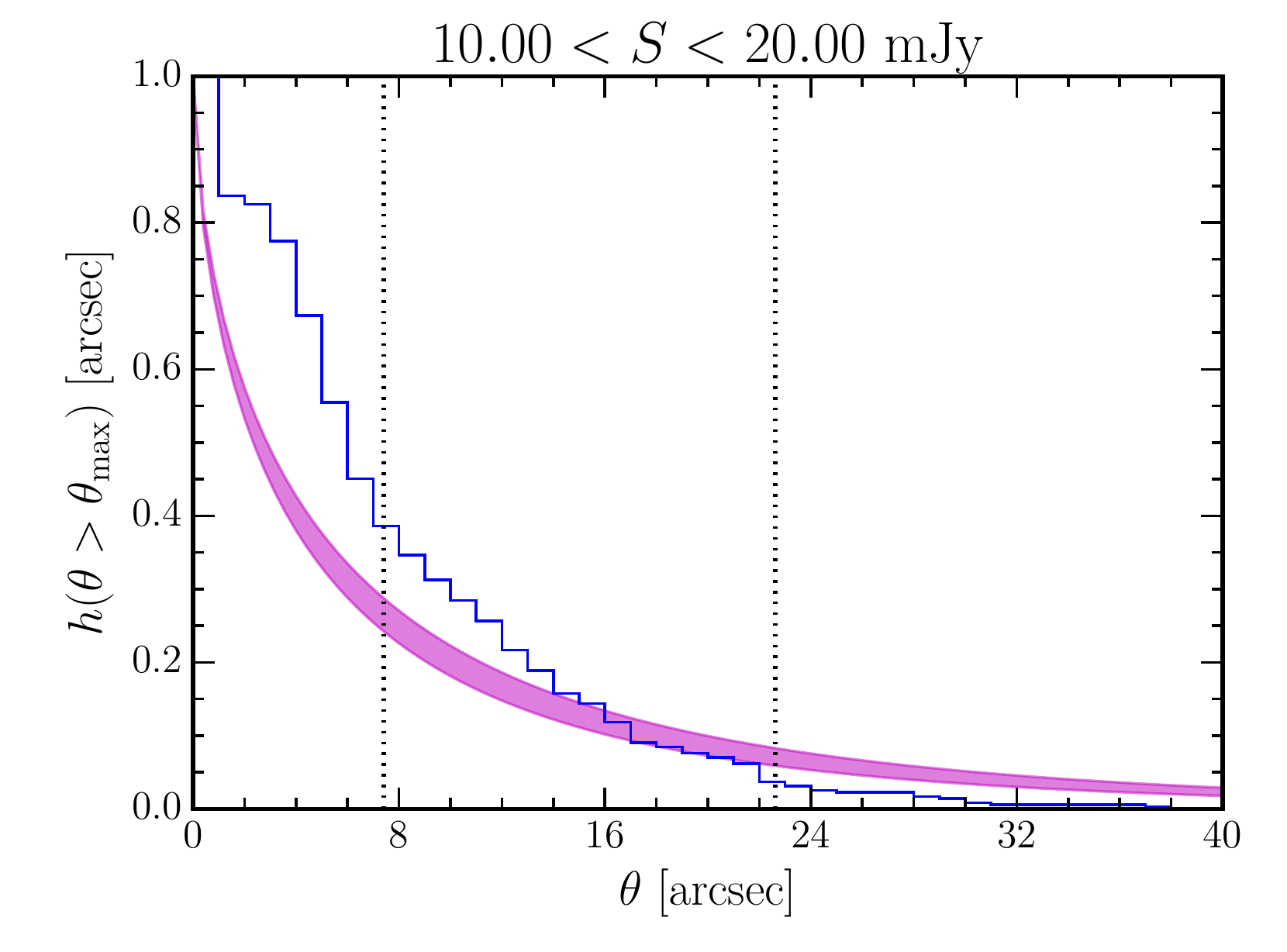}\\
\includegraphics[width=0.4\textwidth, trim=0cm 0cm 0cm 0cm, clip]{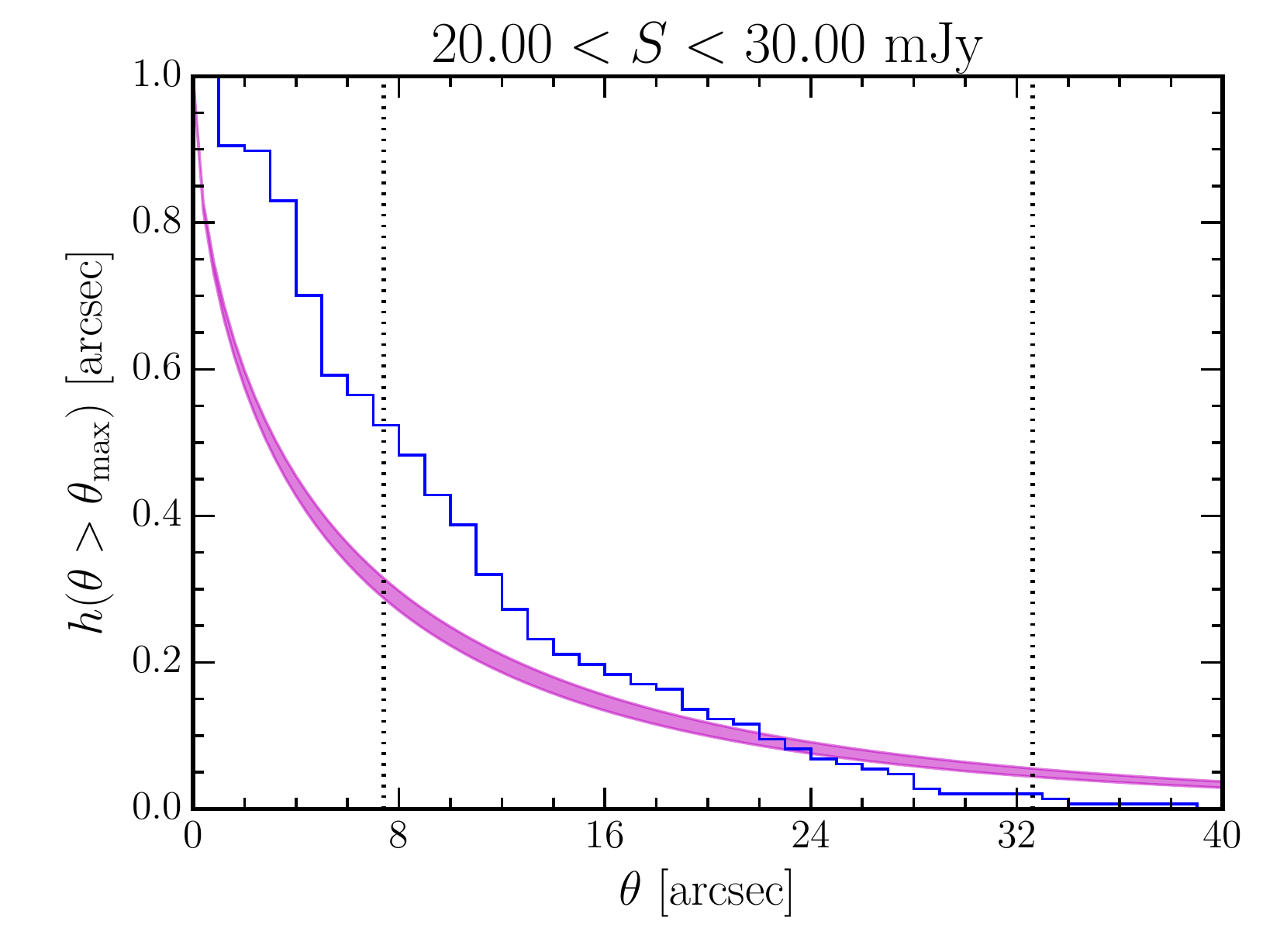}
\includegraphics[width=0.4\textwidth, trim=0cm 0cm 0cm 0cm, clip]{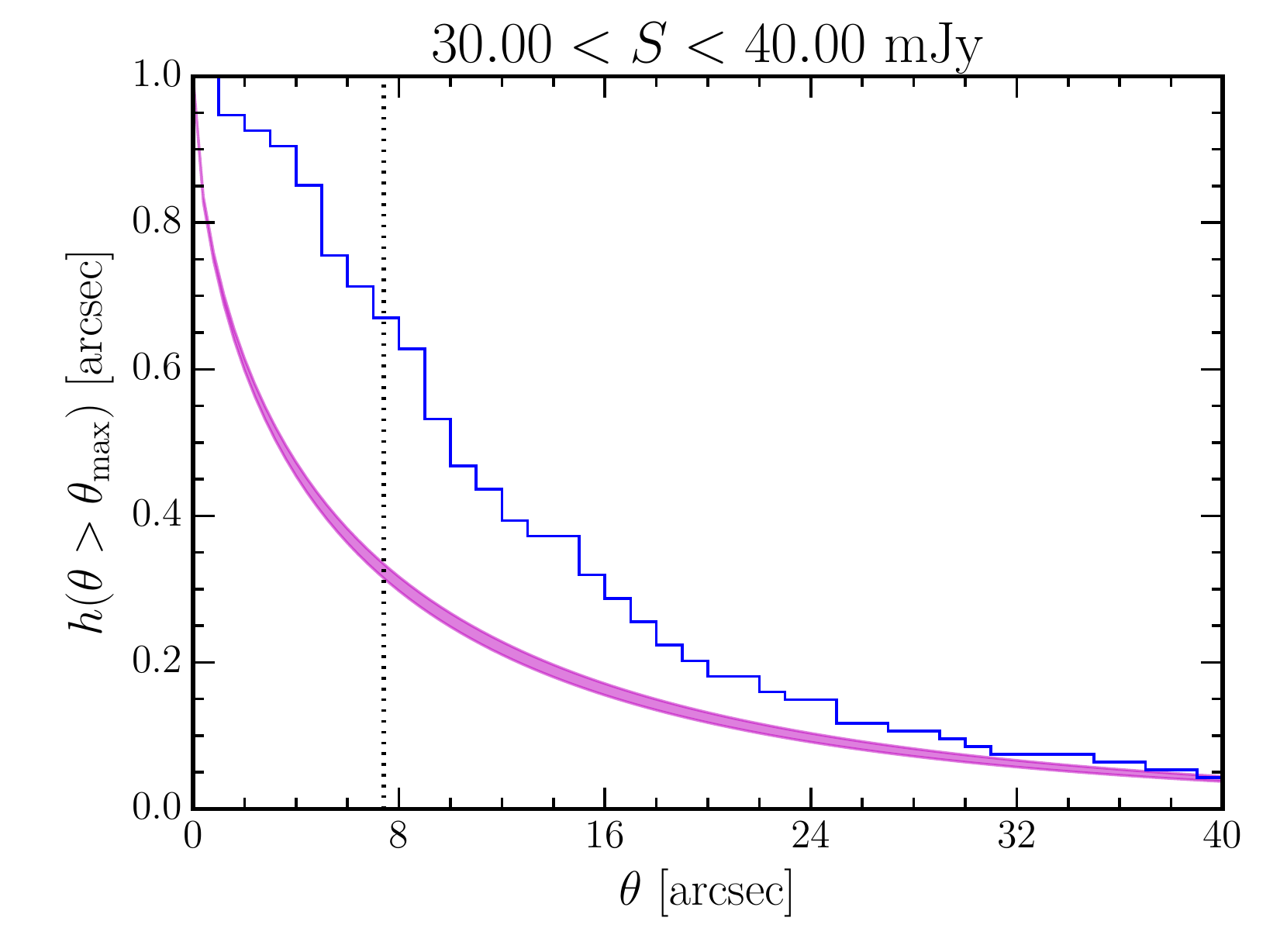}
\caption{The observed, deconvolved, source size distribution (in blue) in four flux density intervals. The magenta curves show the \citet{1990ASPC...10..389W} size distribution for the upper and lower bounds of the flux density bin. In each the vertical dotted lines are  $b_{\mathrm{maj}}$ (on the left) and  the approximate maximum size  a source can have before it drops below the peak-flux detection threshold (on the right). The catalogue will be incomplete for sources larger than the right line. }
\label{fig:ap:sizes}
\end{figure*}

We have investigated the extent to which the extrapolation of the \cite{1990ASPC...10..389W} size distribution might be valid at low frequencies. This is relevant to the resolution bias correction to the source counts, described in Section~\ref{sect:resbias}. We have done this by comparing the true (deconvolved) angular size distribution of the  $150$\,MHz  LOFAR sources to the \cite{1990ASPC...10..389W} distribution, given by
\[h(>\theta_{\rm max}) = \exp[-\ln 2 (\theta_{\rm max}/\theta_{\rm med})^{0.62}], \]
with the median size, in {\arcsec}, as a function of flux density of
\[\theta_{\rm med} = 2 S_{1.4\mbox{\,GHz}}^{0.30}, \] 
where $S$ is in {\mJy} and we have scaled the $1.4$\,GHz flux densities to $150$\,MHz with a spectral index of $-0.8$. The observed and extrapolated size distributions are shown in Fig.~\ref{fig:ap:sizes} for four flux density bins. The low frequency emission appears to be more extended, which would suggest that the actual resolution bias correction should be somewhat larger. Future LOFAR Survey results, in particular, at high resolution using very long baseline interfereometry with the LOFAR international stations, will allow for more detailed studies of the  true angular size distribution of radio sources.

\label{lastpage}
\end{document}

%% file: ext_srcs_v4_extendedimages.tex
\begin{figure*}
\centering
\includegraphics[width=0.245\textwidth, trim=0.5cm 0.5cm 0.5cm 0.cm, clip]{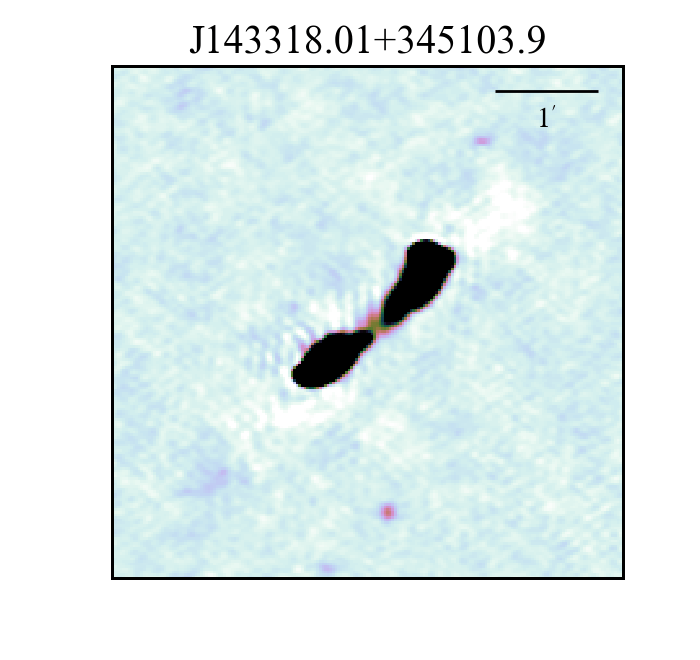}
\includegraphics[width=0.245\textwidth, trim=0.5cm 0.5cm 0.5cm 0.cm, clip]{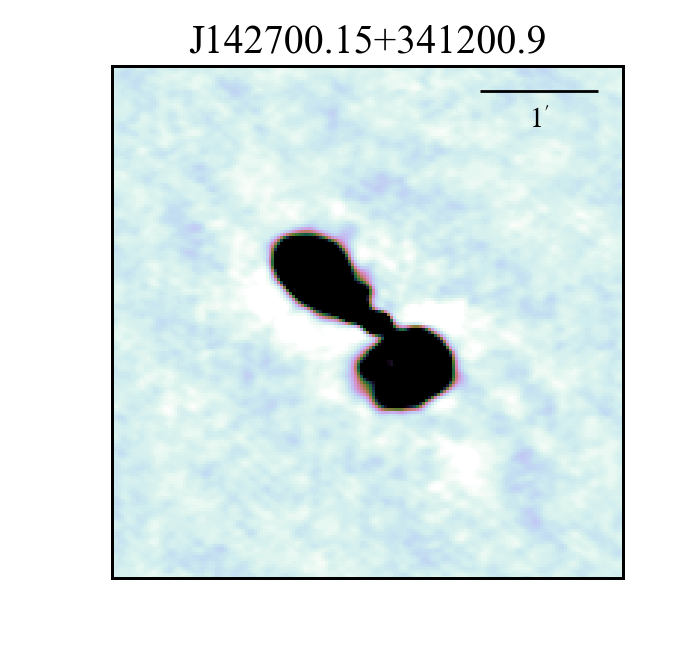}
\includegraphics[width=0.245\textwidth, trim=0.5cm 0.5cm 0.5cm 0.cm, clip]{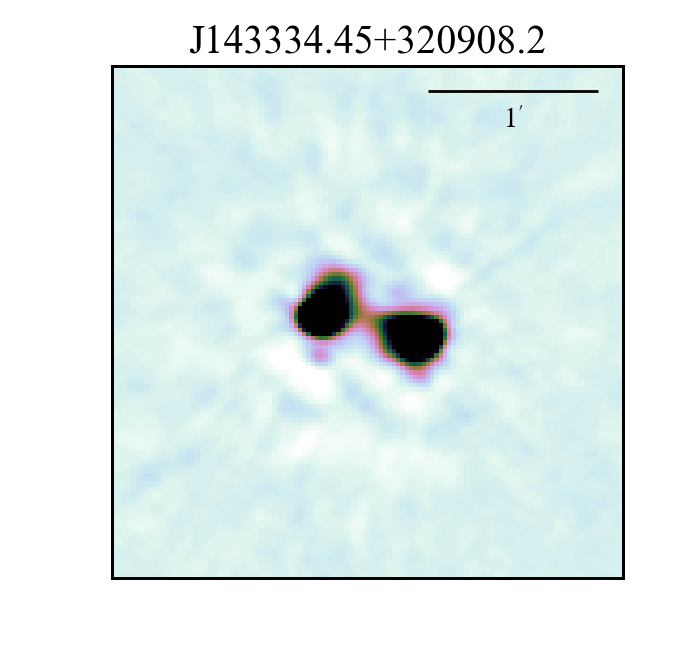}
\includegraphics[width=0.245\textwidth, trim=0.5cm 0.5cm 0.5cm 0.cm, clip]{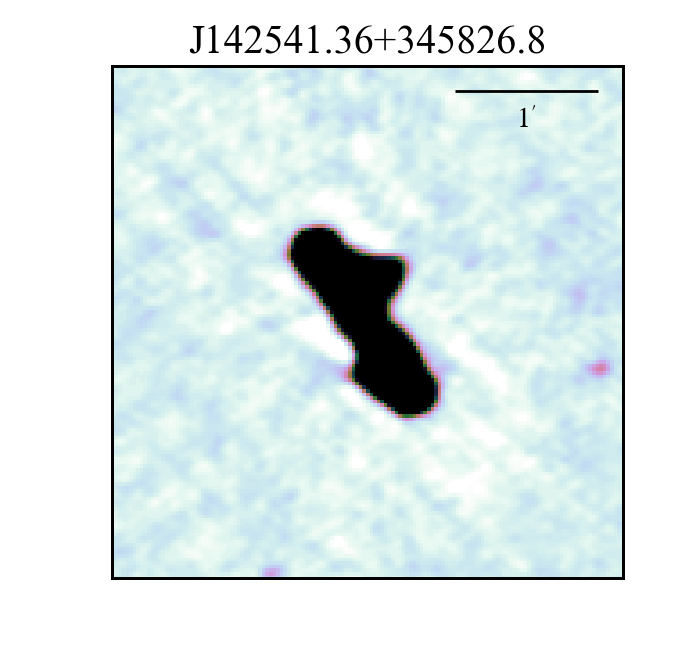}
\\
\includegraphics[width=0.245\textwidth, trim=0.5cm 0.5cm 0.5cm 0.cm, clip]{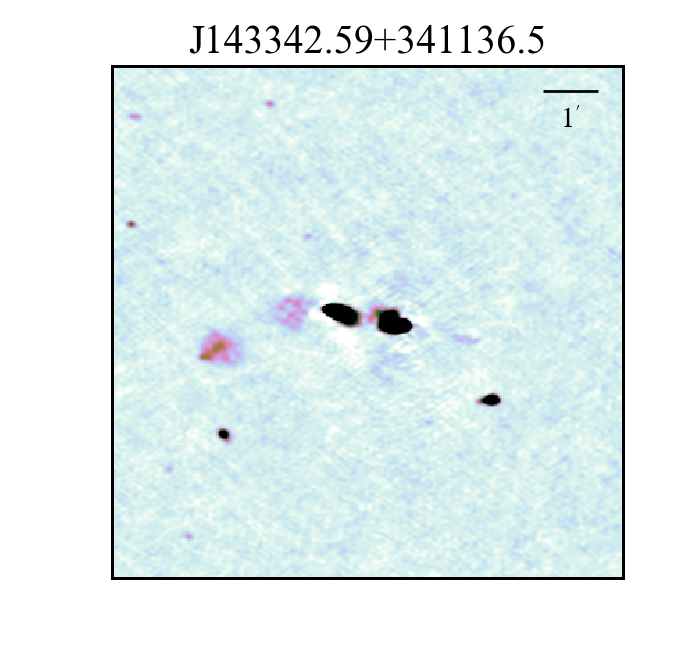}
\includegraphics[width=0.245\textwidth, trim=0.5cm 0.5cm 0.5cm 0.cm, clip]{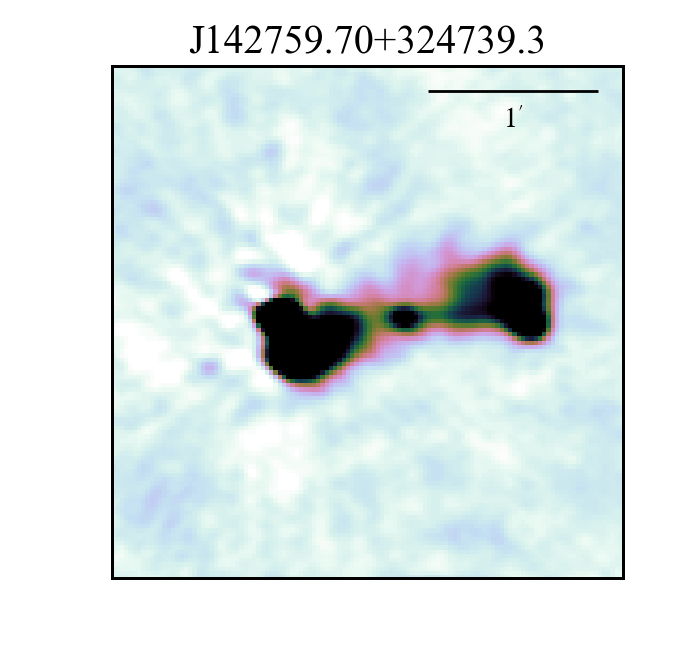}
\includegraphics[width=0.245\textwidth, trim=0.5cm 0.5cm 0.5cm 0.cm, clip]{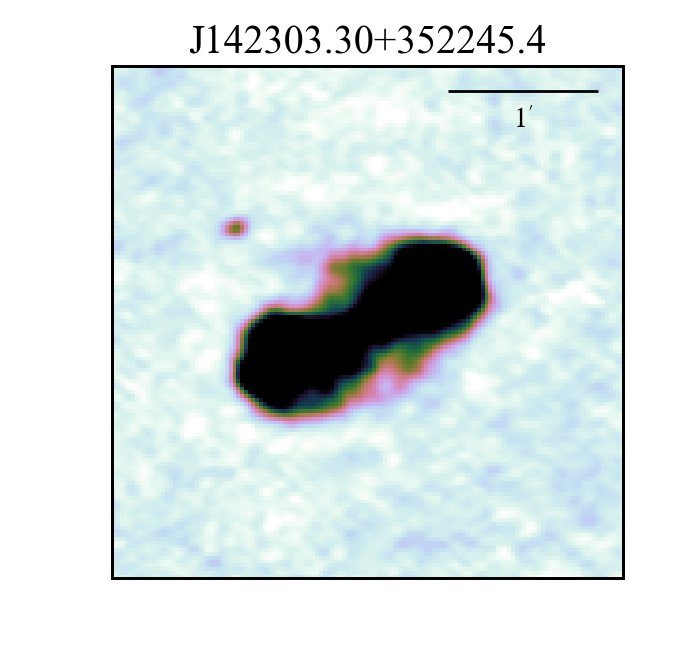}
\includegraphics[width=0.245\textwidth, trim=0.5cm 0.5cm 0.5cm 0.cm, clip]{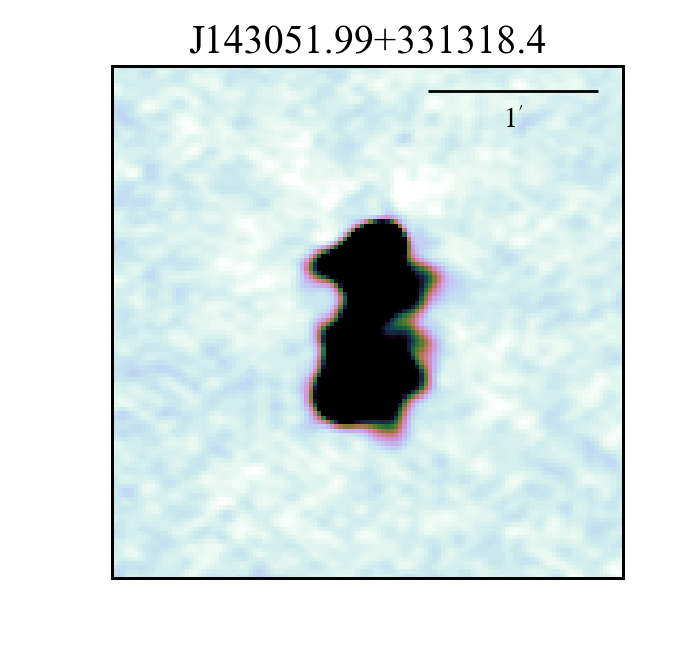}
\\
\includegraphics[width=0.245\textwidth, trim=0.5cm 0.5cm 0.5cm 0.cm, clip]{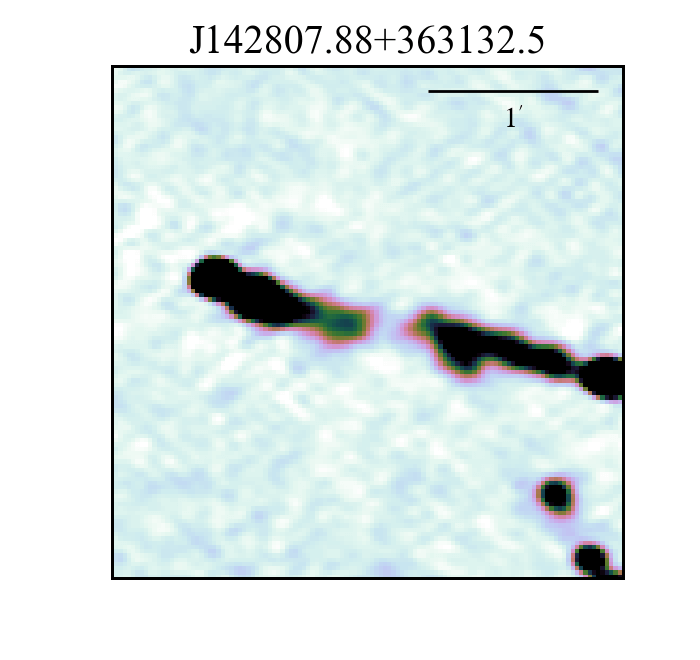}
\includegraphics[width=0.245\textwidth, trim=0.5cm 0.5cm 0.5cm 0.cm, clip]{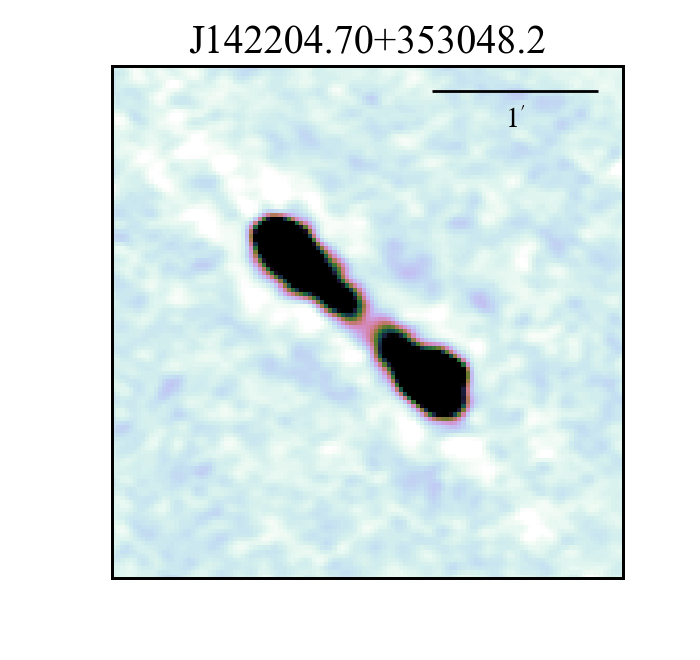}
\includegraphics[width=0.245\textwidth, trim=0.5cm 0.5cm 0.5cm 0.cm, clip]{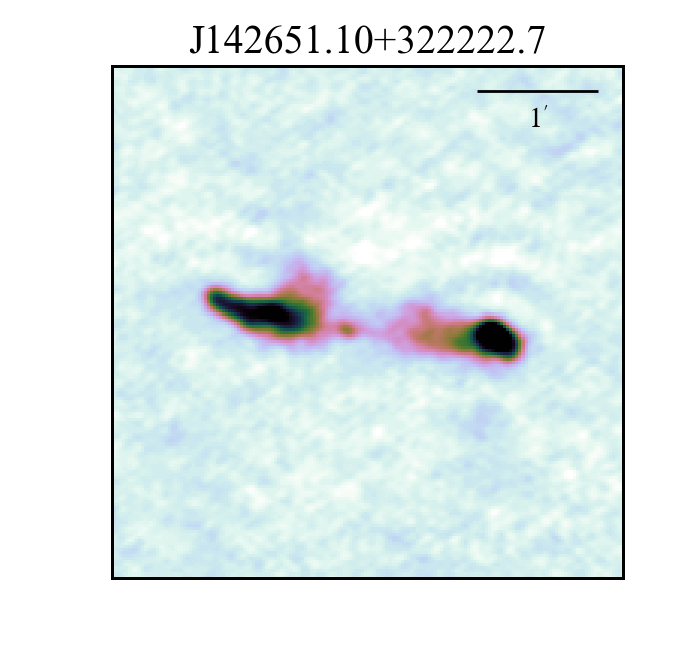}
\includegraphics[width=0.245\textwidth, trim=0.5cm 0.5cm 0.5cm 0.cm, clip]{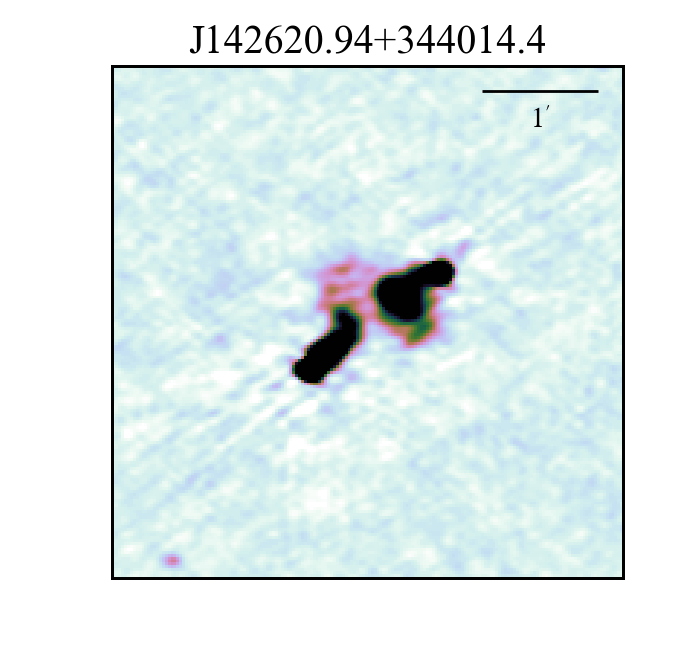}
\\
\includegraphics[width=0.245\textwidth, trim=0.5cm 0.5cm 0.5cm 0.cm, clip]{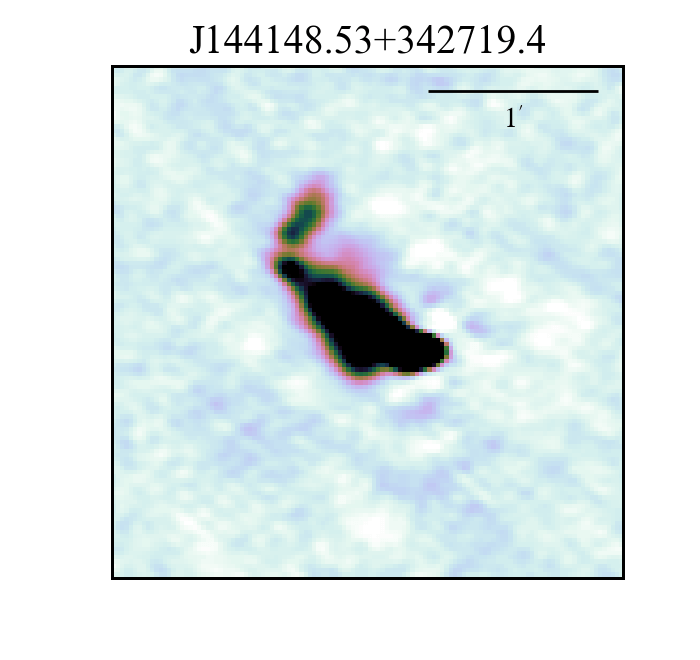}
\includegraphics[width=0.245\textwidth, trim=0.5cm 0.5cm 0.5cm 0.cm, clip]{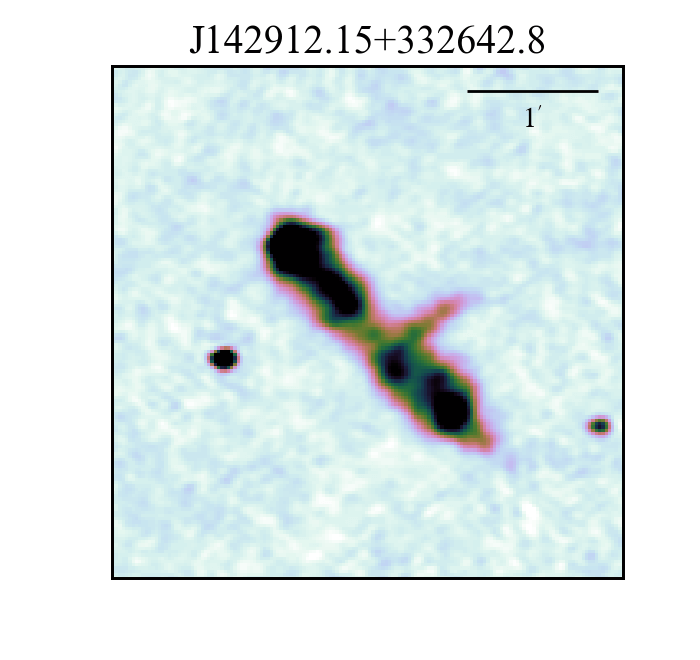}
\includegraphics[width=0.245\textwidth, trim=0.5cm 0.5cm 0.5cm 0.cm, clip]{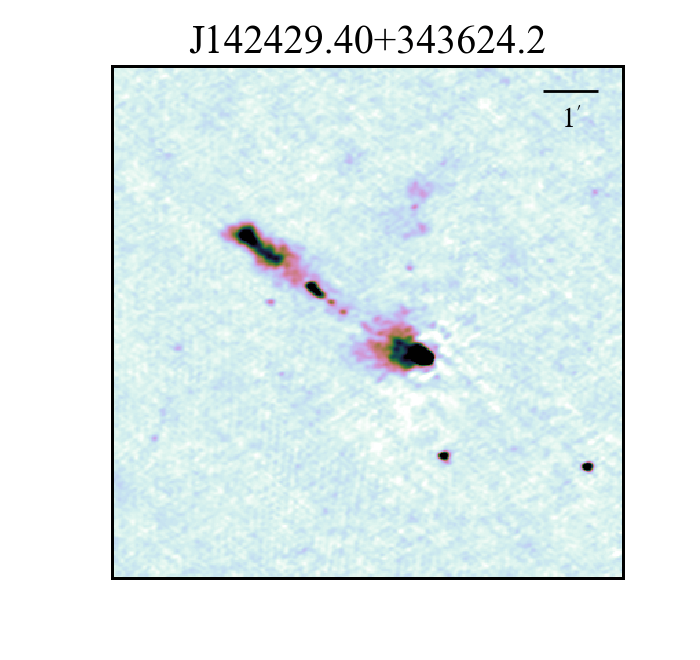}
\includegraphics[width=0.245\textwidth, trim=0.5cm 0.5cm 0.5cm 0.cm, clip]{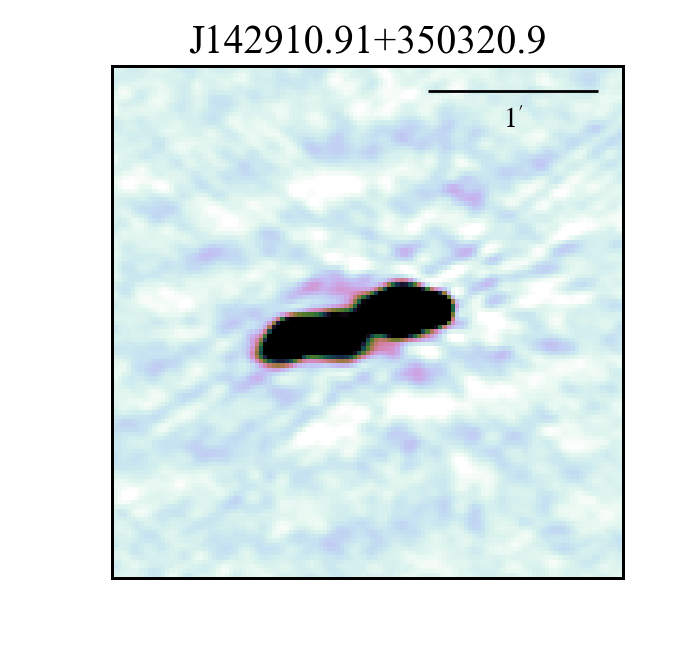}
\\
\includegraphics[width=0.245\textwidth, trim=0.5cm 0.5cm 0.5cm 0.cm, clip]{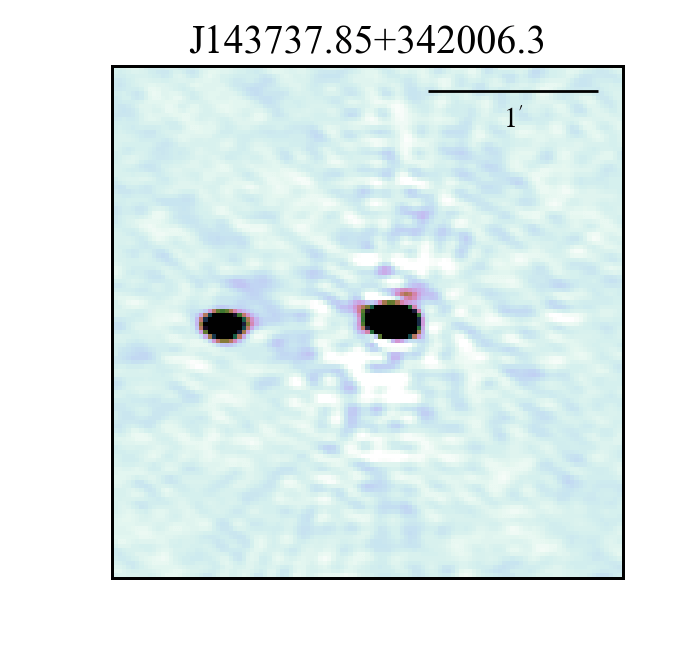}
\includegraphics[width=0.245\textwidth, trim=0.5cm 0.5cm 0.5cm 0.cm, clip]{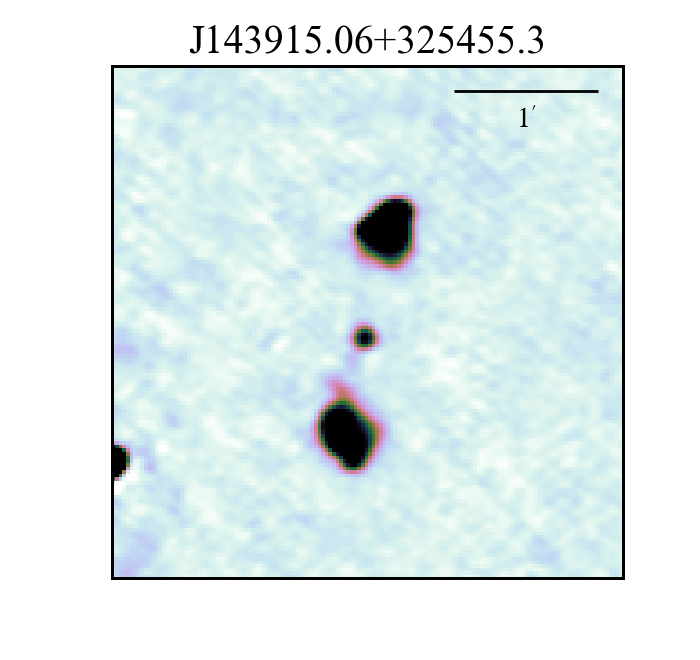}
\includegraphics[width=0.245\textwidth, trim=0.5cm 0.5cm 0.5cm 0.cm, clip]{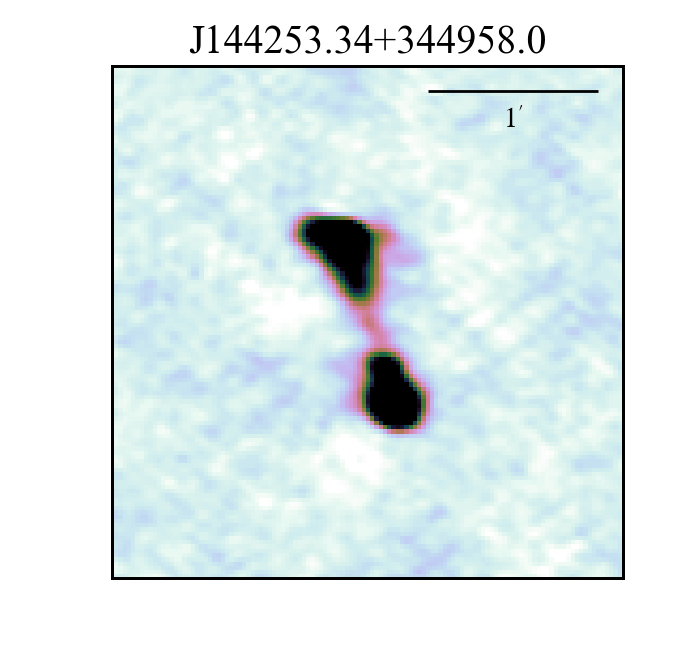}
\includegraphics[width=0.245\textwidth, trim=0.5cm 0.5cm 0.5cm 0.cm, clip]{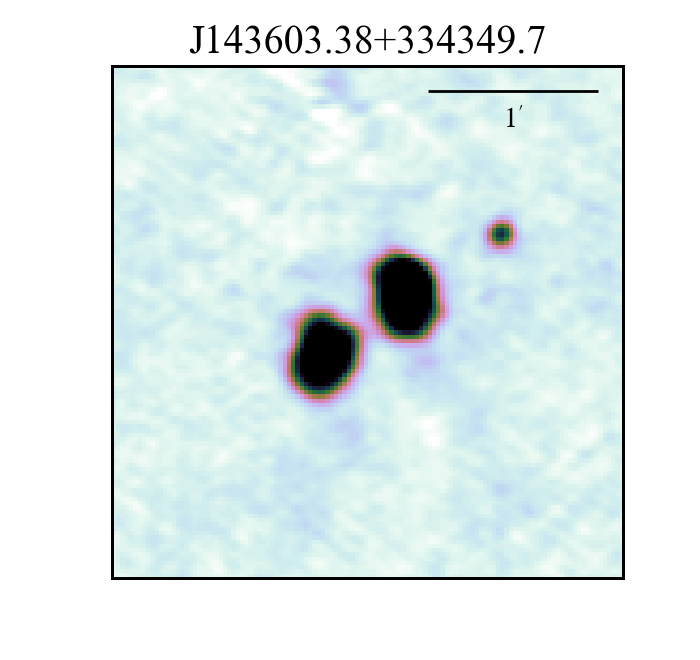}
\\
 \caption{Postage stamps of extended sources identified visually, showing all the sources with approximate LAS$>45${\arcsec}. These include some Giant Radio Galaxies. The greyscale shows the flux density from  $-3\sigma_{local}$ to $30\sigma_{local}$ where  $\sigma_{local}$ is the local $rms$ noise. The scalebar in each image is $1${\arcmin}.}
 \label{fig:app:extendedsources}
\end{figure*}
\begin{figure*}
\centering
\includegraphics[width=0.245\textwidth, trim=0.5cm 0.5cm 0.5cm 0.cm, clip]{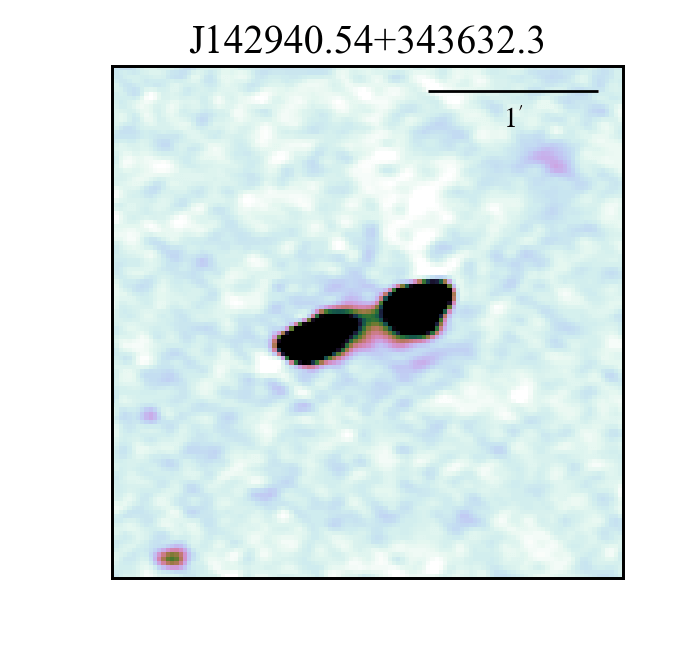}
\includegraphics[width=0.245\textwidth, trim=0.5cm 0.5cm 0.5cm 0.cm, clip]{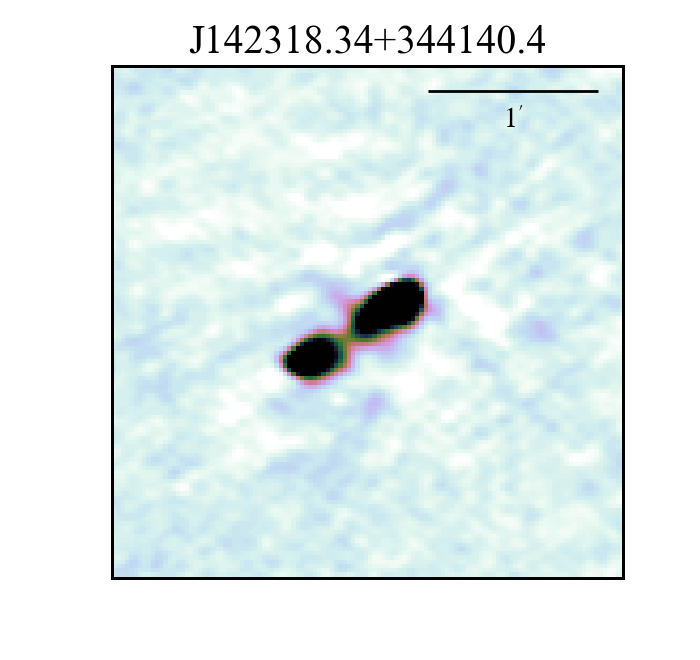}
\includegraphics[width=0.245\textwidth, trim=0.5cm 0.5cm 0.5cm 0.cm, clip]{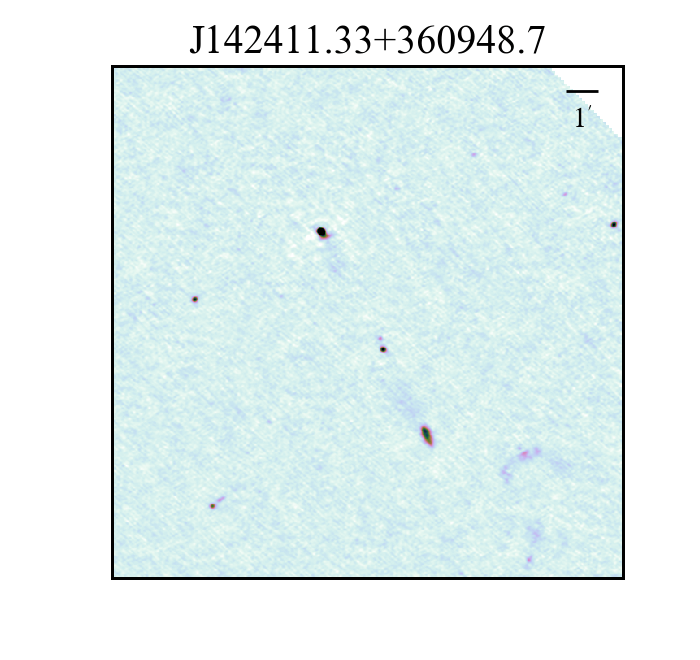}
\includegraphics[width=0.245\textwidth, trim=0.5cm 0.5cm 0.5cm 0.cm, clip]{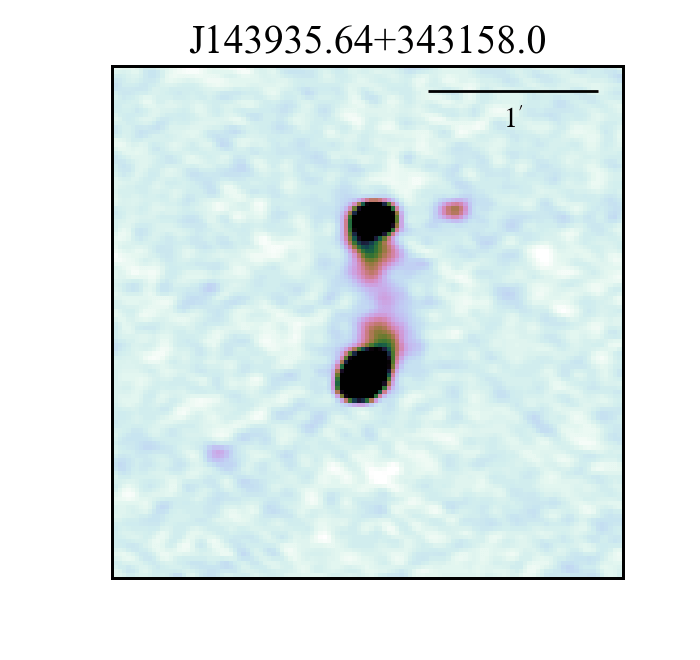}
\\
\includegraphics[width=0.245\textwidth, trim=0.5cm 0.5cm 0.5cm 0.cm, clip]{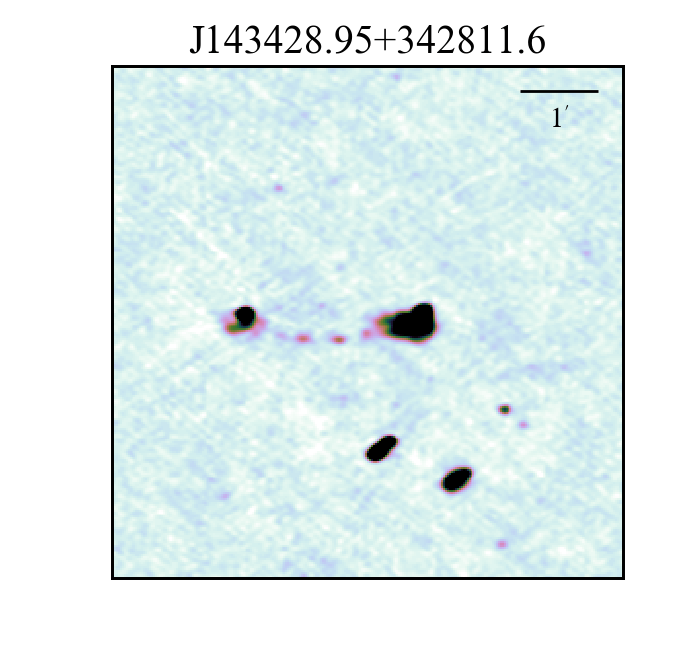}
\includegraphics[width=0.245\textwidth, trim=0.5cm 0.5cm 0.5cm 0.cm, clip]{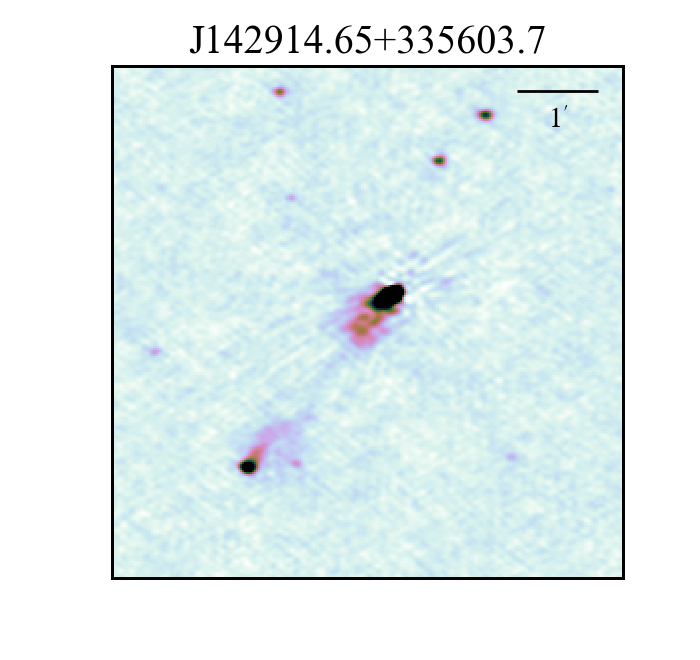}
\includegraphics[width=0.245\textwidth, trim=0.5cm 0.5cm 0.5cm 0.cm, clip]{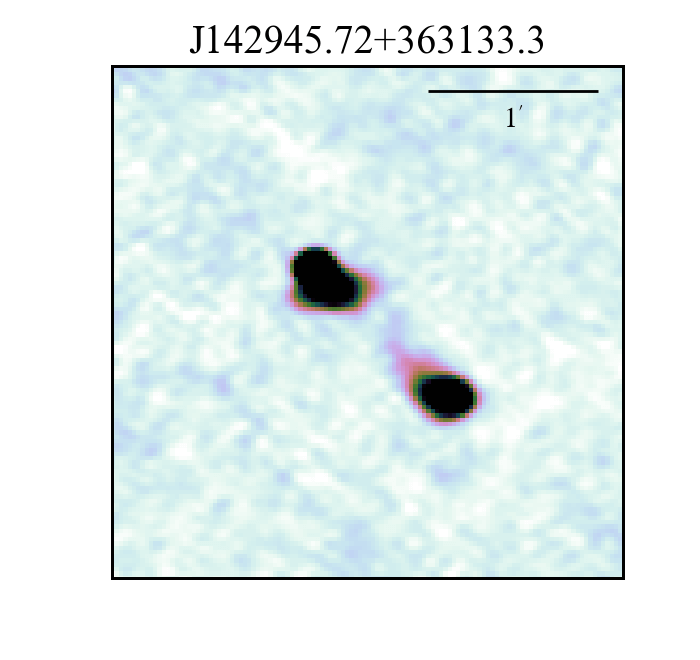}
\includegraphics[width=0.245\textwidth, trim=0.5cm 0.5cm 0.5cm 0.cm, clip]{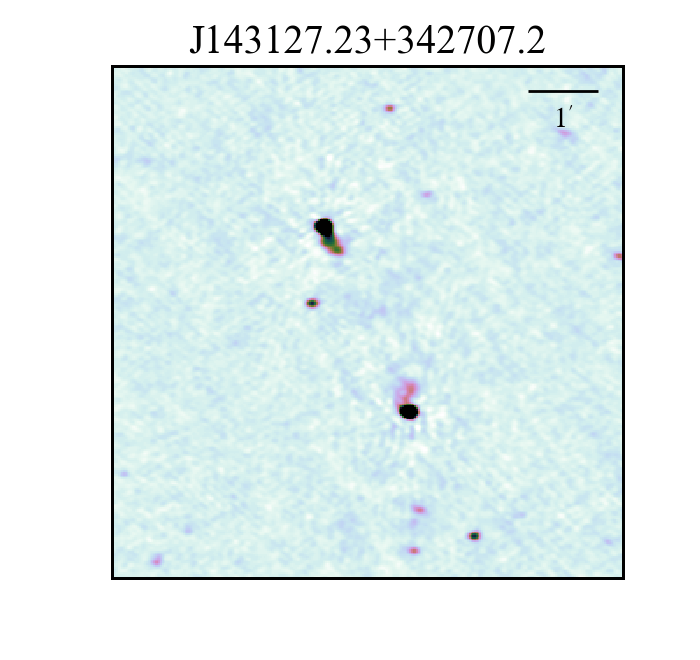}
\\
\includegraphics[width=0.245\textwidth, trim=0.5cm 0.5cm 0.5cm 0.cm, clip]{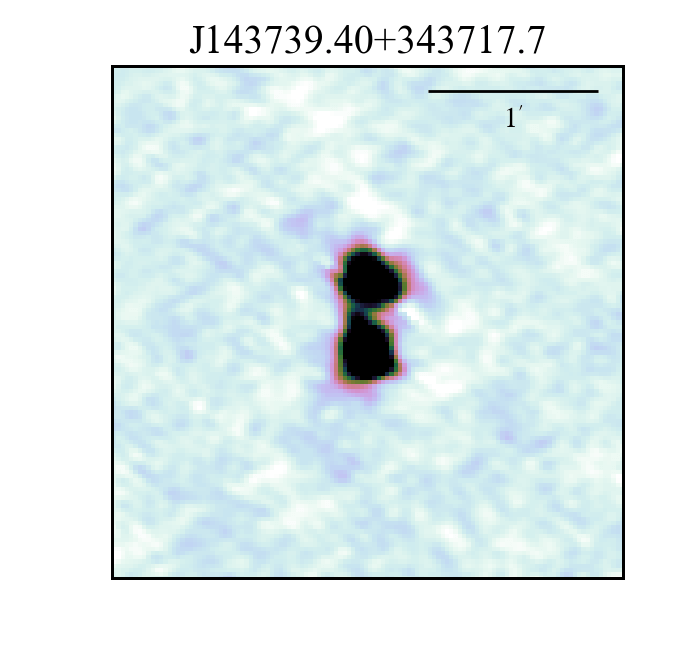}
\includegraphics[width=0.245\textwidth, trim=0.5cm 0.5cm 0.5cm 0.cm, clip]{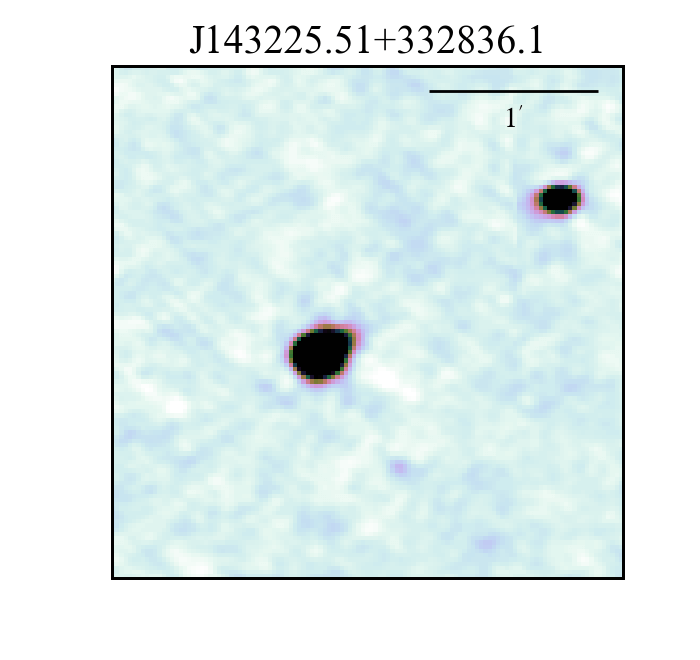}
\includegraphics[width=0.245\textwidth, trim=0.5cm 0.5cm 0.5cm 0.cm, clip]{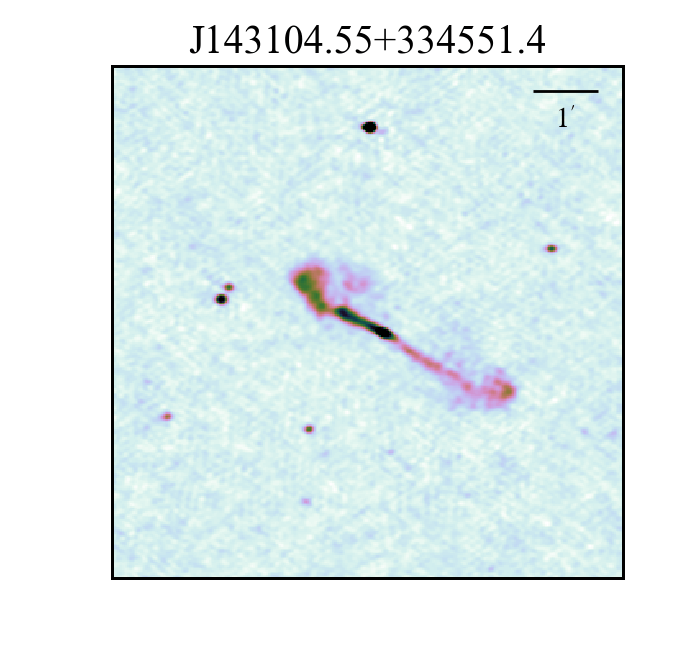}
\includegraphics[width=0.245\textwidth, trim=0.5cm 0.5cm 0.5cm 0.cm, clip]{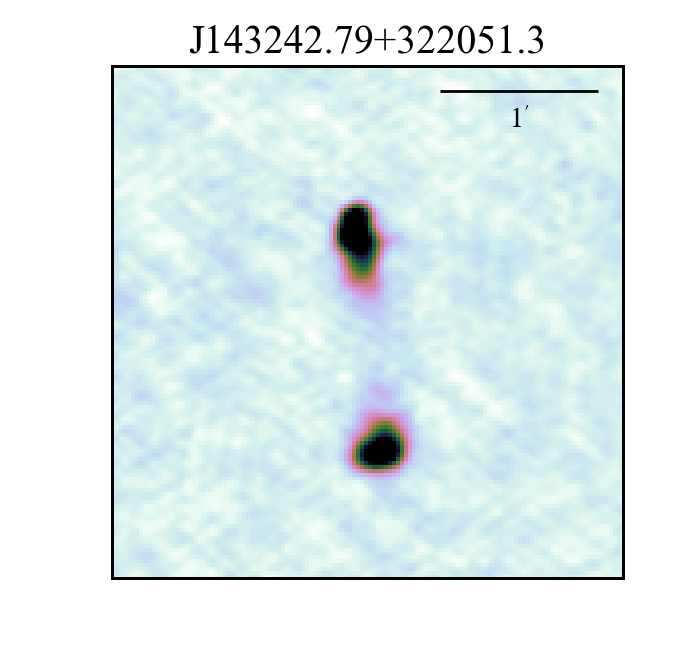}
\\
\includegraphics[width=0.245\textwidth, trim=0.5cm 0.5cm 0.5cm 0.cm, clip]{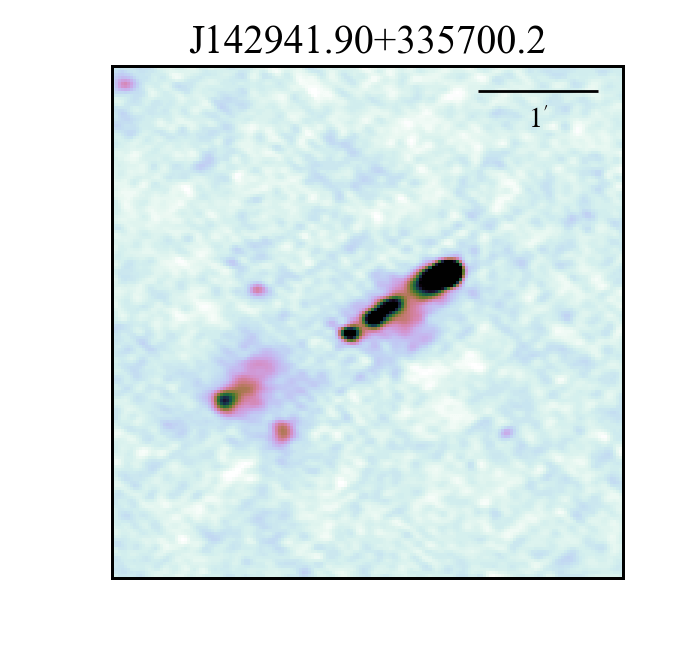}
\includegraphics[width=0.245\textwidth, trim=0.5cm 0.5cm 0.5cm 0.cm, clip]{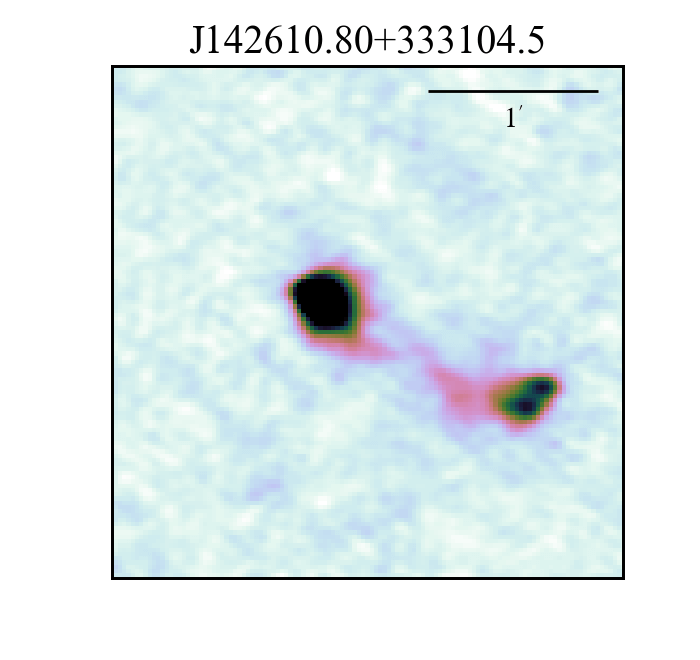}
\includegraphics[width=0.245\textwidth, trim=0.5cm 0.5cm 0.5cm 0.cm, clip]{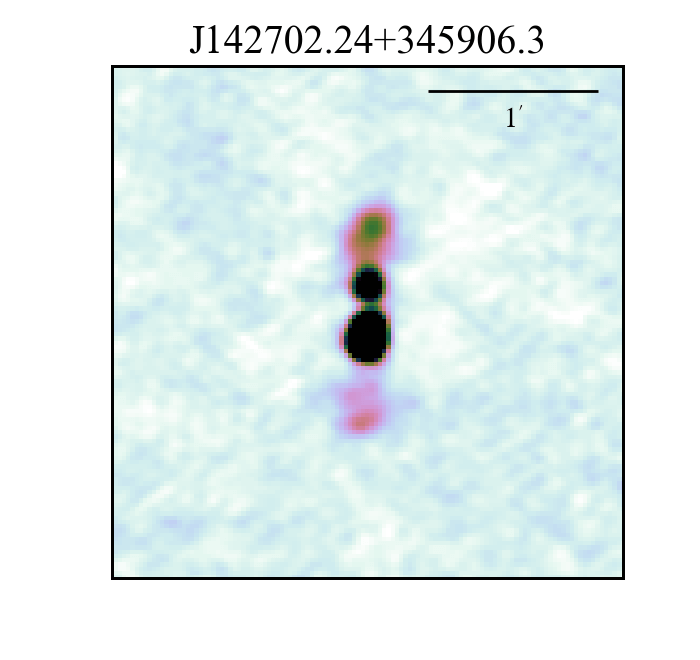}
\includegraphics[width=0.245\textwidth, trim=0.5cm 0.5cm 0.5cm 0.cm, clip]{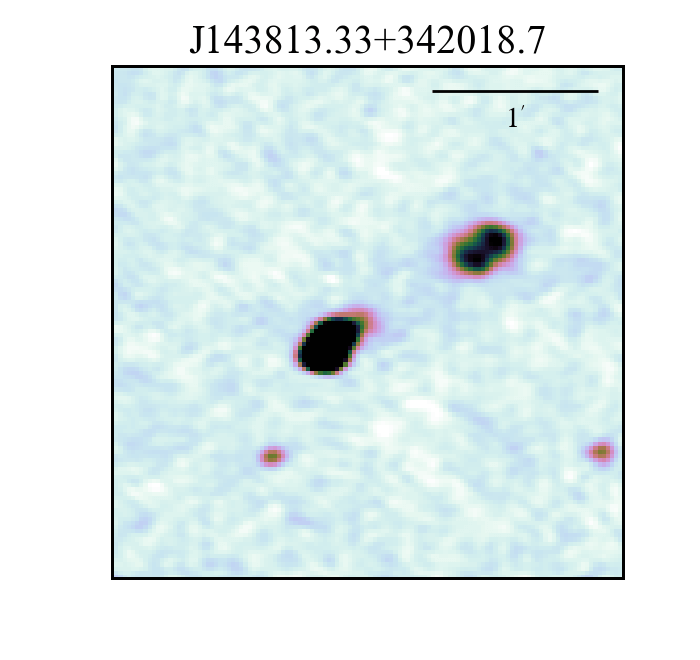}
\\
\includegraphics[width=0.245\textwidth, trim=0.5cm 0.5cm 0.5cm 0.cm, clip]{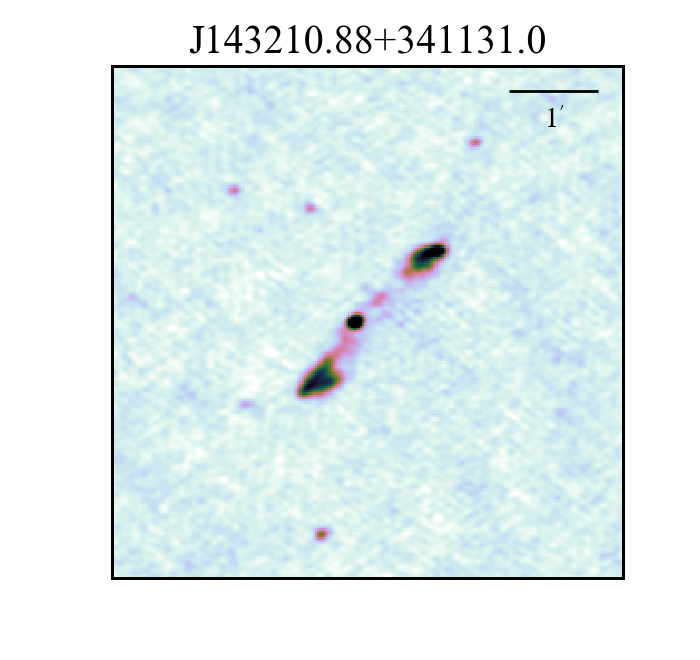}
\includegraphics[width=0.245\textwidth, trim=0.5cm 0.5cm 0.5cm 0.cm, clip]{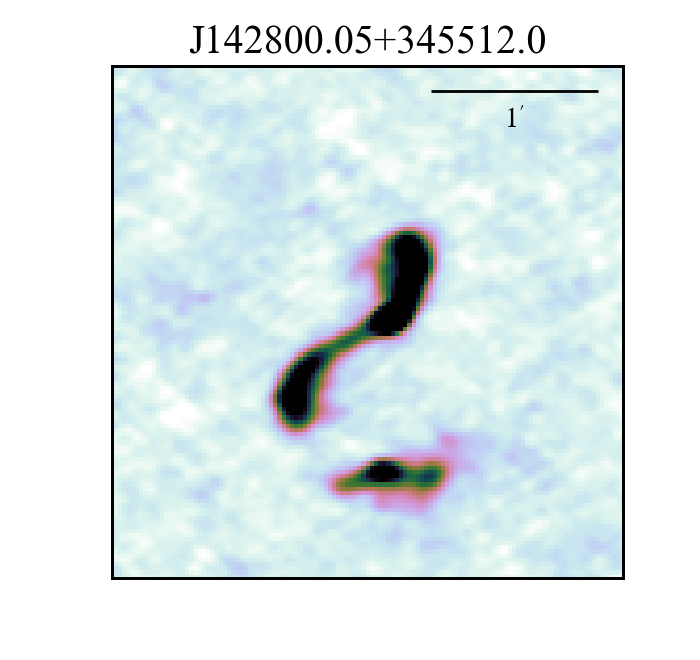}
\includegraphics[width=0.245\textwidth, trim=0.5cm 0.5cm 0.5cm 0.cm, clip]{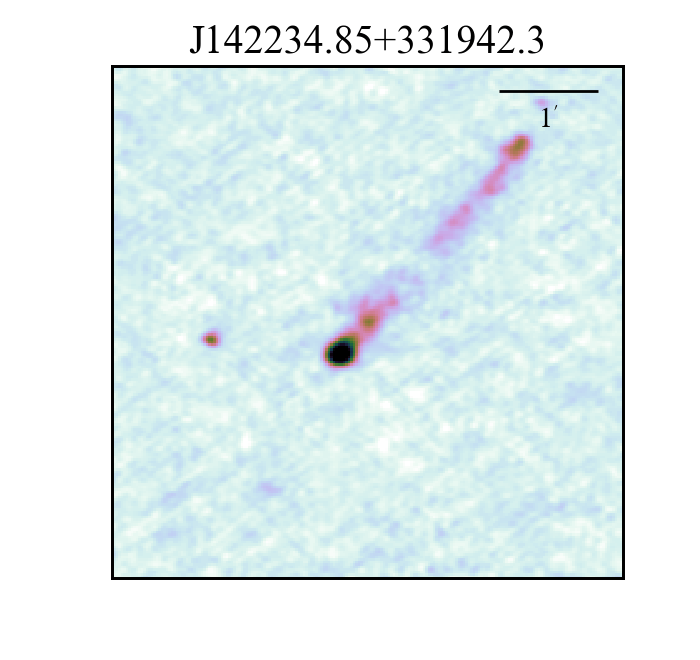}
\includegraphics[width=0.245\textwidth, trim=0.5cm 0.5cm 0.5cm 0.cm, clip]{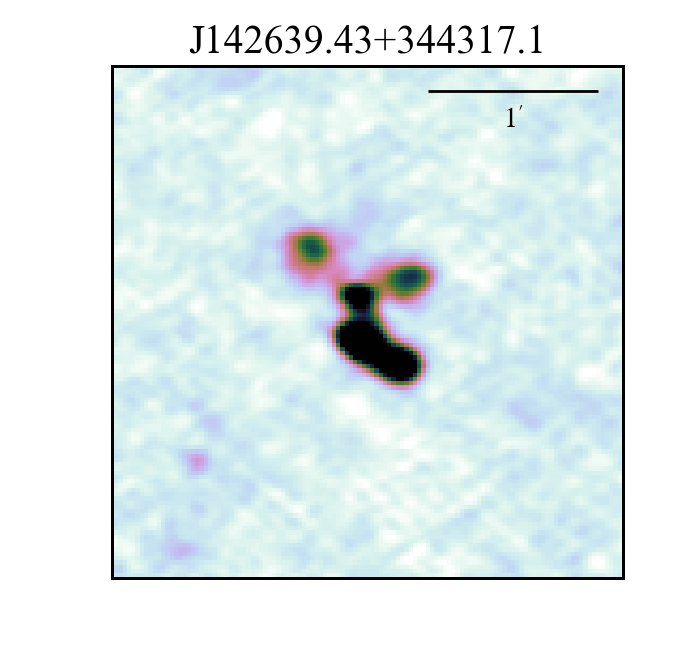}
\\
\contcaption{}
\end{figure*}
\begin{figure*}
\centering
\includegraphics[width=0.245\textwidth, trim=0.5cm 0.5cm 0.5cm 0.cm, clip]{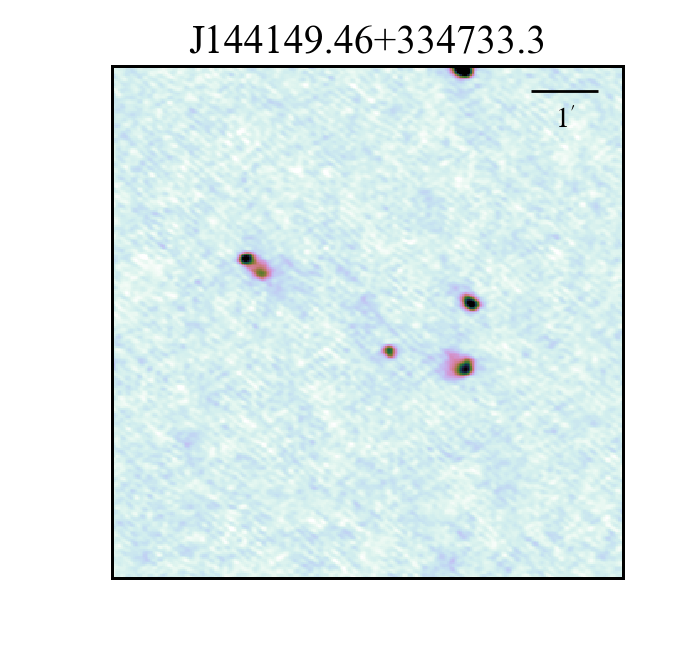}
\includegraphics[width=0.245\textwidth, trim=0.5cm 0.5cm 0.5cm 0.cm, clip]{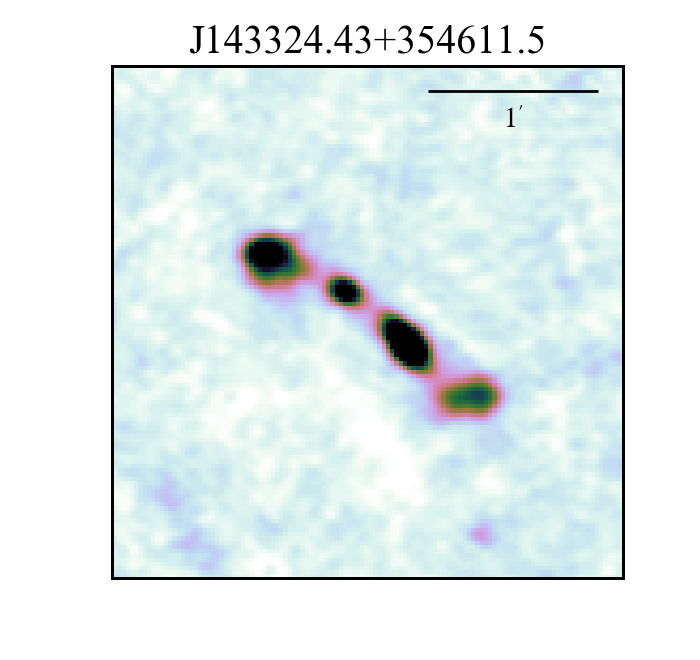}
\includegraphics[width=0.245\textwidth, trim=0.5cm 0.5cm 0.5cm 0.cm, clip]{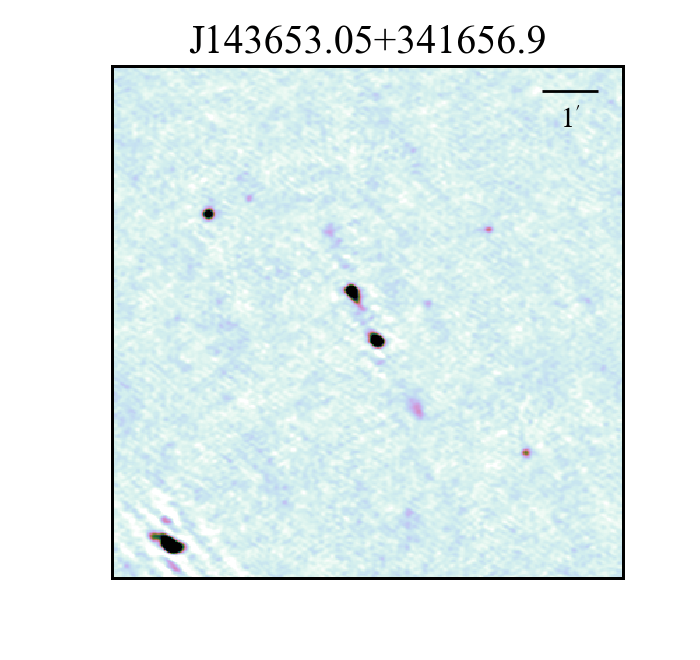}
\includegraphics[width=0.245\textwidth, trim=0.5cm 0.5cm 0.5cm 0.cm, clip]{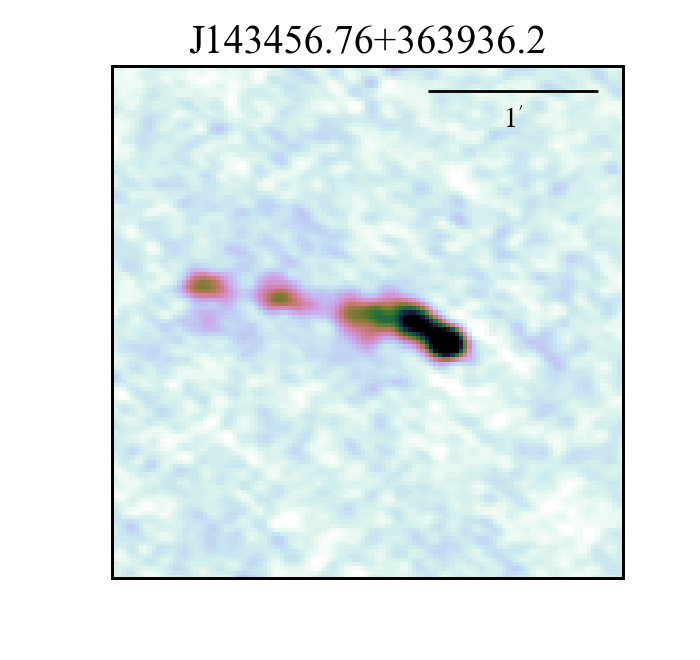}
\\
\includegraphics[width=0.245\textwidth, trim=0.5cm 0.5cm 0.5cm 0.cm, clip]{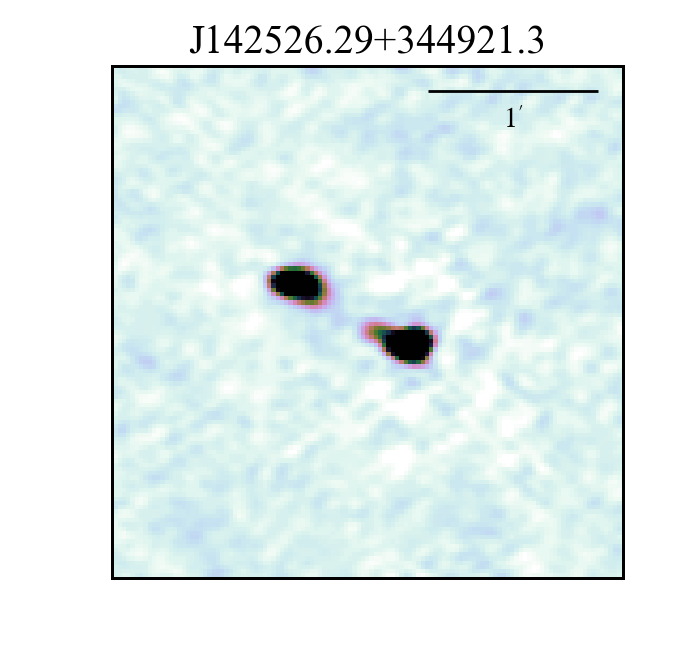}
\includegraphics[width=0.245\textwidth, trim=0.5cm 0.5cm 0.5cm 0.cm, clip]{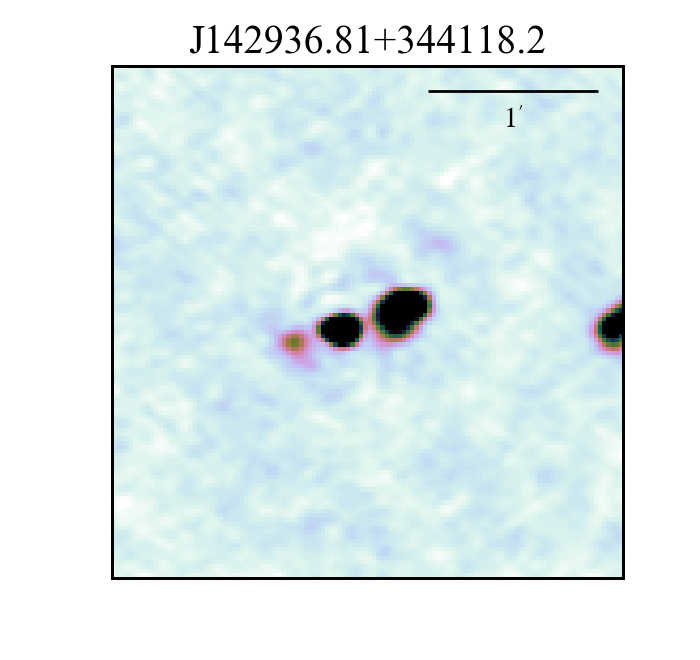}
\includegraphics[width=0.245\textwidth, trim=0.5cm 0.5cm 0.5cm 0.cm, clip]{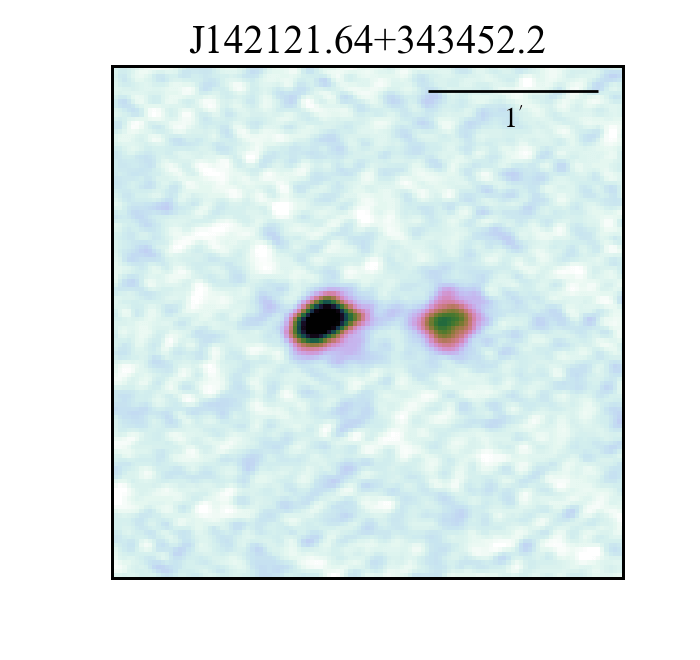}
\includegraphics[width=0.245\textwidth, trim=0.5cm 0.5cm 0.5cm 0.cm, clip]{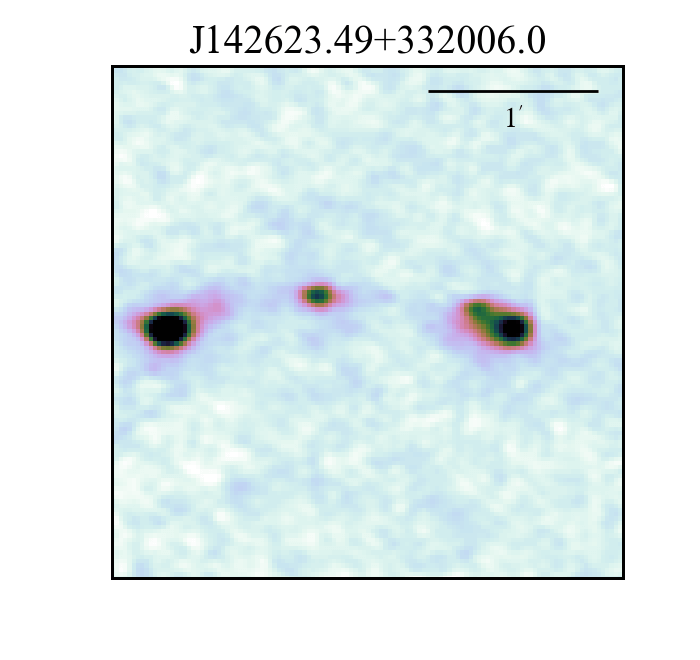}
\\
\includegraphics[width=0.245\textwidth, trim=0.5cm 0.5cm 0.5cm 0.cm, clip]{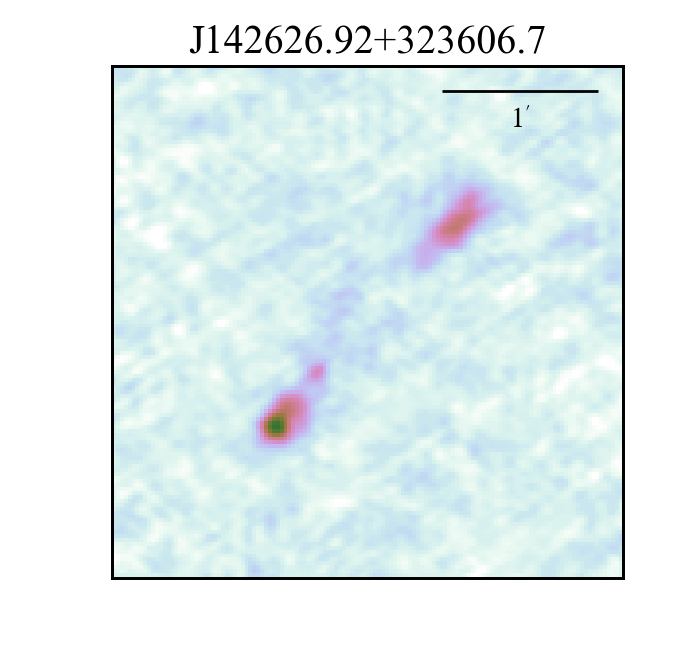}
\includegraphics[width=0.245\textwidth, trim=0.5cm 0.5cm 0.5cm 0.cm, clip]{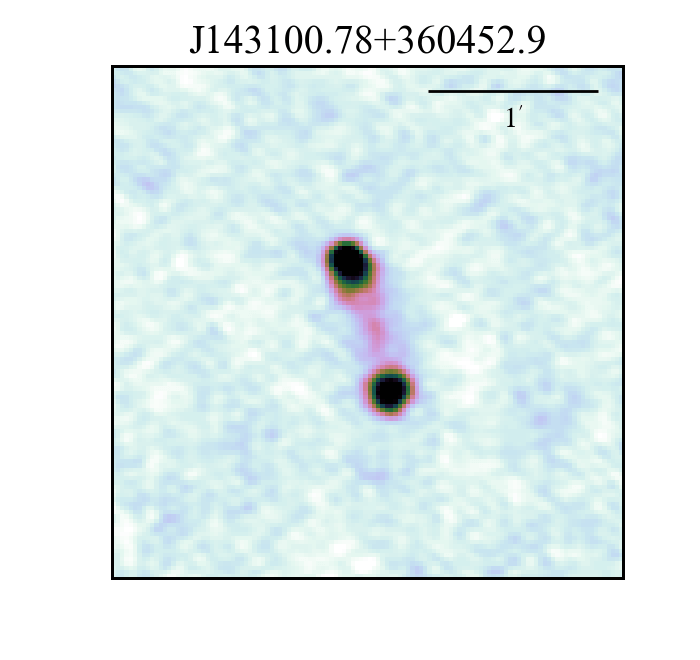}
\includegraphics[width=0.245\textwidth, trim=0.5cm 0.5cm 0.5cm 0.cm, clip]{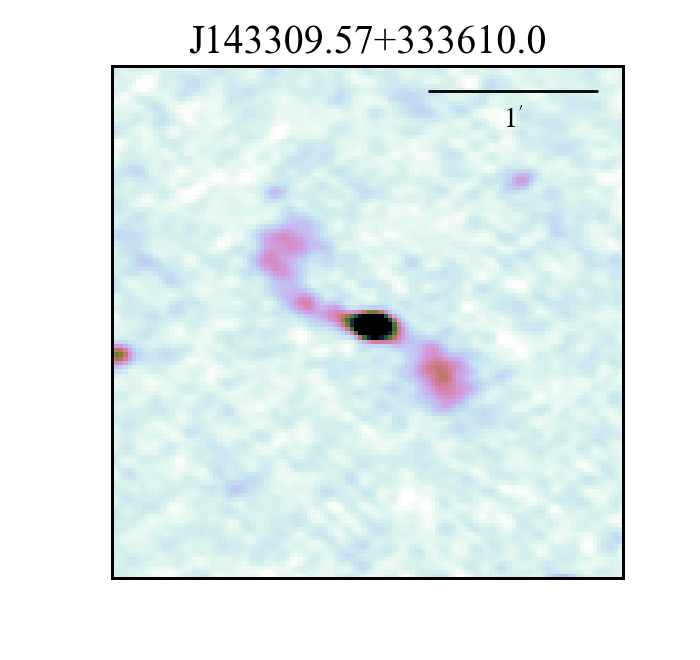}
\includegraphics[width=0.245\textwidth, trim=0.5cm 0.5cm 0.5cm 0.cm, clip]{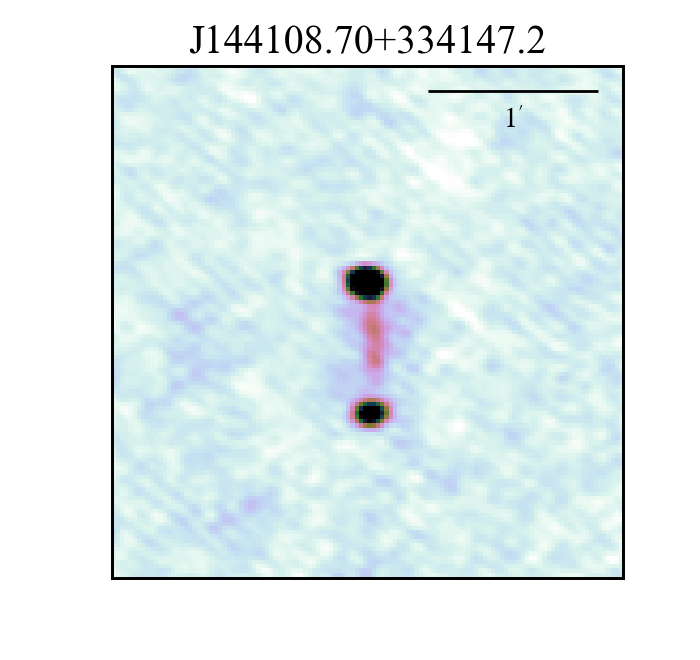}
\\
\includegraphics[width=0.245\textwidth, trim=0.5cm 0.5cm 0.5cm 0.cm, clip]{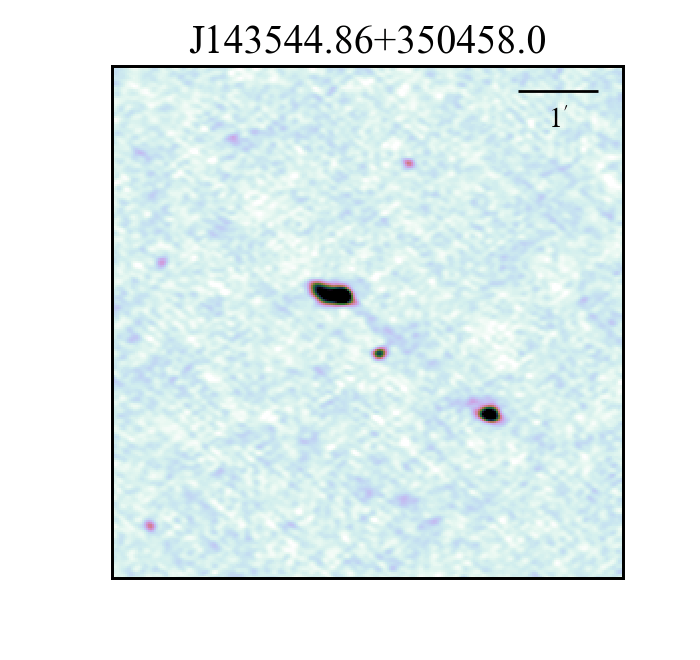}
\includegraphics[width=0.245\textwidth, trim=0.5cm 0.5cm 0.5cm 0.cm, clip]{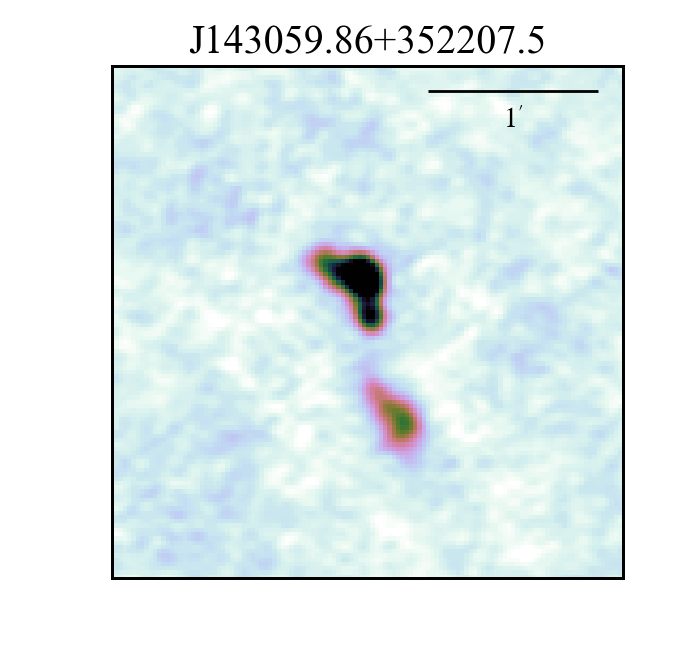}
\includegraphics[width=0.245\textwidth, trim=0.5cm 0.5cm 0.5cm 0.cm, clip]{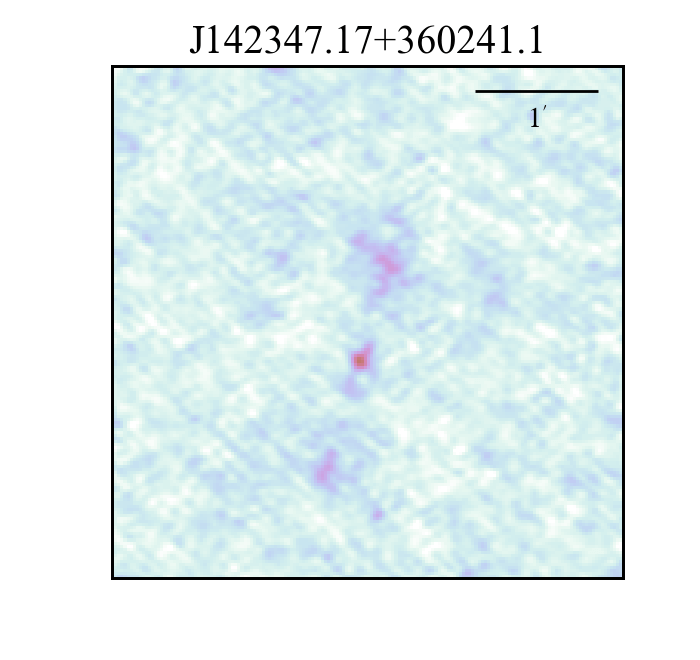}
\includegraphics[width=0.245\textwidth, trim=0.5cm 0.5cm 0.5cm 0.cm, clip]{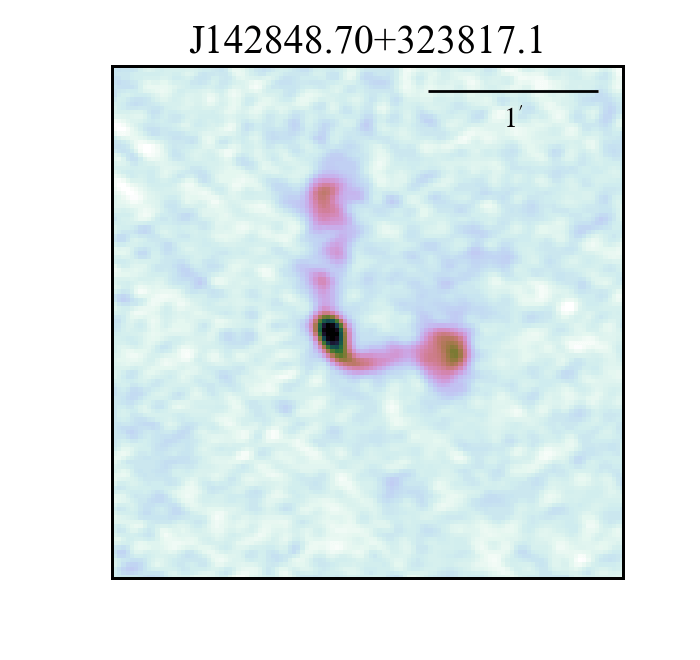}
\\
\includegraphics[width=0.245\textwidth, trim=0.5cm 0.5cm 0.5cm 0.cm, clip]{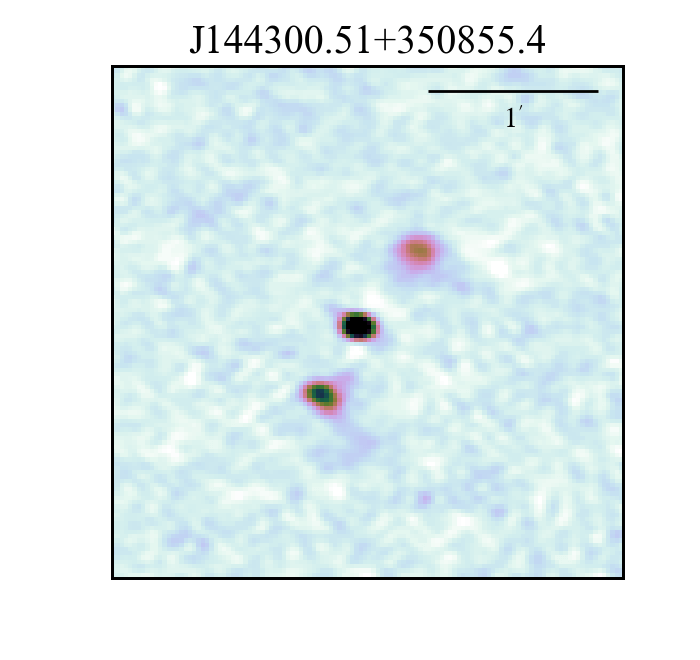}
\includegraphics[width=0.245\textwidth, trim=0.5cm 0.5cm 0.5cm 0.cm, clip]{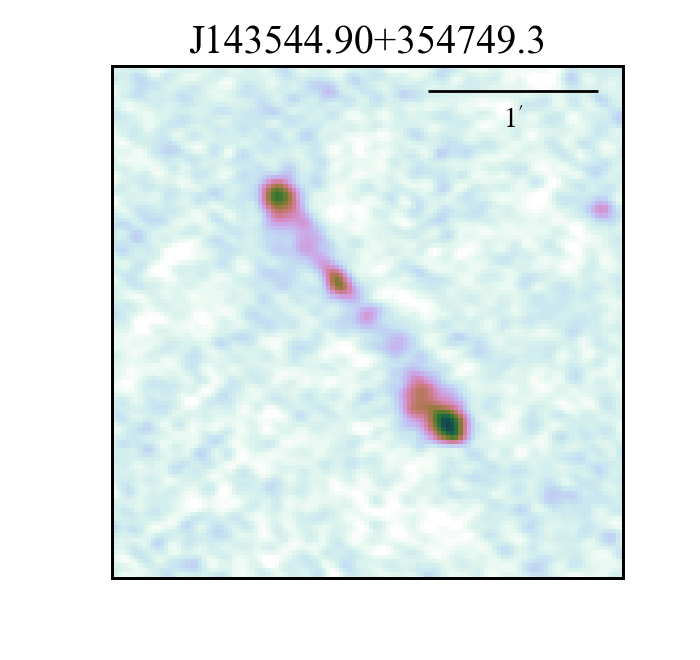}
\includegraphics[width=0.245\textwidth, trim=0.5cm 0.5cm 0.5cm 0.cm, clip]{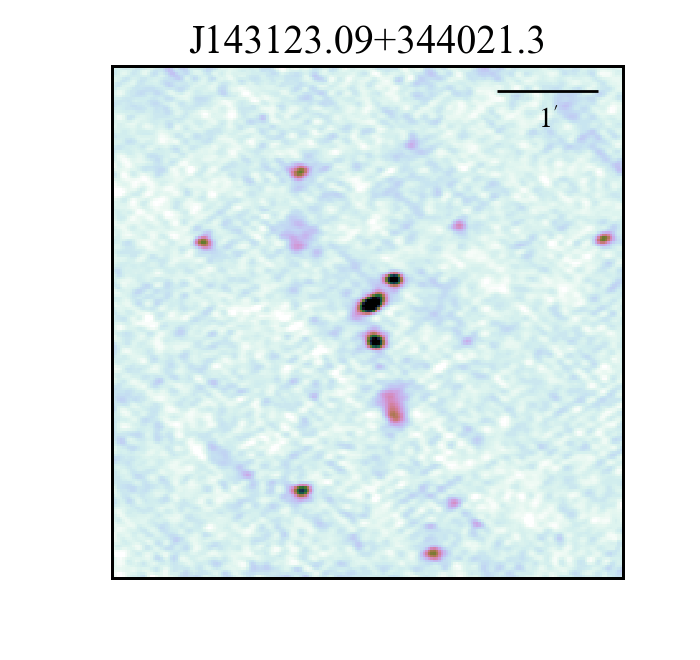}
\includegraphics[width=0.245\textwidth, trim=0.5cm 0.5cm 0.5cm 0.cm, clip]{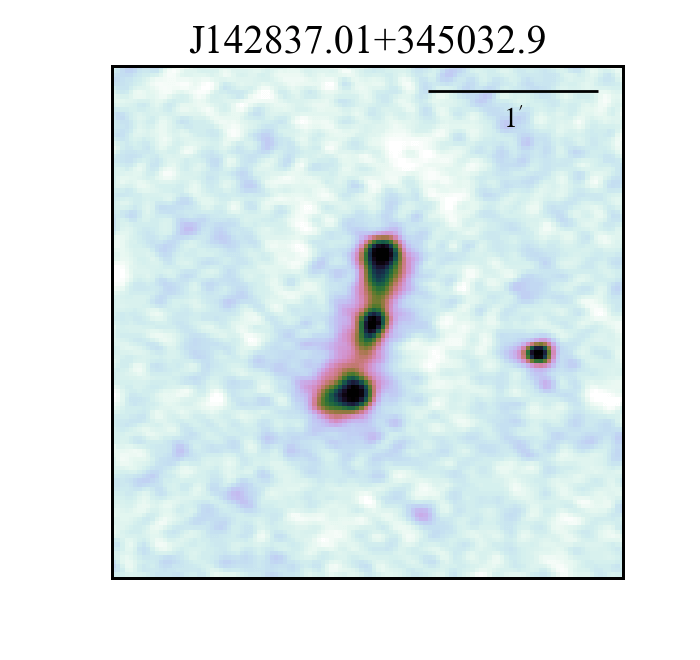}
\\
\contcaption{}
\end{figure*}
\begin{figure*}
\centering
\includegraphics[width=0.245\textwidth, trim=0.5cm 0.5cm 0.5cm 0.cm, clip]{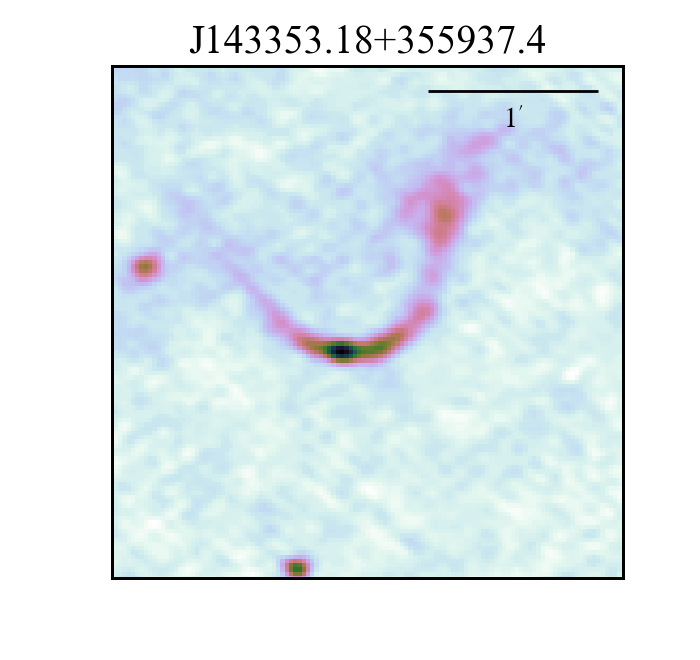}
\includegraphics[width=0.245\textwidth, trim=0.5cm 0.5cm 0.5cm 0.cm, clip]{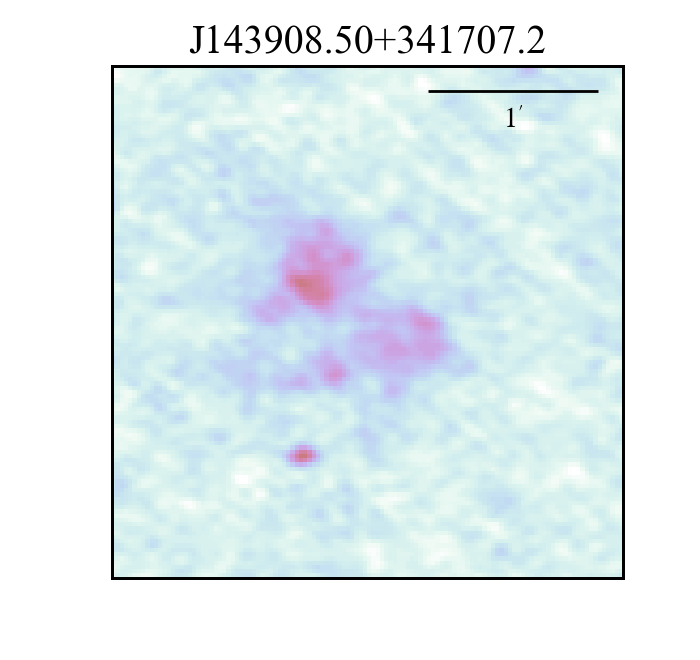}
\includegraphics[width=0.245\textwidth, trim=0.5cm 0.5cm 0.5cm 0.cm, clip]{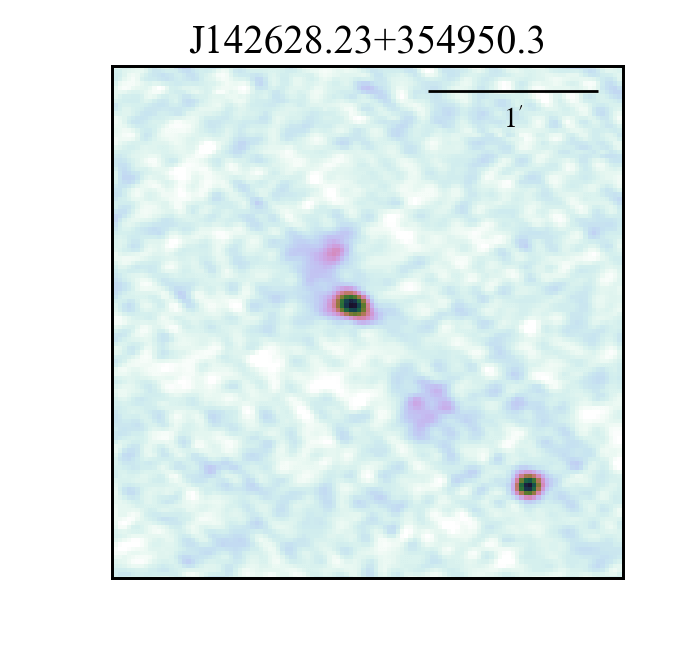}
\includegraphics[width=0.245\textwidth, trim=0.5cm 0.5cm 0.5cm 0.cm, clip]{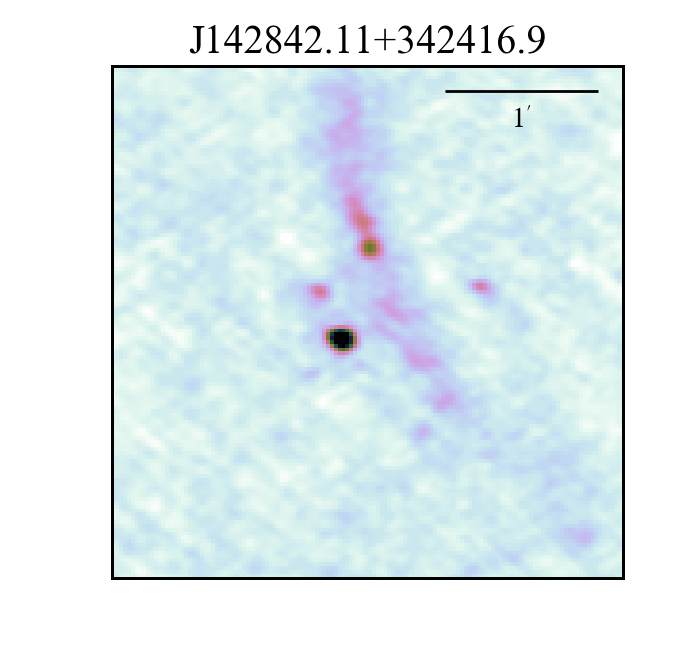}
\\
\includegraphics[width=0.245\textwidth, trim=0.5cm 0.5cm 0.5cm 0.cm, clip]{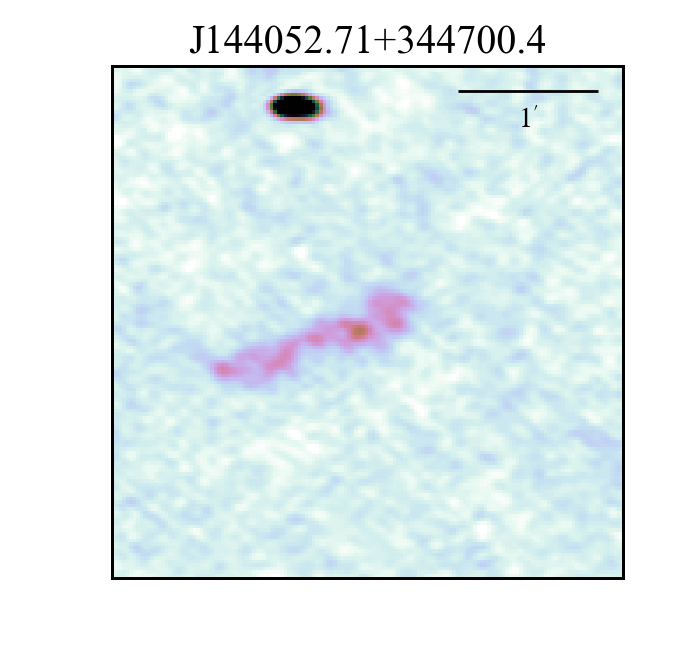}
\\
\contcaption{}
\end{figure*}